%% file: chapter_arxiv.tex
\documentclass[12pt]{extarticle}
\pdfoutput=1
\usepackage[utf8]{inputenc}
\usepackage{amsmath, amssymb, amsfonts}
\usepackage{natbib}
\usepackage{graphicx}
\usepackage{hyperref}
\usepackage{tikz}
\usetikzlibrary{bayesnet}
\usetikzlibrary{positioning}
\usepackage{algorithm}
\usepackage[noend]{algorithmic}
\input{preamble.tex}

\title{Inference of Dynamic Regimes \\ in the Microbiome}
\author{Kris Sankaran and Susan P. Holmes}
\begin{document}
\maketitle

\begin{abstract}
Many studies have been performed to characterize the dynamics and stability of
the microbiome across a range of environmental contexts
\citep{costello2012application, stein2013ecological, faust2015metagenomics}. For
example, it is often of interest to identify time intervals within which certain
subsets of taxa have an interesting pattern of behavior. Viewed abstractly,
these problems often have a flavor not just of time series modeling but also of
regime detection, a problem with a rich history across a variety of
applications, including speech recognition \citep{fox2011sticky}, finance
\citep{lee2009optimal}, EEG analysis \citep{camilleri2014automatic}, and
geophysics \citep{weatherley2002relationship}. However, in spite of the
parallels, regime detection methods are rarely used in microbiome data analysis,
most likely due to the fact that references for these methods are scattered
across several literatures, descriptions are inaccessible outside limited
research communities, and implementations are difficult to come across. Finally,
the correspondence between regime detection for these applications and for the
microbiome is not always obvious.

We distill the core ideas of different regime detection methods, provide example
applications, and share reproducible code, making these techniques more
accessible to microbiome researchers. Specifically, we reanalyze the data of
\cite{dethlefsen2011incomplete}, a study of the effects of antibiotics on the
microbiome, using Classification and Regression Trees (CART)
\citep{breiman1984classification}, Hidden Markov Models (HMMs)
\citep{rabiner1986introduction}, Bayesian nonparametric HMMs
\citep{teh2010hierarchical, fox2008hdp}, mixtures of Gaussian Processes (GPs)
\citep{rasmussen2002infinite}, switching dynamical systems
\citep{linderman2016recurrent}, and multiple changepoint detection
\citep{fan2015empirical}. Along the way, we summarize each method, their
relevance to the microbiome, and the tradeoffs associated with using them.
Ultimately, our goal is to describe types of temporal or regime switching
structure that can be incorporated into studies of microbiome dynamics.
\end{abstract}

\tableofcontents

\section{Introduction}

Microbiome data describe the abundances of bacterial species across multiple
samples. In previous chapters, we have studied latent structure across species
and samples separately. For example, we have developed interactive visualization
techniques to compare subsets of species, and we have applied mixed-membership
models to characterize variation in samples. In contrast, our goal here is to
study latent structure across species and samples simultaneously. This
difference is analogous to the change in perspective obtained by studying a
coclustering rather than two separate clusterings, or an ordination biplot
instead of simply the scores or loadings. We will focus on the case where
samples are collected over time, so that this problem can be understood as one
of detecting dynamic regimes, as explained in Section
\ref{sec:problem_description}

The primary contributions of this chapter are,
\begin{itemize}
\item The relation of the regime detection problem to several statistical
  frameworks, and a comparison of the types of interpretation facilitated by
  each.
\item The development of experiments to evaluate the practical utility of these
  different formulations.
\item A catalog of algorithm pseudocode and complete implementations, to serve as a
  reference for researchers interested in regime detection.
\item The design of and code for static visualizations that can be used to
  evaluate the results of various methods.
\end{itemize}

In Section \ref{sec:problem_description}, we describe the scientific problem of
interest in more detail and provide a high-level statistical formulation. In
Section \ref{sec:baseline}, we describe approaches which are easy to implement,
but that fail to incorporate temporal structure -- these serve as reference
points for evaluating more complex models. Sections
\ref{sec:smooth_temporal_models} and \ref{sec:temporal_mixture_models}
review and apply smoothing and mixture modeling techniques to this problem.

\section{Problem description}
\label{sec:problem_description}

In latent variable modeling, our ultimate goal is a succinct representation of
complex data. In a way, we can think of the reduced representations as a type of
data compression for human interpretation, and as in any (lossy) compression,
there is a choice of what structure to preserve. Different reduced
representations facilitate different comparisons -- for example, clustering
bacteria allows easy comparison of taxonomic categories, while clustering
samples allows a comparison of full community states.

In the regime detection problem, the comparisons we would like to support are
\begin{itemize}
\item For each species, should different time intervals be assigned to different
  dynamic regimes?
\item Should we distinguish subsets of species as having similar patterns of
  behavior, with respect to these regimes?
\end{itemize}

Conceretely, we may expect that over the time course of a study, individual
species may switch between stable or unstable, increasing or decreasing, and
present or absent regimes, either due to natural ecological dynamics or
experimentally induced perturbations. We would like to detect these alternative
regimes automatically.

Further, as we often work with hundreds or thousands of bacterial species at a
time, we would like to group or relate species according to these regimes, so
that (1) we do not need to inspect regime switching behavior for individual
species one by one and (2) we can achieve gains in power by pooling across
species. The resulting subsets of species can be related to available taxonomic
information to draw scientific conclusions about the behavior of different taxa
during different sampling periods. For example, for the data of
\cite{dethlefsen2011incomplete}, we might be interested in conclusions like
``$x$\% of species $y$ exhibited stability during the first half of the
antibiotic time course, while the rest showed decreasing abundances.''

We can frame this analysis problem using the language of latent variables. Let
$\left(t_{i}\right)_{i = 1}^{n}$ be the sampling timepoints, and index species
by $\left(s_{j}\right)_{j = 1}^{p}$. Our goal is to infer a function $\theta$
mapping time by species pairs $\left(t_{i}, s_{j}\right)$ to an associated
latent state, which can be discrete, continuous, or a mixed membership.

We expect the function $\theta$ to be reasonably well-behaved over time.
Further, by comparing $\theta\left(\cdot, s_{j}\right)$ across different species
$j$, we can group or sort species according to their regime membership behavior.

\section{Methods baseline}
\label{sec:baseline}

\subsection{Hierarchical clustering}

As a baseline, Figure \ref{fig:heatmap-euclidean} provides a heatmap where
species are ordered according to a hierarchical clustering tree. Note that
clustering trees are invariant under left-right swaps of node children; to fix a
unique tree, we order branches so that the average abundance of the left subtree
is larger. The resulting ordered heatmap could potentially resolve partitions in
the species by time space. A limitation of this approach is that it does not
provide any clustering for the timepoints, even if blocks across timepoints seem
to appear in the resulting heatmap. Coclustering species and timepoints is not a
sufficient alternative, because the blocking across timepoints must respect the
known temporal order. On the other hand, advantages of this approach are that it
is simple to implement and interpret.

The figure generated by hierarchical clustering is sensitive to several choices,
\begin{itemize}
\item Transformations: Different transformations might be more effective
  representations of the underlying data.
\item Distance: Different distances accentuate different types of variation.
\end{itemize}

For example, some natural transformations are
\begin{itemize}
\item $\asinh$: Raw count data in microbiome data sets tend to
  be heavy tailed, but with a large spike at 0. An $\asinh$ transformation
  behaves like a $\log$-transformation for large counts, but goes through the
  origin -- this downweights coordinates with large counts. This can be seen
  from the representation $\asinh\left(x\right) = \logarg{x + \sqrt{1 -
      x^{2}}}$.
\item Innovations: Rather than clustering the raw series, we can cluster the
  first differenced series. This will cluster series that have similar changes between
  neighboring timepoints, even when their overall values are quite different. This
  can highlight bacterial series with similar dynamics, at the cost of ignoring
  differences in overall abundances.
\item Binarized: We can transform the series $x_{i}$ into $\indic{x_{i} > 0}$.
  This loses substantial information, but measuring differences in
  presence-absence patterns in microbiome studies can be scientifically
  meaningful.
\end{itemize}

On these transformed data, we now need to compute pairwise distances between
series. Three choices that we consider are,
\begin{itemize}
\item Euclidean: This distance is ideal when clusters have a spherical shape.
\item Jaccard: This is a distance between pairs of length $p$ binary sequences
  $x_{i}$ and $x_{j}$ defined as
\begin{align*}
  d\left(x_{i}, x_{j}\right) &= 1 - \frac{\sum_{k = 1}^{p} \indic{x_{ik} = x_{jk} = 1}}{p},
\end{align*}
or one minus the fraction of coordinates that are both 1. The motivation for this
distance is that coordinates that are both 0 should not contribute to similarity
between sequences, especially series that are dominated by 0s. We apply this
distance to the binarized version of the species counts.
\item Mixture: Since any convex combination of distances is still a distance, we
  can define mixtures of distances that reflect several characteristics of the
  data.
\end{itemize}

In general, we have no quantitative approach for comparing the clusterings
obtained by different distances. Instead, we compare the resulting heatmaps,
noting how different patterns are identified by different approaches.

\subsubsection{Example}
\label{sec:hclust_example}

\begin{figure}
  \centering
  \includegraphics[width=0.9\textwidth]{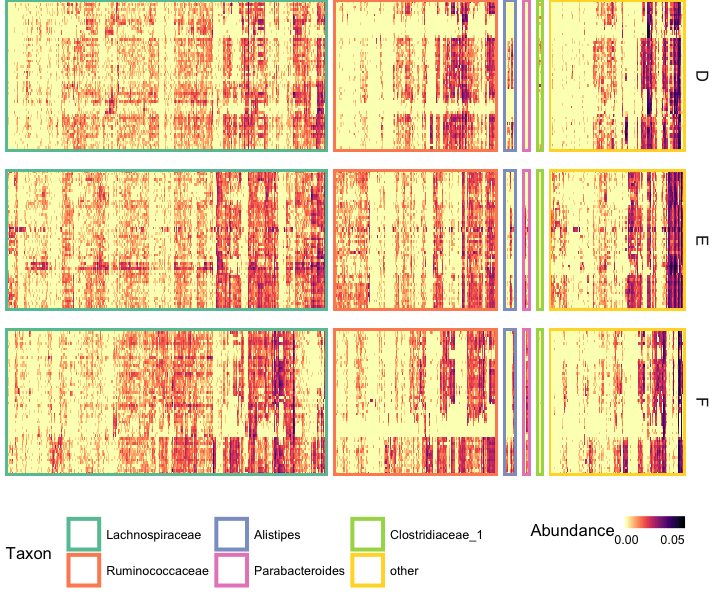}
  \caption{The three main row panels correspond to the three study subjects.
    Each column gives the time series of abundances for a single species, with
    time evolving from bottom to top and abundance indicated by heatmap shade
    intensity. Species are grouped into boxes according to their taxonomic
    family, and within boxes they are ordered according to their position on the
    hierarchical clustering tree.
    \label{fig:heatmap-euclidean} }
\end{figure}

\begin{figure}
  \centering
  \includegraphics[width=0.9\textwidth]{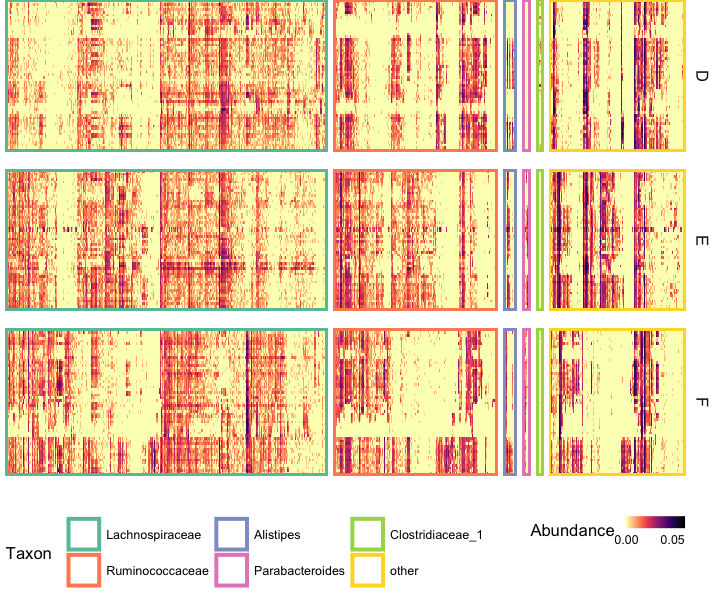}
  \caption{The analog of Figure \ref{fig:heatmap-euclidean} for the Jaccard
    distance. Note that species with very different abundances may be grouped
    together, as long as they have the same patterns of presence and absence.
    \label{fig:heatmap-jaccard} }
\end{figure}

Now we use hierarchical clustering the data of \cite{dethlefsen2011incomplete}.
The heatmaps in Figures \ref{fig:heatmap-euclidean} through
\ref{fig:heatmap-jaccard} describe which species have similar behaviors
according to Euclidean and Jaccard distances, respectively. See Supplementary
Figures \ref{fig:heatmap-mix} through \ref{fig:heatmap-innovations-bin} for
variations on the distance and transformation. The most important takeaways from
these figures are that some groups of bacteria are more strongly affected by the
antibiotic treatment, and there is variation in the time it takes to recover.

As expected, different distances group series according to the features that
define the distance. For example, the Jaccard distance, used in Figure
\ref{fig:heatmap-jaccard}, groups series with similar zero patterns, even if
their associated abundances are very different. On the other hand, the Euclidean
distance tends to group series with similar averages, and there is less blocking
by presence-absence structure. Note that patterns of response to antibiotics
form subclusters within taxonomic families. This suggests that while phylogeny
certainly relates to patterns of abundance, variation occurs at levels of
granularity more subtle than family level.

\begin{figure}
  \centering
  \includegraphics[width=0.9\textwidth]{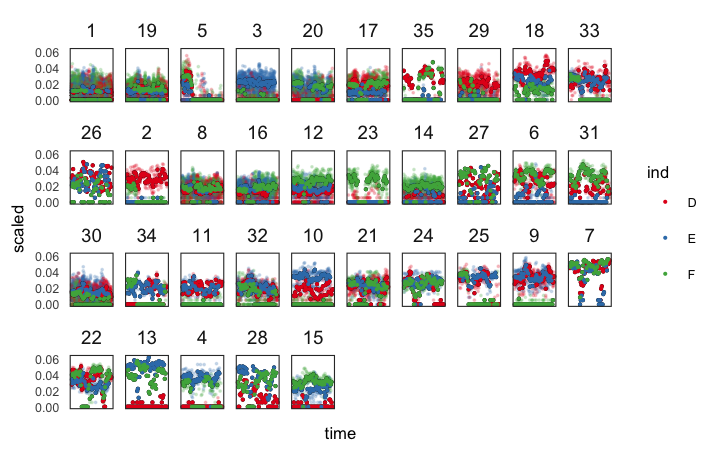}
  \caption{Centroids of the subclusters identified by the heatmap. Panels
    correspond to subclsuters defined by cutting the hierarchical clustering
    tree. Different individuals are represented by different colors. The solid
    points are the time series of averages of positive species abundances for a
    given cluster, per individual. Semitransparent points are the raw values
    used to compute these averages. These centroids provide context for
    interpreting the leaves in the original hierarchical clustering.
    \label{fig:centroid-euclidean-conditional} }
\end{figure}

\begin{figure}
  \centering
  \includegraphics[width=0.9\textwidth]{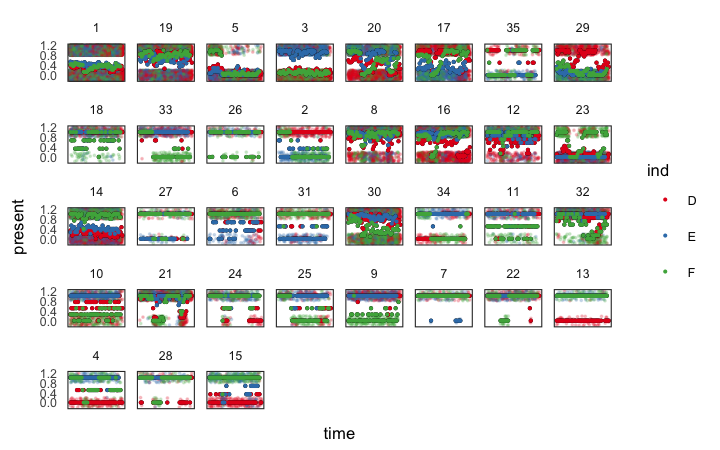}
  \caption{The analog of Figure \ref{fig:centroid-euclidean-conditional},
    looking at the running averages of presence and absence of species within
    different subclusters. The values of the solid points can be interpreted as
    the probability that a species within that cluster will be present, as a
    function of time and subject.
    \label{fig:centroid-euclidean-presence} }
\end{figure}

To summarize behavior within clusters, we display the centroids in Figures
\ref{fig:centroid-euclidean-conditional} \ref{fig:centroid-euclidean-presence}.
Evidently, some of the clustering structure is due to the presence of species
within only some of the subjects. Further, differential responses can be seen
within some of the panels. For example, cluster 17 in Figure
\ref{fig:centroid-euclidean-conditional} includes bacteria that are affected by
the first antibiotics time course, but only for subjects D and F, and which are
only affected in subject D during the second time course.

The fact that the data include many zeros makes the mean a somewhat misleading
cluster summary. Instead, we can decompose the summary into a presence-absence
component and a conditional-on-presence component. The presence-absence
component computes the proportion of bacteria that are present at any given
timepoint, while the conditional-on-presence component computes the time
averages among bacteria with strictly positive abundance.

\subsection{CART}

While placing similar time series near to one another suggests time windows
where the series behave similarly to one another, the hierarchical clustering
approach presented above does not explicitly partition timepoints into windows
with distinct behaviors. In this section, we consider an alternative that
provides such a partitioning.

The main idea is that a regression tree that uses $X$ to predict $y$ provides a
partition of the space in which $X$ lives, where $y$ has lower variation within
than between partitions \citep{breiman1984classification}. We apply the
hierarchical clustering approach from the previous section to obtain an ordering
$s_{j} \in \{1, \dots, n_{\text{species}}\}$ across species. We write the
timepoint for the $i^{th}$ sample as $t_{i}$ Then, we model the count for the
$j^{th}$ species in the $i^{th}$ sample as $y_{ij} \approx f\left(s_{j},
t_{i}\right)$ where $f \in \mathcal{T}$, the space of decision trees. The output
of interest is the fitted partition on $\left(s_{j}, t_{i}\right)$.

For completeness, we review the CART algorithm. Following \citep{stat315bnotes},
we can describe it in terms of (1) the structural model, (2) the score criterion
for choosing between models, and (3) the search strategy used to produce the
final fit.

The structural model $\mathcal{F}$ is the class of functions that can be
expressed as
\begin{align*}
f\left(x\right) &= \sum_{m = 1}^{M} c_{m} \indic{x \in R_{m}},
\end{align*}
where $\left(R_{m}\right)_{m = 1}^{M}$ is a partition of the covariate space
and $c_{m} \in \reals$ are constants associated with each partition element.

For regression, the criterion is the expected generalization squared-error
between $y_{i}$ and the $c_{m}$ associated with the partition element in which
$x_i$ lies -- we will denote this by $c\left(x_{i}\right)$. In classification,
the corresponding criterion is the expected missclassification error. More
precisely, we calculate empirical risks,
\begin{align*}
  \frac{1}{n} \sum_{i = 1}^{n} \left(y_{i} - c\left(x_{i}\right)\right)^{2},
\end{align*}

for regression and
\begin{align*}
  \frac{1}{n} \sum_{i = 1}^{n} L_{y_{i}, c\left(x_{i}\right)} \indic{y_{i} \neq c\left(x_{i}\right)},
\end{align*}
for classification, where $L_{kk^{\prime}}$ is the loss induced by
missclassifying $k$ as $k^{\prime}$. Since the empirical risk on the training
data underestimates generalization error, these estimates are constructed on
test sets.

To obtain a fitted $\hat{f} \in \mathcal{F}$, the algorithm must identify a
partition $\left(R_{m}\right)$ and constants $c_{m}$ that (approximately)
minimize the score criterion. Given a fixed $R_{m}$, fitting $c_{m}$ is
straightforwards, since the score decouples across partition elements -- in
regression the minimizers are averages of the $y_{i}$ within partition elements,
while in classification they are majority votes. On the other hand, finding the
optimal $\left(R_{m}\right)$ is an intractable combinatorial optimization
problem, and the search strategy resorts to a greedy approach in hope of a
reasonable local minimizer.

More precisely, the final partition is found by recursively splitting the input
space and then pruning away splits that seem less relevant. At the first step,
the partition consists of a single $R_{1}$, equal to the entire covariate space,
and it is split on covariate $j^{\ast}$ at position $t_{j1}^{\ast}$, chosen to
solve the optimization
\begin{align*}
  \left(j^{\ast}, t_{j1}^{\ast}\right) &= \arg \min_{\substack{j = 1, \dots, p \\ t_{j1} \in \reals}} \sum_{i \in R_{1, l}} \left(y_{i} - \bar{y}_{l}\right)^{2} + \sum_{i \in R_{1, r}} \left(y_{i} - \bar{y}_{r}\right)^{2}
\end{align*}
where $R_{1, l}$ and $R_{1, r}$ (for ``left'' and ``right'') are a splitting of
$R_{1}$ along feature $j$ feature at position $t_{j1}$.

This procedure is iterated recursively. That is, at the $m^{th}$ iteration, the
next split solves nearly the same optimization,

\begin{align*}
  \left(m^{\ast}, j^{\ast}, t_{jm}^{\ast}\right) = \arg \min_{\substack{j = 1, \dots, p \\ t_{jm} \in \reals}}
  \sum_{i \in R_{m, l}} \left(y_{i} - \bar{y}_{l}\right)^{2} +
  \sum_{i \in R_{m, r}} \left(y_{i} - \bar{y}_{r}\right)^{2},
\end{align*}

where the main difference is that we must now choose which of the $m$ previously
defined partition elements to split.

This is done for some prespecified number of splits, $M$. This partition is
often unecessarily highly resolved, and it can improve generalization
performance to introduce a pruning step. Let $C_{m}$ denote the ``cost'' of a
(potentially intermediate) partition element $R_{m}$, defined by
\begin{align*}
  C_{m} &= \begin{cases}
    \hat{r}_{m} + k & \text{if $m$ was never split} \\
    \sum_{m^\prime} C_{m^{\prime}} & \text{if $m$ was split into two $m^{\prime}$}
    \end{cases}
\end{align*}
where $\hat{r}_{m} = \sum_{x_{i} \in R_{m}} \left(y_{i} - c_{m}\right)^{2}$. $k$
can be interpreted as the amount of improvement to the score criterion that the
split must provide in order to be accepted, and it is fixed in advance. The
final partitioning is obtained by choosing to split or merge each intermediate
partition element $R_m$ so that $C_{m}$ is minimized. Specifically, if
$\hat{r}_{m} + k < \sum_{m^{\prime}} C_{m^{\prime}}$, then all descendant nodes
(subpartition elements) are merged into $R_{m}$. Otherwise, the left and right
splits are accepted. These choices are made sequentially, from the bottom up.

\subsubsection{Example}
\label{subsubcart_example}

\begin{figure}
  \centering
  \includegraphics[width=0.8\textwidth]{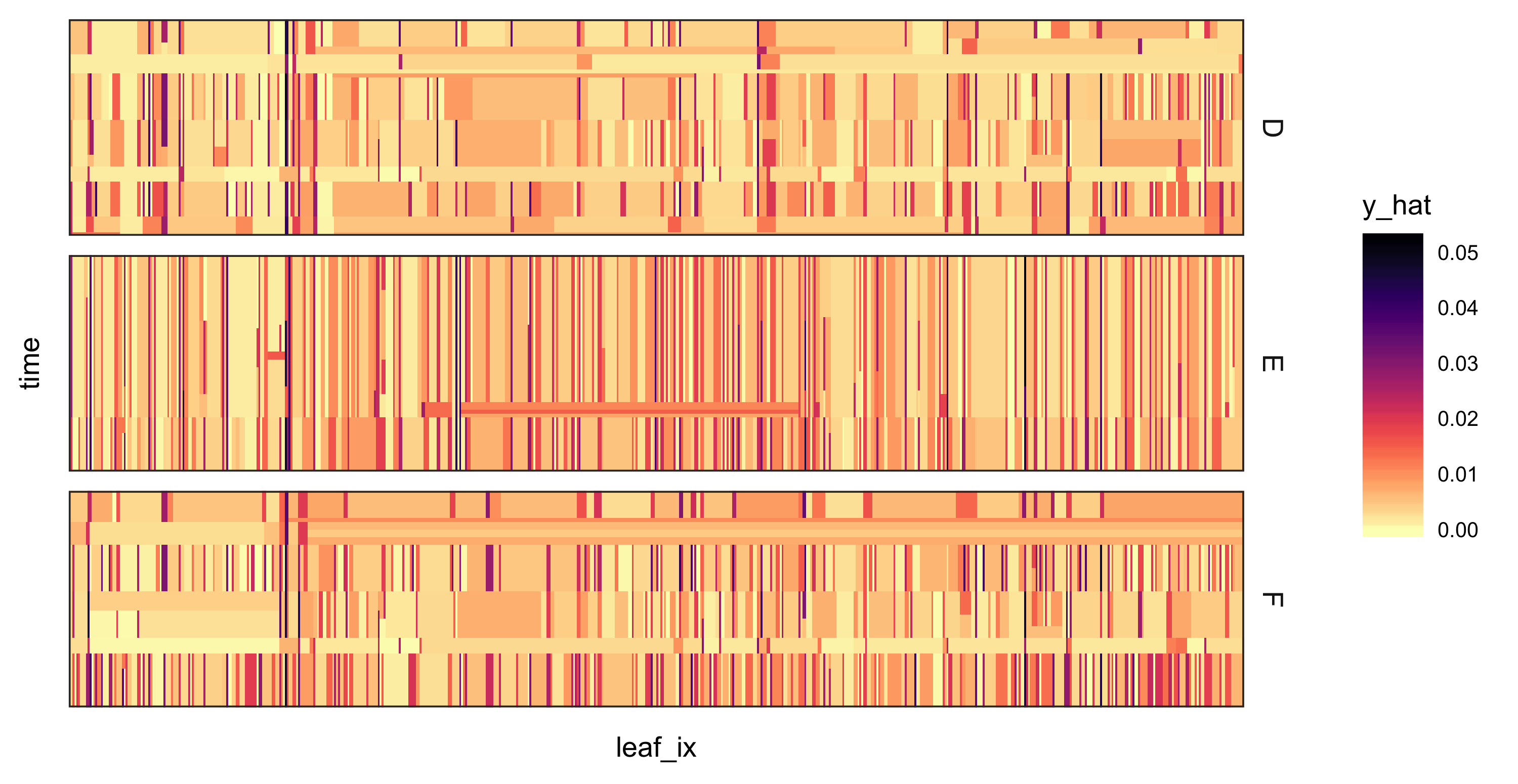}
  \caption{As in the heatmaps produced by hierarchical clustering, the three
    vertical panels represent three subjects, while rows and columns correspond
    to species and temporally-ordered samples, respectively. The rows have been
    ordered according to a hierarchical clustering. The shading within partition
    blocks corresponds to the fitted $\hat{c}_{m}$ from the regression tree.
    \label{fig:rpart_complex}
  }
\end{figure}

Example output from this approach is provided in Figure \ref{fig:rpart_complex}.
Note the presence of two light vertical stripes in subjects D and F -- these
correspond to the two antibiotic treatment regimes. In general, ``tall'' blocks
are more common than ``wide'' ones. This reflects the fact that timepoints
within bacteria tend to have more similar abundances, compared to multiple
bacteria at a single timepoint, even when those bacteria have been ordered by a
hierarchical clustering.

One interesting detail is the delayed, or sometimes nonexistent, recovery after
the first antibiotic time course among a cluster of bacteria near the bottom-left of
the panel for subject F. This long-term impact of antibiotics on bacterial
populations, at least in one subject, was a central finding in
\cite{dethlefsen2011incomplete}. Observe that a similar pattern is visible after
the second antibiotic time course among subject D, also for species near the
bottom-left of the panel.

Related views are given in Supplemental Figures \ref{fig:rpart_simple} through
\ref{fig:rpart_conditional}. Supplemental Figures \ref{fig:rpart_binary_simple}
and \ref{fig:rpart_conditional} decompose the CART prediction problem into a
binary part and a conditional-on-present component, using the hurdle heuristic.

Two limitations become clear in this example application. First, partitioning
across species seems hard to obtain simultaneously across all subjects -- in
Figure \ref{fig:rpart_complex}, there seem to be no ``wide'' blocks for Subject
E. This is a consequence of first ordering all species according to a
hierarchical clustering based on all subjects. A potential solution would be to
cluster the species separately across subjects, trading off the ability to match
the same species across several panels in order to better study species blocking
within subjects.

Second, these global views of the data make it difficult to inspect individual
species and sample identities. Potential solutions are (1) link necessary
supplemental (e.g., taxonomic) information within the static view, through a
shaded taxonomic-family stripe, as in Figure \ref{fig:heatmap-euclidean} (2)
construct an interactive version of these plots, where hovering over a partition
element provides focused information about that element (e.g., the species
identities it contains), in the spirit of our \texttt{centroidview}
package\footnote{\url{https://github.com/krisrs1128/centroidview}}.

\section{Smooth temporal models}
\label{sec:smooth_temporal_models}

In Section \ref{sec:baseline}, we described distance and regression-based
techniques to approaching the questions outlined in Section
\ref{sec:problem_description} of identifying which subsets of species have
similar abundance profiles during which windows of time. In particular, we have
avoided direct modeling of abundances across species or over time. In this
section, we adopt this alternative direct modeling approach, using the distance
and regression-based methods of Section \ref{sec:baseline} as a reference for
what analysis is possible with minimal effort.

We first review two fundamental approaches to probabilistic temporal modeling
which are used as building blocks for the more advanced methods in Section
\ref{sec:temporal_mixture_models}: Linear Dynamical Systems in Section
\ref{subsec:linear_dynamical_systems}) and we consider Gaussian Processes in
Section \ref{subsec:gaussian_processes}). These approaches are designed for
single time series, or collections of independent ones. However, models that
consider collections of related time series can be constructed from these, by
introducing latent variables.

\subsection{Gaussian Processes}
\label{subsec:gaussian_processes}

\begin{figure}
  \centering
  \begin{tikzpicture}
    \node[obs] (y) {$y$};
    \node[latent, below=0.7 of y, xshift=-0.4cm] (f) {$f$};
    \node[obs, left=of y, xshift=-0.4cm]  (x) {$x$};
    \node[const, right=of y, xshift=-0.2cm]  (sigma) {$\sigma^2$};
    \node[const, left=0.4 of f] (theta) {$\theta$};

    \edge {x, f, sigma} {y} ; %
    \edge {theta} {f} ; %

    \plate {yx} {(x)(y)} {$n$} ;
  \end{tikzpicture}

  \caption{Gaussian Process plate diagram.\label{fig:gp_plate} }
\end{figure}
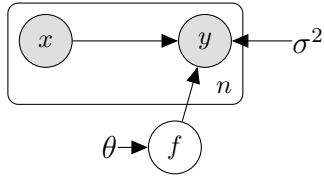

Gaussian Processes (GPs) provide a prior over classes of stationary, smoothly
varying functions. Their appeal as a building block for probabilistic models
lies in the fact that they are nonparametric -- they can adapt to more complex
functions as more data arrives -- while still admitting tractable inference. One
of the simplest GPs models observations $\left(x_{i}, y_{i}\right) \in
\reals^{p}\times \reals$ as
\begin{align*}
  y_{i} &= f\left(x_{i}\right) + \eps_{i}
\end{align*}
where $f \sim GP\left(m, \kappa\right)$, meaning that for any collections of
covariates $\left(x_{1}, \dots, x_{n}\right)$, we have
\begin{align*}
  \left(f\left(x_{1}\right), \dots, f\left(x_{n}\right)\right) &\sim
  \Gsn\left( \begin{pmatrix} m\left(x_{1}\right) \\ \vdots \\ m\left(x_{n}\right) \end{pmatrix}, \begin{pmatrix} \kappa\left(x_{1}, x_{1}\right) & \dots & \kappa\left(x_{1}, x_{n}\right) \\ \vdots & & \vdots \\ \kappa\left(x_{n}, x_{1}\right) & \dots &\kappa\left(x_{n}, x_{n}\right) \end{pmatrix}\right).
\end{align*}
$m$ and $\kappa$ are called the mean and covariance functions, respectively. We
will denote the covariance matrix of $\kappa$ applied to pairs of $x_{i}$ by
$K\left(x, x\right)$. A plate diagram representing this model is provided in
Figure \ref{fig:gp_plate}.

It is common to initially center the data before analysis, in which case we can
assume $m \equiv 0$. Further, any positive-definite covariance function $\kappa$
can be used -- a common choice is the Gaussian covariance,
\begin{align*}
\kappa_{\sigma_{f}, M}\left(x_{p}, x_{q}\right) &= \sigma_{f}^{2}\exp{-\frac{1}{2}\left(x_{p} - x_{q}\right)^{T}M\left(x_{p} - x_{q}\right)},
\end{align*}
where $M$ can be $\frac{1}{l^{2}}I_{n}$, which assumes similar degrees of
smoothness across all coordinates, $\diag\left(l\right)^{-2}$, which allows
different smoothness along different axes, or $\Lambda \Lambda^{T} +
\diag\left(l\right)^{-2}$, which allows variation along directions that are not
axis-aligned. While the Gaussian covariance provides a reasonable default, it is
good practice to adapt the covariance function to the data problem at hand,
accounting for seasonality, multiple scales of variation, or the presence of
outliers, see Section 5.4.3 of \citep{rasmussen2006gaussian} for an in depth
example. Note that the covariance function is responsible for the GPs poor
ability to model transient events, compared to smooth trajectories. However,
multiscale behavior can be modeled by introducing mixtures, as we will see in
Section \ref{sec:temporal_mixture_models}.

In this model, the posterior of $f$ is available analytically, by virtue of
Gaussian-Gaussian conjugacy. Consider evaluating the function $f$ at new points
$x_{\ast}$, denoted by $f_{\ast} := f\left(x_{\ast}\right)$. The joint of $y$
and $f^{\ast}$ is
\begin{align*}
  \begin{pmatrix}
    y \\ f_{\ast}
  \end{pmatrix} &= \Gsn\left(
 \begin{pmatrix}
   0 \\ 0
 \end{pmatrix} ,
\begin{pmatrix}
  K\left(x, x\right) + \sigma^{2}I_{n} & K\left(x, x_{\ast}\right) \\
  K\left(x_{\ast}, x\right) & K\left(x_{\ast}, x_{\ast}\right)
\end{pmatrix}
  \right),
\end{align*}
which yields the posterior,
\begin{align*}
  f_{\ast} \vert y &\sim \Gsn\left(\Earg{f_{\ast} \vert y}, \Covarg{f_{\ast} \vert y}\right),
\end{align*}
where
\begin{align*}
  \Earg{f_{\ast} \vert y} &= K\left(x_{\ast}, x\right)\left(K\left(x, x\right) + \sigma^{2}I_{n}\right)^{-1}y \\
  \Covarg{f_{\ast} \vert y} &= K\left(x_{\ast}, x_{\ast}\right) - K\left(x_{\ast}, x\right)\left(K\left(x, x\right) + \sigma^{2}I_{n}\right)^{-1}K\left(x, x_{\ast}\right).
\end{align*}
Note the $n\times n$ matrix inversion in the covariance calculation. This is the
source of the $O\left(n^{3}\right)$ complexity of using standard GPs, though a
variety of fast approximations have been proposed, exploiting the sparse,
banded, or block structure often present in covariance matrices
\citep{quinonero2007approximation}.

In this computation, we have assumed the the kernel hyperparameters $\theta$ are
known\footnote{For example, in the Gaussian covariance case, this has the form
  $\theta = \left(\sigma_{f}^{2}, M\right)$}. In reality, these must be inferred
from the data. Two standard approaches are based on maximizing (1) the marginal
likelihood of $y$ and (2) the cross-validated predictive likelihood. The first
approach leverages the fact that the marginal likelihood,
\begin{align*}
  \log p\left(y \vert x; \theta\right) &= -\frac{n}{2}\log 2\pi - \log\absarg{K_{\theta}\left(x, x\right) + \sigma^{2}I_{n}} - \frac{1}{2}y^{T}\left(K_{\theta}\left(x, x\right) + \sigma^{2}I_{n}\right)^{-1}y
\end{align*}
and its gradients over $\theta$ have a closed form, and so can be optimized.

The cross-validation approach instead maximizes the average predicted log probability,
\begin{align*}
\sum_{i = 1}^{n} \log p\left(y_{i} \vert x, y_{-i}; \theta\right),
\end{align*}
which can also be found analytically, by conditioning the marginal for $y$.


\subsection{Linear Dynamical Systems}
\label{subsec:linear_dynamical_systems}

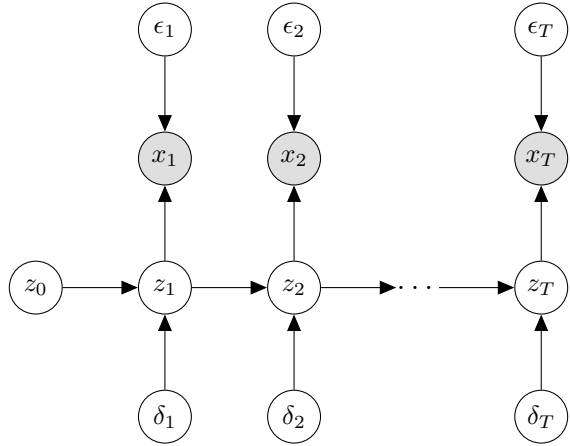
\begin{figure}
  \centering
  \begin{tikzpicture}
    \node[latent] (z1) {$z_{1}$};
    \node[latent, left=of z1] (z0) {$z_{0}$};
    \node[obs, above=of z1] (x1) {$x_{1}$};
    \node[latent, above=of x1] (eps1) {$\eps_{1}$};
    \node[latent, below=of z1] (delta1) {$\delta_{1}$};
    \node[latent, right=of z1] (z2) {$z_{2}$};
    \node[obs, above=of z2] (x2) {$x_{2}$};
    \node[latent, above=of x2] (eps2) {$\eps_{2}$};
    \node[latent, below=of z2] (delta2) {$\delta_{2}$};
    \node[const, right=of z2] (zdots) {$\dots$};
    \node[latent, right=of zdots] (zT) {$z_{T}$};
    \node[latent, below=of zT] (deltaT) {$\delta_{T}$};
    \node[obs, above=of zT] (xT) {$x_{T}$};
    \node[latent, above=of xT] (epsT) {$\eps_{T}$};

    \edge {z1} {x1} ;
    \edge {z2} {x2} ;
    \edge {z0} {z1} ;
    \edge {z1} {z2} ;
    \edge {z2} {zdots} ;
    \edge {zdots} {zT} ;
    \edge {zT} {xT} ;
    \edge {eps1} {x1} ;
    \edge {eps2} {x2} ;
    \edge {epsT} {xT} ;
    \edge {delta1} {z1} ;
    \edge {delta2} {z2} ;
    \edge {deltaT} {zT} ;
  \end{tikzpicture}
  \caption{Plate diagram corresponding to a Gaussian LDS. $z_{t}$ denotes the
    underlying Markov state sequence, while $x_t$ are observed
    emissions. \label{fig:lds_graphical} }
\end{figure}

Linear Dynamical System models (LDSs) treat an observed time series as a
transformation of temporally evolving latent states.
There have been many proposals that allow general transformation and state
evolution behavior \citep{hostetler1983nonlinear, wan2000unscented}, but a
fundamental starting point is the Linear-Gaussian dynamical system,
\begin{align*}
  x_{t} &= C z_{t} + \eps_{t} \\
  z_{t} &= A z_{t - 1} + \delta_{t} \\
  \eps_{t} &\sim \Gsn\left(0, R\right) \\
  \delta_{t} &\sim \Gsn\left(0, Q\right).
\end{align*}
The $z_{t}$'s are a Markov chain of latent states, while the $x_{t}$'s represent
the observed emissions from it. $A$ and $Q$ govern the dynamics of the
underlying process, while $C$ and $R$ describe the emission structure. The
associated graphical model is provided in Figure \ref{fig:lds_graphical}.

There are two conceptual components to fitting this model,
\begin{itemize}
\item Inference: Even if $\Theta = \{A, C, Q, R\}$ were known, there
  is often value in estimating the latent $z_{i}$.
\item Learning: Typically, the parameters $\Theta$ are unknown, and must
  themselves be learned from the data.
\end{itemize}

Further, inference can be approached in several different ways, depending on
problem context and constraints. Among the most common approaches are
\begin{itemize}
\item Filtering: Update beliefs of the current latent state in an online
  fashion. Quantitatively, the goal is to estimate the distribution
  $p\left(z_{t} \vert x_{1:t}\right)$.
\item Smoothing: Use the full history to estimate beliefs of latent states at
  each time. This means to compute $p\left(z_{t} \vert x_{1:T}\right)$ for each
  $t = 1, \dots, T$.
\item Forecasting: Predict the next few latent states given all observations so
  far. For example, we might be interested in $p\left(z_{t + 1} \vert
  x_{1:t}\right)$.
\end{itemize}

We first describe inference in detail, before explaining how inference and
learning can be alternated to fit models on real data. Due to the
Linear-Gaussian assumption, the conditionals required by filtering and smoothing
are still Gaussian; however, their means and covariances are not immediately
apparent. It turns out that, for both filtering and smoothing, they can be
efficiently computed via dynamic programming recursions. In the literature,
these are referred to as the Kalman Filtering and Rauch-Tung-Striebel (RTS)
smoothing recursions.

Consider the filter, where the goal is calculation of $p\left(z_{t} \vert
x_{1:t}\right)$. We will assume $z_{1} \sim \Gsn\left(0, Q\right)$. We
could alternatively fix it to a constant -- the point is that the distribution
for $z_{1}$ must be known.

For $t > 1$, define the \textit{one-step-ahead} and \textit{updated} means and
covariances by
\begin{align*}
  \mu_{t \vert t - 1} &:= \Earg{z_{t} \vert x_{1:t - 1}} \\
  \Sigma_{t \ vert t - 1} &:= \Covarg{z_{t} \vert x_{1:t - 1}} \\
  \mu_{t} &:= \Earg{z_{t} \vert x_{1:t}} \\
  \Sigma_{t} &:= \Covarg{z_{t} \vert x_{1:t}}.
\end{align*}
The means $\mu_t$ and covariances $\Sigma_{t}$ are the main quantities of
interest in filtering. The filtering algorithm is detailed in in Algorithm
\ref{alg:kalman_filter}. It can be thought of as a forwards pass through the
observed sequence, updating means and covariances along the way.

The derivation is as follows. For the predict step, use the tower property and
the law of total variance,
\begin{align*}
  \Earg{z_{t} \vert x_{1:t - 1}} &= \Earg{\Earg{z_{t} \vert z_{t - 1}, x_{1:t - 1}}} \\
  & = \Earg{Az_{t - 1}\vert x_{1:t - 1}} \\
  &= A\mu_{t - 1}, \\
  \Covarg{z_{t} \vert x_{1:t - 1}} &= \Covarg{\Earg{z_{t} \vert z_{t - 1}, x_{1:t - 1}}} + \Earg{\Covarg{z_{t} \vert z_{t - 1}, x_{1:t - 1}}} \\
  &= \Covarg{Az_{t - 1} \vert z_{t - 1}, x_{1:t - 1}} + \Earg{Q} \\
  &= A\Sigma_{t - 1}A^{T} + Q.
\end{align*}
For the update step, use Gaussian conjugacy to obtain the required means and
covariances and the matrix inversion lemma to express them more concisely.
More precisely,
\begin{align*}
  p\left(z_{t} \vert x_{1:t}\right) \propto p\left(x_{t} \vert z_{t}\right)p\left(z_{t} \vert x_{1:\left(t - 1\right)}\right),
\end{align*}
and both densities on the right are Gaussian -- the first is the likelihood of
the observed $x_{t}$ while the second is the predict-step density derived above.
By Bayes' rule, the posterior is Gaussian with increased precision,
\begin{align}
  \label{eq:sigma_t_inv}
\Sigma_{t}^{-1} &= \Sigma_{t \vert t - 1}^{-1} + C^{T}R^{-1}C
\end{align}
and shrunken mean,
\begin{align}
  \label{eq:mu_t}
\mu_{t} &= \Sigma_{t}CR^{-1}x_{t} + \Sigma_{t}\Sigma_{t \vert t - 1}^{-1} \mu_{t \vert t - 1}.
\end{align}

Recall the matrix inversion lemma,
\begin{align*}
\left(A + UCV\right)^{-1} &= A^{-1} + A^{-1}U\left(C^{-1} + VA^{-1}U\right)^{-1}VA^{-1},
\end{align*}
and apply it to equation \ref{eq:sigma_t_inv} to find
\begin{align*}
  \Sigma_{t} &= \Sigma_{t \vert t - 1} - \Sigma_{t \vert t - 1}C^T\left(R + C \Sigma_{t \vert t - 1}C^{T}\right)^{-1}C\Sigma_{t \vert t - 1} \\
  &= \left(I - K_{t}C\right)\Sigma_{t \vert t - 1},
\end{align*}
according to the definition of $K_{t}$ in Algorithm
\ref{alg:kalman_filter}.

Substituting this expression, we can simplify equation \ref{eq:mu_t},
\begin{align*}
  \mu_{t} &= \left(I - K_{t} C\right)\Sigma_{t \vert t - 1}\left(C R^{-1}x_{t} + \Sigma_{t \vert t - 1}^{-1} \mu_{t \vert t - 1}\right) \\
  &= \left(I - K_{t}C\right)\mu_{t \vert t - 1} + \left(I - K_{t} C\right)\Sigma_{t \vert t - 1}C R^{-1} x_{t} \\
  &= \mu_{t \vert t - 1} + K_{t}\left(x_{t} - C\mu_{t \vert t- 1}\right),
\end{align*}
where for the simplification on the second half of the second line,
\begin{align*}
  \left(I - K_{t}C\right)\Sigma_{t \vert t - 1} C R^{-1} x_{t} &= \left(I - \left(\Sigma_{t \vert t - 1}^{-1} + C^{T} R C\right)^{-1}C^{T}R^{-1}C\right)\Sigma_{t \vert t - 1}C R^{-1} x_{t} \\
  &= K_{t}x_{t},
\end{align*}
we again used the matrix inversion lemma.

\begin{algorithm}
   \caption{The Kalman filtering predict-update recursions.}
   \label{alg:kalman_filter}
\begin{algorithmic}
  \STATE {\bfseries Input:} Model parameters $\Theta = \{A, C, Q, R\}$ and
    observed sequence $x_{1:T}$.
    \STATE $\mu_{0} \leftarrow 0, \Sigma_{0} \leftarrow Q$ \hfill initialize distribution of
    $z_{0}$.
    \FOR{$t = 1 \dots T$}
    \STATE $\mu_{t \vert t - 1} \leftarrow A\mu_{t - 1}$ \hfill predict step
    \STATE $\Sigma_{t \vert t - 1} \leftarrow A \Sigma_{t - 1} A^{T} + Q$
    \STATE $\mu_{t} \leftarrow \mu_{t \vert t - 1} + K_{t}\left(x_{t} - C\mu_{t \vert t - 1}\right)$ \hfill update step
    \STATE $\Sigma_{t} \leftarrow \left(I - K_{t}C\right)\Sigma_{t \vert t - 1}$
    \STATE where $K_{t} \leftarrow \left(\Sigma_{t \vert t - 1}^{-1} + C^{T}R C\right)^{-1} C^{T}R^{-1}$
    \ENDFOR
    \STATE {\bfseries Output:} Filtered means and covariances $\left(\mu_{t},
    \Sigma_{t}\right)_{t = 1}^{T}$.
\end{algorithmic}
\end{algorithm}
While the Kalman filter makes a forwards pass over the observations to estimate
the latent state means and covariances as data are made available, the RTS
smoother can be understood as a pair of forwards-backwards sweeps that propogate
information from all observations to the latent state estimates obtained by
filtering. Define the quantities,
\begin{align*}
\mu_{t \vert T} &= \Earg{z_{t} \vert x_{1:T}} \\
\Sigma_{t \vert T} &= \Covarg{z_{t} \vert x_{1:T}},
\end{align*}
which completely determine the smoothed distributions $p\left(z_{t} \vert
x_{1:T}\right)$ of interest. The forwards pass in RTS smoothing is identical to
the Kalman filter, and pseudocode for the backwards pass is provided in
Algorithm \ref{alg:kalman_smoother}.

\begin{algorithm}
   \caption{The Kalman smoothing backwards pass.}
   \label{alg:kalman_smoother}
\begin{algorithmic}
  \STATE {\bfseries Input:} Model parameters $\Theta = \{A, C, Q, R\}$,
    observed sequence $x_{1:T}$, and filtering quantities $\left(\mu_{t},
    \Sigma_{t}, \mu_{t \vert t - 1}, \Sigma_{t \vert t - 1}\right)$
    \STATE $\mu_{T \vert T} \leftarrow \mu_{T}, \Sigma_{T \vert T} \leftarrow
    \Sigma_{T}$ \hfill initialize distribution of $z_{T}$ from the Kalman
    filter.
    \FOR{$t = T - 1 \dots 1$}
    \STATE $\mu_{t \vert T} \leftarrow \mu_{t} + J_{t}\left(\mu_{t + 1 \vert T} - \mu_{t + 1 \vert t}\right)$
    \STATE $\Sigma_{t \vert T} \leftarrow \Sigma_{t} + J_{t}\left(\Sigma_{t + 1 \vert T} - \Sigma_{t + 1 \vert t}\right)J_{t}^{T}$
    \STATE where $J_{t} \leftarrow \Sigma_{t \vert t}A_{t + 1}^{T} \Sigma_{t + 1\vert t}^{-1}$
    \ENDFOR
    \STATE {\bfseries Output:} Smoothed means and covariances $\left(\mu_{t \vert T},
    \Sigma_{t \vert T}\right)_{t = 1}^{T}$.
\end{algorithmic}
\end{algorithm}

To see why this update works, first consider the joint behavior of neighboring
times,

\begin{align*}
   \left(z_{t}, z_{t + 1}\right) \vert x_{1:t} &\sim \Gsn\left(
\begin{pmatrix}
  \mu_{t \vert t} \\
  \mu_{t + 1 \vert t}
\end{pmatrix},
\begin{pmatrix}
  \Sigma_{t \vert t} & \Sigma_{t} A^{T} \\
  A \Sigma_{t} & \Sigma_{t + 1 \vert t}
\end{pmatrix}\right),
\end{align*}
because
\begin{align*}
  \Covarg{z_{t}, z_{t + 1} \vert x_{1:t}} &= \Covarg{z_{t}, Az_{t} + \eps_{t + 1} \vert x_{1:t}} \\
  &= \Sigma_{t}A^{T}.
\end{align*}

Conditioning, we obtain
\begin{align*}
  z_{t} \vert z_{t + 1}, x_{1:t} &\sim \Gsn\left(
  z_{t} \vert \mu_{t \vert t} + J_{t}\left(z_{t + 1} - \mu_{t + 1 \vert t}\right),
  \Sigma_{t \vert t} - J_{t} \Sigma_{t + 1 \vert t} J_{t}^{T}
  \right).
\end{align*}
Since $x_{t}$ is independent of $x_{(t + 1): T}$ conditional on $z_{t + 1}$,
this is enough to compute the full smoothed means and covariances. By the tower
property,
\begin{align*}
  \mu_{t \vert T} &= \Earg{z_{t} \vert x_{1:T}} \\
  &= \Earg{\Earg{z_{t} \vert z_{t + 1}, x_{1 : T}} \vert x_{1:T}} \\
  &= \Earg{\Earg{z_{t} \vert z_{t + 1}, x_{1:t}} \vert x_{1:T}} \\
  &= \mu_{t \vert t} + J_{t}\left(\mu_{t + 1 \vert T} - \mu_{t + 1 \vert t}\right)
\end{align*}
and similarly by the law of total covariance,
\begin{align*}
  \Sigma_{t\vert T} &= \Covarg{z_{t} \vert x_{1:T}} \\
  &= \Covarg{\Earg{z_{t} \vert z_{t + 1}, x_{1:T}}} + \Earg{\Covarg{z_{t} \vert z_{t + 1}, x_{1:T}}} \\
  &= \Covarg{\Earg{z_{t} \vert z_{t + 1}, x_{1:t}}} + \Earg{\Covarg{z_{t} \vert z_{t + 1}, x_{1:t}}} \\
  &= \Sigma_{t \vert t} + J_{t}\left(\Sigma_{t + 1 \vert T} - \Sigma_{t + 1 \vert t}\right) J_{t}^{T}
\end{align*}

\subsubsection{Zero-inflation in dynamical systems}
\label{subsubsec:zero_inflation_dynamical}

After transformations, many microbiome data sets can be viewed as a mixture
between a continuous, nonnegative component and a spike at 0. Such data has a
rich history in statistics, and has been approached using hurdle and tobit
modeling ideas \citep{min2002modeling}. In this section, we take a detour from
our review of standard temporal modeling methods, to describe a variation of the
standard LDS, called the dynamic tobit model (DTM), that is designed for the
zero-spiked density situation.

The basic idea of the DTM is to truncate an LDS below some threshold, so that
the continuous positive data correspond to excursions of the LDS above some
threshold, while the zeros correspond to parts of the path below the threshold.
That is, the observed data $y_t$ are modeled according to
\begin{align*}
  y_{t} \vert x_t, \tau &= \left(x_t - \tau\right)\indic{x_t > \tau}  \\
  x_t \vert z_t &\sim \Gsn\left(x_t \vert C z_{t}, R\right) \\
  z_t \vert z_{t - 1} &\sim \Gsn\left(z_t \vert A z_{t - 1}, Q\right).
\end{align*}
Note that last two equations are exactly those defining an LDS, but that $x_t,
z_t$, and $\tau$ are all unobserved data in this setting. As before, $A, C, R,
Q$ are model parameters to be learned.

There are several approaches to inference in this model, including particle
filtering \citep{doucet2000rao}, Monte Carlo EM \citep{manrique1998simulation},
and Gibbs sampling \citep{de1997scan, wei1999bayesian}. In many of these
approaches, recursions like those used in the Kalman filter can be used to
perform efficient sampling and estimation.

To give an exaple, we review one method here, the scan sampler of
\cite{de1997scan}. For simplicity, we will assume that $\tau$ is known and
equal to zero. The scan sampler iterates filter and smoother-type calculations
for $x_t$, at each step sampling from a distribution with mean $\Earg{x_t \vert
  x_{-t}}$ and variance $\Covarg{x_t \vert x_{-t}}$. In the Gaussian case
presented above, this provides the exact posterior, though \cite{de1997scan}
describe its utility for general exponential families.

To initialize the scan sampler for the DTM, first set $x_t = y_t$ for all $t$,
then perform a Kalman filter and smoother step. This gives estimates of the
following quantities, using notation analogous to that in the previous section,
\begin{align*}
  \mu_{t \vert t - 1}^{x} &= \Earg{x_t \vert x_{0:\left(t - 1\right)}} \\
  \mu_{t \vert t - 1}^{z} &= \Earg{z_t \vert z_{0:\left(t - 1\right)}} \\
  \Sigma_{t \vert t - 1}^{x} &= \Covarg{x_t \vert x_{0:\left(t - 1\right)}} \\
  \Sigma_{t \vert t - 1}^{z} &= \Covarg{z_t \vert z_{0 : \left(t - 1\right)}}.
\end{align*}

The idea of the scan sampler is to iteratively compute $M_t$ and $u_t$ such that
\begin{align*}
  M_t^{-1} &= \Covarg{x_t \vert x_{-t}} \\
  M_t^{-1}u_t &= x_t - \Earg{x_t \vert x_{-t}}
\end{align*}
so that sampling can cycle forwards and backwards through coordinates of
$\left(x_t\right)$, drawing $x_t^{\prime} \sim \Gsn\left(\Earg{x_t \vert
  x_{-t}}, \Covarg{x_t \vert x_{-t}}\right) = \Gsn\left(x_t - M_t^{-1}u_t,
M_t^{-1}\right)$. Details of the derivation can be found in \citep{de1997scan},
we simply describe the updates necessary to compute $M_t$ and $u_t$, along with
an example application. In an initial backwards pass through the data, we
compute
\begin{align*}
  u_t &= \left(\Sigma_{t \vert t - 1}^{x}\right)^{-1}\mu_{t \vert t - 1}^{x} - K_t^{T}r_t \\
  M_t &= \left(\Sigma_{t \vert t - 1}^{x}\right)^{-1} + K_t^{T} N_t K_t,
\end{align*}
where $K_t$ is the Kalman gain matrix at time $t$ and
\begin{align*}
  r_t &= C^T u_{t + 1} + A^{T}r_{t + 1} \\
  N_t &= C^{T} \Sigma_{t + 1 \vert t}C + L_{t + 1}^{T}N_{t + 1}L_{t + 1}.
\end{align*}
The $M_t$ remain fixed after this initial scan, and only sampling proceeds only
by cycling through the $u_t$. During the forwards scan, $u_t$ is updated
according to
\begin{align*}
  u_t \xleftarrow u_t - \left(\Sigma_{t \vert -t}^{x} C - K_t^{T}N_t\right) b_t \\
  b_{t + 1} &= L_{t} b_t - K_t M_t^{-1}u_t
\end{align*}
where $L_t = A - K_t C$, while for backwards scans, the analogous updates are
\begin{align*}
  u_t \xleftarrow u_t - K_t^{T} b_t \\
  b_{t - 1} &= L_t^T b_t - C_t^{T}M_t^{-1}u_t.
\end{align*}
Each time a $u_t$ is computed, it is used to update the unobserved $x_t$ by
sampling from $\Gsn\left(x_t - M_t^{-1}u_t, M_t^{-1}\right)$.

\begin{figure}
  \centering
  \includegraphics[width=\textwidth]{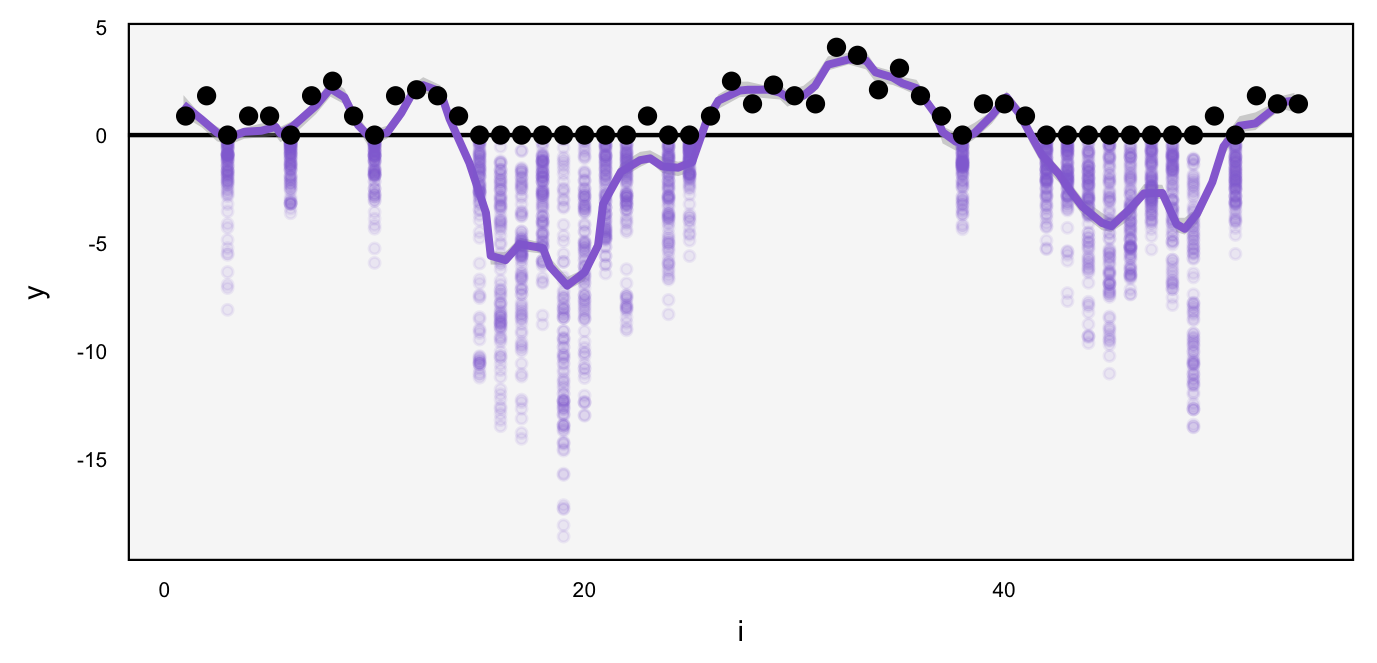}
  \caption{Posterior samples from a DTM applied to a bacterial abundance time
    series, obtained using the scan sampler. We fix $A, C, Q, R$ for the
    dynamical system component to arbitrary, stable values. Black points
    represent observed data, while blue points are draws from the scan sampler
    posterior. The blue line is a loess smooth of the posterior samples for
    $x_t$.
    \label{fig:abt_scan}}
\end{figure}

Code implementing these updates is available at
\url{https://github.com/krisrs1128/tsc_microbiome/tree/master/src/scan}. We have
also provided an example application to a real microbial abundance time series,
whose results are provided in Figure \ref{fig:abt_scan}.

\section{Temporal mixture models}
\label{sec:temporal_mixture_models}

In the models considered in Section \ref{sec:smooth_temporal_models}, all
time windows exhibit comparable behavior. From the problem description of
Section \ref{sec:problem_description} however, we would expect changes in the
behavior of the system during different temporal regimes. In this section, we
describe probabilistic models that generate this sort of behavior. The unifying
probabilistic ``trick'' that helps accomplish this goal is the introduction of
latent variables specying which regime the system is in at any given timepoint.

\subsection{Hidden Markov Models}
\label{subsec:hmms}

The idea behind Hidden Markov Models (HMMs) is similar to that of LDSs, with the
exception that the $z_{t}$ are no longer thought of as continuous latent
variables. Instead, each $z_{t}$ is assigned one of a few discrete states,
which switch between one another according to a Markov chain.

An informal analogy is how I choose to listen to different music tracks. For a
few days, I might be mostly listening to sonatas by Beethoven and Schubert,
because I'm in a latent state that prefers German piano sonatas. I might be more
likely to transition from here to French chamber music than, say, LA punk. So,
my underlying style preferences evolve according to a Markov chain $z_{t}$,
which is in turn reflected by the actual tracks $x_{t}$ that I listen to.

More formally, we suppose $z_{1} \sim \pi_{1}$ comes from some initial
distribution $\pi_{1} \in \simplex^{K - 1}$ and has some $K \times K$ transition
probability matrix $P$ whose rows lie in $\simplex^{K - 1}$, the $K - 1$
dimensional simplex. Conditional on $z_{t}$, $x_{t}$ is emitted according to the
density $p_{\theta_{z_{t}}}\left(x\right)$, one of $K$ emission densities
$p_{\theta_{1}}\left(x\right), \dots, p_{\theta_{K}}\left(x\right)$. Concisely,
\begin{align*}
  x_{t} \vert z_{t} &\sim p_{\theta_{z_{t}}} \\
  z_{t} \vert z_{t - 1} &\sim P_{z_{t - 1}} \text{ for } t > 1 \\
  z_{1} &\sim \pi_{1}
\end{align*}
As in LDSs, fitting this model can be broken into inference and learning steps,
which estimate the distributions of the $z_{t}$ conditional on observed data and
which fit the model parameters $\{P, \pi, \theta_{1}, \dots, \theta_{K}\}$,
respectively. Alternating these two steps optimizes the expected complete data
loglikelihood in an EM algorithm.

The $E$-step, which infers latent states, is referred to as the
forwards-backwards algorithm. This procedure parallels the Kalman filter
(forwards) and smoother (backwards) algorithm for inference in LDSs. First, we
find a simple expression for the analog of the filtered densities, based on the
Markov property for the $z_{t}$'s,
\begin{align*}
  p\left(z_{t} \vert x_{1:t}\right) &\propto p\left(x_{t} \vert z_{t}, x_{1:t - 1}\right) p\left(z_{t} \vert x_{1:t - 1}\right) \\
  &= p\left(x_{t} \vert z_{t}\right) p\left(z_{t} \vert x_{1:t - 1}\right) \\
  &= p_{\theta_{z_{t}}}\left(x_{t}\right)p\left(z_{t} \vert x_{1:t - 1}\right).
\end{align*}
The first term is known by assumption, while the second can be computed from the
previous filtered probabilities,
\begin{align*}
  p\left(z_{t} \vert x_{1:t - 1}\right) &= \sum_{k = 1}^{K} p\left(z_{t} \vert z_{t - 1} = k\right)p\left(z_{t - 1} = k \vert x_{1:t - 1} \right).
\end{align*}

The forwards pass of the Forwards-Backwards algorithm alternates these two steps
(``update'' and ``predict''), after initializing $p\left(z_{1}\right) =
\pi_{1}$, as described in Algorithm \ref{alg:hmm_forwards}.

\begin{algorithm}
   \caption{Safe log-sum-exp}
   \label{alg:normalize_log}
   \begin{algorithmic}
     \STATE {\bfseries Input:} $x \in \reals^{n}$
     \STATE $m \leftarrow \max\left(x\right)$
     \STATE {\bfseries Output:} $\text{lse}\left(x\right) \leftarrow x - \left(m + \log\left(\sum\exp{x_{i} - m}\right)\right)$
   \end{algorithmic}
\end{algorithm}

\begin{algorithm}
   \caption{Forwards pass for HMM Inference}
   \label{alg:hmm_forwards}
\begin{algorithmic}
  \STATE {\bfseries Input:} Model parameters $\{\pi_{1}, P, \theta_1, \dots, \theta_K\}$,
  observed sequence $x_{1:T}$
  \STATE $\log \tilde{p}\left(z_{1}\vert x_{1}\right) \leftarrow
  \log\pi_{1} + \begin{pmatrix} \log p_{\theta_{1}}\left(x_{1}\right) \\ \vdots \\ \log p_{\theta_{k}}\left(x_{1}\right) \end{pmatrix}$ \hfill initialize distribution of $z_{1}$
  \STATE $\log p\left(z_{1} \vert x_{1}\right) \leftarrow \log \tilde{p}\left(z_{t} \vert x_{1}\right) - \text{lse}\left(\log \tilde{p}\left(z_{t} \vert x_{1}\right)\right)$\hfill Normalize with Algorithm \ref{alg:normalize_log}
  \FOR{$t = 2 \dots T$}
  \STATE $\log \tilde{p}\left(z_{t} \vert x_{1:t}\right) \leftarrow P \exp{\log p\left(z_{t - 1} \vert x_{1:t - 1}\right)} + \begin{pmatrix}  \log p_{\theta_{1}}\left(x_{t}\right) \\ \vdots \\ \log p_{\theta_{K}}\left(x_{t}\right) \end{pmatrix}$
  \STATE $\log p\left(z_{1} \vert x_{1:t}\right) \leftarrow \log \tilde{p}\left(z_{t} \vert x_{1:t}\right) - \text{lse}\left(\log \tilde{p}\left(z_{t} \vert x_{1:t}\right)\right)$\hfill
  \ENDFOR
  \STATE {\bfseries Output:} Filtered log densities $\log p\left(z_{t} \vert x_{1:t}\right)$.
\end{algorithmic}
\end{algorithm}

\begin{algorithm}
  \begin{algorithmic}
   \caption{Backwards pass for HMM Inference}
   \label{alg:hmm_backwards}
   \STATE {\bfseries Input:} Model parameters $\{\pi_{1}, P, \theta_1, \dots, \theta_k\}$, observed sequence $x_{1:T}$
   \STATE $\log p\left(x_{T + 1} \vert z_T\right) \leftarrow 0_{K}$ \hfill Initialize messages.
   \FOR{$t = T  - 1\dots 1$}
   \FOR{$k = 1 \dots K$}
   \STATE $\log p\left(x_{\left(t + 1\right): T} \vert z_{t} = k\right) \leftarrow \text{lse}\left(\log p\left(x_{\left(t + 2\right):T} \vert z_{t + 1}\right) + \begin{pmatrix} \log p_{\theta_{1}}\left(x_{t}\right) \\ \vdots \\ p_{\theta_{K}}\left(x_t\right)\end{pmatrix} + \log P_{k}\right)$
   \ENDFOR
   \ENDFOR
   \STATE {\bfseries Output:} Smoothed densities $\log p\left(z_{t} \vert x_{1:T}\right)$.
  \end{algorithmic}
\end{algorithm}

The backwards pass computes the analog of the smoothed densities, based on the
observation\footnote{This is the same observation used in the derivation of the
  Kalman smoother.} that $x_{t}$ is independent of the future conditional on
$z_{t + 1}$,
\begin{align*}
  p\left(z_{t} \vert x_{1:T}\right) &\propto p\left(x_{t + 1:T} \vert z_{t}\right) p\left(z_{t} \vert x_{1:t}\right).
\end{align*}
The second term on the right hand side is available from the forwards pass. The
first term is computed during the backwards pass, which initializes $p\left(x_{T
  + 1} \vert z_{T}\right) = 1$ and then accumulates the ``future observation''
densities from right to left,
\begin{align*}
  p\left(x_{t + 1: T} \vert z_{t}\right) &= \sum_{k = 1}^{K} p\left(z_{t + 1} = k, x_{t + 1 : T} \vert z_{t}\right) \\
  &= \sum_{k = 1}^{K} p\left(x_{t + 2: T}\vert z_{t + 1} = k, x_{t + 1}, z_{t}\right) p\left(z_{t + 1} = k, x_{t + 1} \vert z_{t}\right) \\
  &= \sum_{k = 1}^{K} p\left(x_{t + 2:T}\vert z_{t + 1} = k\right)p_{\theta_{z_{t + 1}}}\left(x_{t + 1}\right) P_{z_{t}, k}.
\end{align*}
The backwards pass is summarized in Algorithm \ref{alg:hmm_backwards}.

In the M-step, the parameters $\left(\theta_{k}\right)$, $\pi_0$ and
$P_{k, k^{\prime}}$ must be learned. Estimation of the per-regime parameters
can be done separately for each $k$. For example, if HMM has Gaussian emissions,
$p_{\theta_{k}}\left(x_{t}\right) = \Gsn\left(x_{t} \vert \mu_{k},
\Sigma_{k}\right)$, then
\begin{align*}
  \hat{\mu}_{k} &= \frac{\sum_{t = 1}^{T} \indic{z_{t} = k}x_{t}}{\sum_{t = 1}^{T}\indic{z_{t} = k}} \\
  \hat{\Sigma}_{k} &= \frac{\sum_{t = 1}^{T} \indic{z_{t} = k}\left(x_{t} - \hat{\mu}_{k}\right)\left(x_{t} - \hat{\mu}_{k}\right)^{T}}
      {\sum_{t = 1}^{T} \indic{z_{t} = k}},
\end{align*}
while, regardless of the emission structure, the Markov chain parameters can be
estimated via
\begin{align*}
  \hat{\pi}_{k, k^{\prime}} &= \frac{\sum_{t = 1}^{T} \indic{z_{t} = k, z_{t + 1} = k^{\prime}}}{\sum_{t = 1}^{T} \indic{z_{t} = k}}.
\end{align*}

\subsubsection{Example}
\label{subsubsec:hmm_example}

In the context of regime detection in the microbiome, we imagine a few
underlying states shared across all bacteria, with means ranging from high abundance to
complete absence. Within each species, states evolve independently according to
a Markov chain. Note that, while we do share across species when estimating
underlying parameters $\left(\theta_{k}\right)$, we do not tie together latent
indicators $z_{it}$ across species. Specifically, we use a DAG that factors
across all species -- given $\theta$, all species abundances follow independent
HMMs.

\begin{figure}
  \centering
  \includegraphics[width=\textwidth]{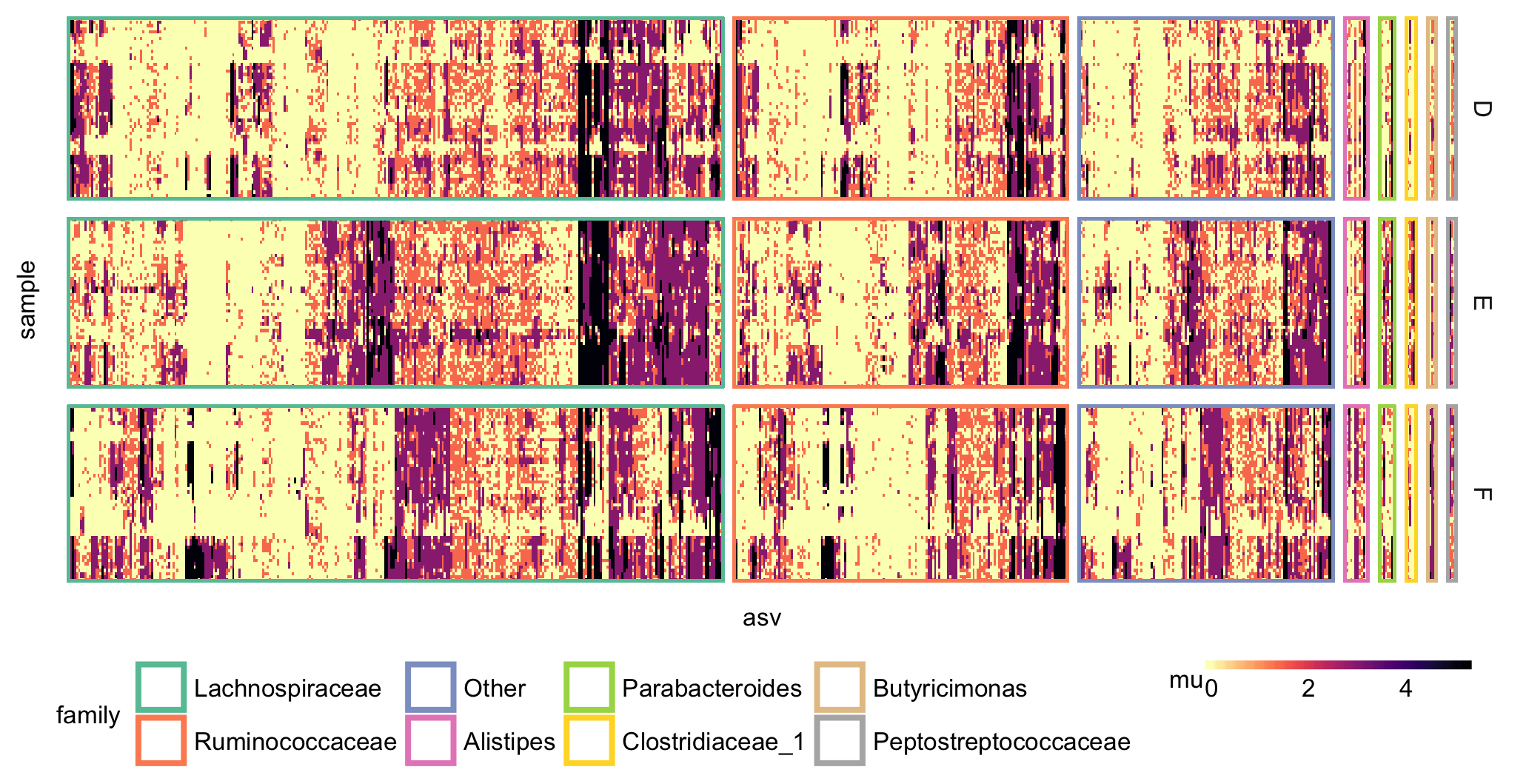}
  \caption{The modal sequences according to the HMM estimated through
    EM. \label{fig:hmm_mode} }
\end{figure}

\begin{figure}
  \centering
  \includegraphics[width=0.9\textwidth]{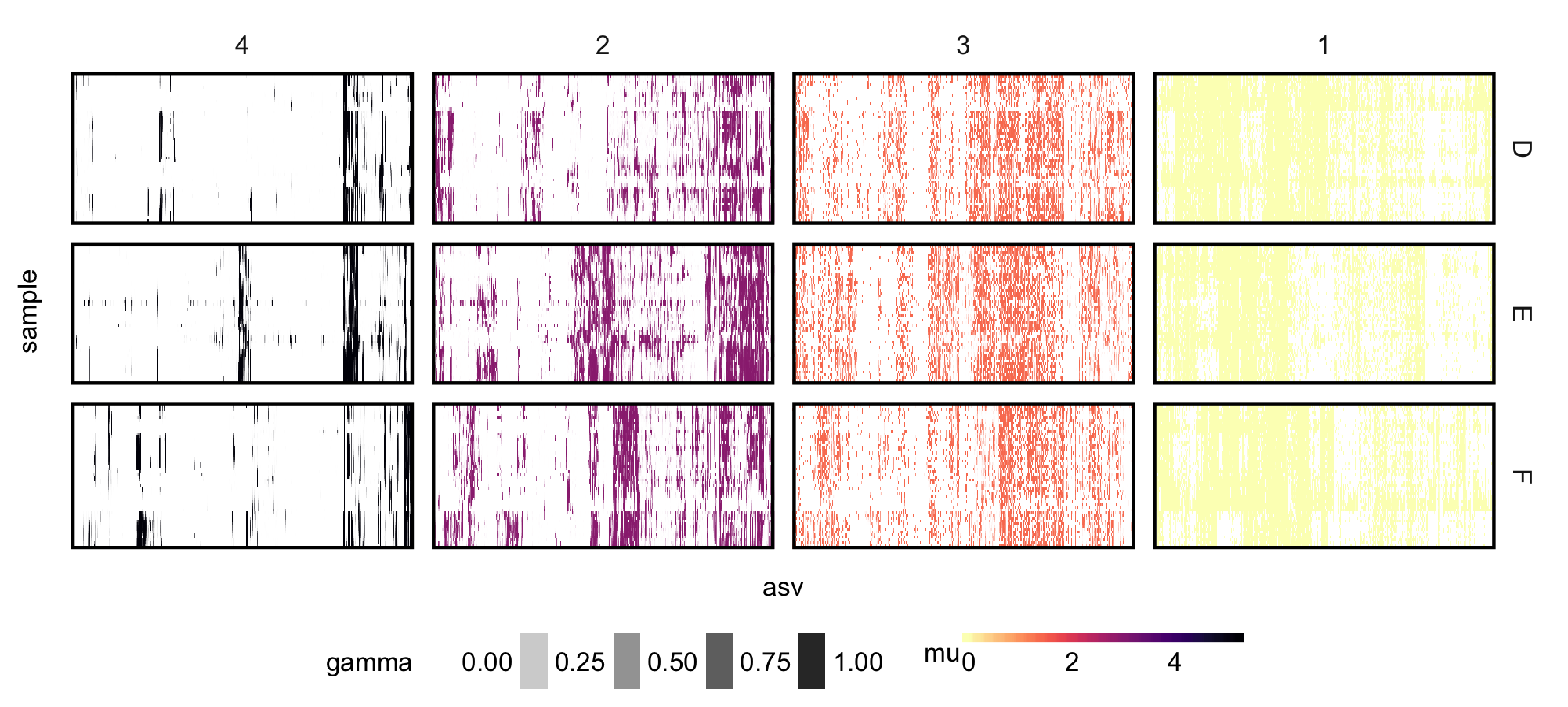}
  \caption{Fitted probabilities for each of the $K = 4$ states in the HMM
    estimated by EM. Different rows of panels correspond to different
    individuals, while different columns are different states. The colors across
    columns represent the emission means for the associated states, while the
    transparency of a cell in a certain column corresponds to the probability
    that cell was generated by that column's state. \label{fig:hmm_probs} }
\end{figure}

Therefore, in the $E$-step (the forwards-backwards algorithm) the $z_{it}$ can
be estimated in parallel across all species. The $M$-step is modified so that
mixture component parameters and transition probabilities are estimated
simultaneously across all species, for example, the new mean update for cluster
$k$ in an HMM with Gaussian emissions becomes
\begin{align*}
\hat{\mu}_{k} &= \frac{\sum_{i = 1}^{n}\sum_{t = 1}^{T} \indic{z_{t} = k}x_{it}}{\sum_{i = 1}^{n}\sum_{t = 1}^{T} \indic{z_{it} = k}}.
\end{align*}
We employ this approach on the antibiotics data described in Section
\ref{subsubcart_example}. The results are summarized in Figure
\ref{fig:hmm_mode} and Supplementary Figure \ref{fig:hmm_probs}, which display
the modes and raw probabilities of the smoothed probabilities $p\left(z_{it}
\vert x_{i, 1:T}\right)$ after applying EM. As in Figure
\ref{fig:centroid-euclidean-conditional} and its counterparts made through
hierarchical clustering, rows correspond to individual species, while columns
are samples sorted by time. The three main panels are the three subjects in the
study. Rows $i$ are ordered according to a hierarchical clustering on Euclidean
distances between sequences $\left(\hat{z}_{i1}, \dots, \hat{z}_{iT}\right)$,
where $\hat{z}_{it} = \arg \max_{k} p\left(z_{it} = k \vert x_{1:T}\right)$. The
intensity of the shading of each cell corresponds to the average value
$\hat{\mu}_{k}$ for that mode.

We can think of Figure \ref{fig:hmm_mode} as a smoothed version of the raw
heatmaps made through hierarchical clustering. By coarsening the raw values into
cluster centers, certain patterns are more easily visible. For example, in
addition to effect of the two antibiotics time courses across all subjects, we
note that

\begin{itemize}
\item The effect of the first antibiotic time courses is more prolonged in
  subject F, and in some cases species disappear for most of the remainder of
  the experiment. Some of these species seem to have also disappeared from
  subject D, but most others recovered in both subjects D and E.
\item The phenomenon in which some species become \textit{more} abundant during
  the time courses -- presumably filling in newly emptied niches -- is stronger
  in subject E than the others. Further, some species that exhibit this behavior
  in subject E respond differently within subjects D and F, becoming less
  abundant rather than more.
\end{itemize}

Note that, while smoothing makes the antibiotic effect clearer among moderately
and highly abundant species, variation among rarer species, to which all
timepoints are assigned to the rare abundance state, is obscured. This could be
remedied by increasing $K$, at the cost of increasing the complexity of
interpretation for abundant species.

Instead of displaying the sequence of modes $\hat{z}_{it}$, Supplemental Figure
\ref{fig:hmm_probs} provides the individual probabilities $p\left(z_{it} = k
\vert x_{1:T}\right)$ across $k$. The vertical panels distinguish between
different $k$, while transparency represents the size of the probability --
smaller probabilities are more transparent. The colors are associated with
centroid means, as before. In this application, most probabilities seem near
one, so this view adds little to that in Figure \ref{fig:hmm_mode}. However,
this does suggest the possibility of overfitting, which we address by
introducing priors on $\Theta$ in Section \ref{sec:sticky_hmms}.

The estimated transition probabilities between states, ordered in terms of
increasing $\hat{\mu}_{k}$, are displayed below.
\begin{verbatim}
      1     2     3     4
1 0.810 0.159 0.028 0.003
2 0.389 0.527 0.083 0.002
3 0.077 0.094 0.810 0.019
4 0.027 0.007 0.056 0.910
\end{verbatim}
Unsurprisingly, most transitions remain in the same state, and any departures
are generally restricted to states with similar $\hat{\mu}_{k}$s. Generally,
there appear to be more transitions downwards than upwards, and the sum of
transitions into state 1 is higher than the sums for the rest. This can likely
be attributed to the antibiotic effect, which reduces abundances of species that
recover at differential rates.

As in Figure \ref{fig:centroid-euclidean-conditional}, the primary drawback of
this display is the difficulty in linking observed species trajectories to
species or taxonomic detail. To easily link the smoothed HMM estimates to raw
time series values and species descriptions would require either splitting the
view across taxonomies or introducing interactivity.

\subsubsection{Sticky HMMs}
\label{sec:sticky_hmms}

Sticky HMMs are an extension of ordinary HMMs designed to induce additional
``stickiness'' in state transitions -- states should be encouraged to remain
unchanged over longer stretches of time. To practically implement this basic
idea, a Bayesian view is useful, and we describe inference in detail. An
application to the antibiotics data follows this explanation.

Consider the standard Bayesian HMM, which places $\Dir\left(\alpha\right)$
priors on the rows of the transition probability matrix $P_{1}, \dots, P_{K}$
and conjugate priors on $\Theta$. The main idea of the sticky HMM is to
introduce a ``stickiness'' parameter $\kappa$ to the Dirichlet
prior\footnote{Here, $e_{k} := \left(0, \dots, 0, 1, 0, \dots, 0\right)$ is the
  vector with a $1$ in the $k^{th}$ coordinate and zeros everywhere else.}:
$P_{k} \sim \Dir\left(P_{k} \vert \alpha + \kappa e_{k}\right)$. This means that
draws from the prior on $P$ will have larger weight along the diagonal,
depending on the size of $\kappa$, and since diagonal elements correspond to
self-transitions, chains drawn from the prior will be ``sticky.''

To draw samples from $p\left(\left(z_{it}\right), P, \Theta \vert
\left(x_{it}\right)\right)$, we consider a block Gibbs sampler that parallels EM
\citep{fruhwirth2006finite}. In place of the $E$-step, we draw from the
conditional $p\left(\left(z_{it}\right) \vert \left(x_{it}\right), P,
\Theta\right)$ using a variation of the forwards-backwards algorithm, and in
place of the $M$-step, we draw from $p\left(\Theta, \vert P,
\left(z_{it}\right), \left(x_{it}\right)\right)$ and $p\left(P \vert \Theta,
\left(z_{it}\right), \left(x_{it}\right)\right)$. We detail this below,
following the derivation of \citep{fox2009bayesian}.

First consider sampling $p\left(\left(z_{it}\right) \vert \left(x_{it}\right),
P, \Theta\right)$. Since each sequence is independent of all others, given
$\Theta$, we can sample the $z_{it}$ separately across each $i$, so for the
explanation below, we drop this subscript. By the DAG structure of the HMM, the
probability a single sequence can be decomposed according to
\begin{align}
  \label{eq:sticky_block_z}
  p\left(z_{1:T} \vert x_{1:T}, \Theta, \pi, P\right) &= p\left(z_{1} \vert x_{1:T}, \Theta, \pi, P\right) \prod_{t = 2}^{T} p\left(z_{t} \vert z_{t - 1}, x_{1:T}, \Theta, \pi, P\right).
\end{align}
If we could calculate each of these individual probabilities, then one mechanism
for sampling the entire sequence $z_{1:T}$ would be to first sample $z_{1}$,
then sample $z_{2}$ given $z_{1}$, etc. Note that the first term has a different
structure than the rest (it doesn't depend on any previous $z_{t}$), so we
analyze it first,
\begin{align*}
  p\left(z_{1} \vert x_{1:T}, \Theta, \pi, P\right) &\propto p\left(z_{1} \vert \Theta, \pi, P\right) p\left(x_{1} \vert z_{1}, \Theta, \pi, P\right) p\left(x_{2:T} \vert z_{1}, \Theta, \pi, P\right) \\
  &\propto \pi_{z_{1}} p\left(x_{1} \vert \theta_{z_{1}}\right) p\left(x_{2:T} \vert z_{1}, \Theta, \pi, P\right)
\end{align*}
where we used the fact that $x_{1} \independent x_{2:T} \vert z_{1}$. The first
two terms are easy to sample, and we will show below that terms of the form
$p\left(x_{t + 1:T} \vert z_{t}, \Theta, \pi, P\right)$ can be calculated
efficiently using a backwards-pass type recursion.

Consider now the terms in the product of equation \ref{eq:sticky_block_z}. By
Bayes rule,
\begin{align*}
  p\left(z_{t} \vert z_{t - 1}, x_{1:T}, \Theta, \pi, P\right) &\propto p\left(x_{1:t - 1} \vert z_{t - 1}, z_{t}, \Theta, \pi, P\right)
  p\left(x_{t} \vert z_{t}, \Theta, \pi, P\right)
  p\left(x_{t + 1 : T} \vert z_{t}, \Theta, \pi, P\right) \\
  &\propto p\left(x_{1:t - 1} \vert z_{t - 1}, z_t, \Theta, \pi, P\right) p\left(x_t \vert \theta_{z_t}\right) p\left(x_{t + 1 : T}\vert z_{t}, \Theta, \pi, P\right).
\end{align*}
The second term is a likelihood, and the third will be calculated by the
backwards-pass. Note that the first term can be further reduced to
\begin{align*}
  p\left(x_{1:t - 1} \vert z_{t - 1}, z_t, \Theta, \pi, P\right) &\propto p\left(z_{t} \vert x_{1:t - 1}, z_{t - 1}, \Theta, \pi, P\right)p\left(x_{1:t - 1} \vert z_{t - 1}, \Theta, \pi, P\right) \\
  &\propto p\left(z_{t} \vert z_{t - 1}, \Theta, \pi, P\right) \\
  &= P_{z_{t - 1}, z_{t}},
\end{align*}
since we care only about the terms involving $z_{t}$, as we are sampling
$p\left(z_{t} \vert z_{t - 1}, x_{1:T}, \Theta, \pi, P\right)$.

We next describe how to compute the terms $p\left(x_{t + 1:T} \vert z_t, \Theta,
\pi, P\right)$ efficiently, as promised above. Consider storing these terms in a
$T \times K$ matrix, whose $tk^{th}$ element is $p\left(x_{t:T} \vert z_{t - 1} =
k, \Theta, \pi, P\right)$. The bottom row is initialized according to
\begin{align*}
 p\left(x_T \vert z_{T - 1}, \Theta, \pi, P\right)  &= \sum_{z_{T} = 1}^{K} p\left(x_{T} \vert \theta_{z_{T}}\right) P_{z_{T - 1}, z_{T}}.
\end{align*}
Then the recursion computes the $t^{th}$ row from the $t + 1^{st}$ according to
\begin{align*}
  p\left(x_{t + 1:T} \vert z_t, \Theta, \pi, P\right) &= \sum_{z_{t + 1} = 1}^{K} p\left(x_{t + 1 : T} \vert z_{t}, z_{t + 1} \Theta, \pi, P\right) P_{z_{t}, z_{t + 1}} \\
  &= \sum_{z_{t + 1} = 1}^{K} p\left(x_{t + 1} \vert \theta_{z_{t + 1}}\right) p\left(x_{t + 2 : T} \vert z_{t + 1}, \Theta, \pi, P\right) P_{z_{t}, z_{t + 1}}.
\end{align*}

Next consider sampling from the conditional $p\left(P \vert \left(x_{it}\right),
\left(z_{it}\right), \Theta\right)$ of transition probabilities between the $K$
states. The main observation is that we can use Dirichlet-Multinomial conjugacy
jointly across all sequences. That is, defining
\begin{align*}
 n_{k \cdot} &:= \begin{pmatrix} \sum_{i = 1}^{n} \sum_{t = 1}^{T - 1} \indic{z_{it} = k, z_{i,t + 1} = 1} \\ \vdots \\ \sum_{i = 1}^{n} \sum_{t = 1}^{T - 1} \indic{z_{it} = k, z_{i,t + 1} = K} \end{pmatrix},
\end{align*}
we have
\begin{align*}
n_{k\cdot} &\sim \Mult\left(n_{k\cdot} \vert \sum_{i = 1}^{n} \sum_{t = 1}^{T - 1} \indic{z_{it} = k}, P_{k}\right)
\end{align*}
and therefore
\begin{align}
  \label{eq:sticky_hmm_p_update}
  P_{k} \vert \left(z_{it}\right) &\sim \Dir\left(P_{k} \vert \alpha + \kappa e_{k} + n_{k\cdot}\right).
\end{align}
Since given $\left(z_{it}\right)$, $P_{k}$ is independent of all other
parameters and data, this completely specifies this Gibbs sampling step.

To sample $p\left(\Theta \vert \left(x_{it}\right), \left(z_{it}\right), \pi,
P\right)$, we use assumed conjugacy separately across each class $k$. For
example, if $\theta_{k} = \left(\mu_{k}, \Sigma_{k}\right)$ and
\begin{align*}
  x_{it} \vert \left[z_{it} = k\right] &\sim \Gsn\left(x_{it} \vert \mu_{k}, \Sigma_{k}\right),
\end{align*}
and if
\begin{align*}
  \mu_{k} &\sim \Gsn\left(\mu_{0} \vert \mu_{k}, \Sigma_{0}\right) \\
  \Sigma_{k} &\sim \text{InvWish}\left(\Sigma_{k} \vert \nu, \Delta\right)
\end{align*}
then the usual Normal-Inverse Gamma posterior can be used, after subsetting to
those samples $\left(x_{it}\right)^{(k)}$ with $z_{it} = k$,
\begin{align*}
  \mu_{k} \vert \left(x_{it}\right)^{(k)} &\sim \Gsn\left(\theta_{k} \vert \bar{\Sigma}\left(\Sigma_0^{-1}\mu_{0} + \Sigma_{k}^{-1}\right) \sum_{i, t} x_{it}\indic{z_{it} = k}, \bar{\Sigma}\right) \\
  \Sigma_{k} \vert \left(x_{it} \right)^{(k)} &\sim  \text{InvWish}\left(\nu + \sum_{i, t} \indic{z_{it} = k}, \nu \Delta + \left(x^{(k)} - \mathbf{1} \mu_{k}^{T}\right)\left(x^{(k)} - \mathbf{1} \mu_{k}^{T}\right)^{T}\right).
\end{align*}
An implementation of this sampler is available at
\url{https://github.com/krisrs1128/tsc\_microbiome/blob/master/src/hmm/bayes\_hmm.R}.

\subsubsection{Example}
\label{subsubsec:sticky_hmm_example}

\begin{figure}
  \centering
  \includegraphics[width=\textwidth]{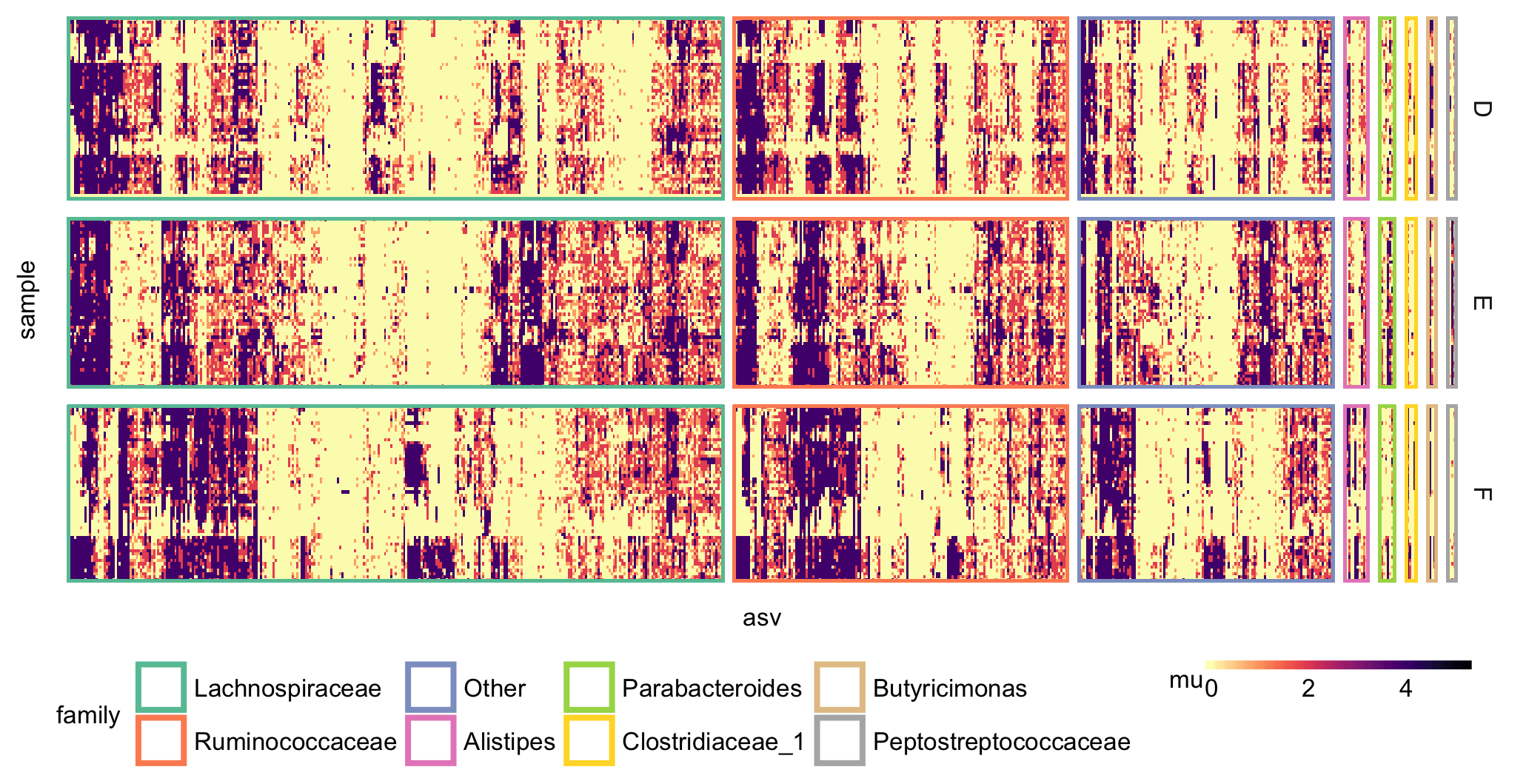}
  \caption{The analog of Figure \ref{fig:hmm_mode} for the sticky HMM,
    representing means of modal states estimated at given timepoints for
    particular species. Note that, compared to Figure \ref{fig:hmm_mode},
    species have been reordered according to a hierarchical clustering on these
    means. \label{fig:bayes_mode} }
\end{figure}

An application of the sticky HMM to antibiotics data is provided in Figure
\ref{fig:bayes_mode}. As in Section \ref{subsubsec:hmm_example}, we choose $K =
4$ and a Gaussian emission model. After manual experimentation, we set $\kappa =
4$. We have reordered species according to a hierarchical clustering on the new
estimated means. The interpretation suggested by this figure is generally quite
similar to those for the nonsticky HMM. As before, the antibiotic treatments are
visible across all subjects, especially D and F, with differential recovery
rates among certain species, and some species in subject D that seem to increase
in abundance during the first antibiotic time course and one timepoint during
the interim. In contrast to the HMM fitted with EM, it is easier to pick out a
block of species among the Ruminoccocus that are strongly affected by the first,
but not second, antibiotics time courses, especially in subject F. On the other
hand, species that increase in subject D during the first time course are now
split across a few blocks, when they had been all grouped together in Figure
\ref{fig:hmm_mode}. Overall, the clustering seems somewhat more interpretable in
the sticky HMM, though we have not quantitatively compared different metrics for
two clustering approaches.

As before, we can study the estimated transition matrices between pairs of
states. These are printed below, with states sorted from lowest to highest
emission means.
\begin{verbatim}
       3     4     1     2
3  0.775 0.113 0.044 0.068
4  0.160 0.414 0.170 0.256
1  0.073 0.300 0.196 0.431
2  0.025 0.083 0.083 0.809
\end{verbatim}

Unlike EM, Gibbs sampling can provide a sense of the uncertainty of these
parameter estimates. The standard errors of the cells for these transition
probabilities are printed below.
\begin{verbatim}
       3     4     2     1
3  0.005 0.007 0.002 0.004
4  0.010 0.017 0.011 0.017
1  0.009 0.014 0.031 0.011
2  0.002 0.006 0.006 0.004
\end{verbatim}

Counterintuitively, while the sticky HMM was proposed to model state
persistence, the probabilities of self-transitions estimated here are lower than
those estimated for the ordinary HMM. One explanation is that the Dirichlet
priors placed on the rows of the transition matrix in the sticky HMM serve as
regularizers. Decreasing the Dirichlet concentration parameter would likely
return us to more extreme self-transition estimates from EM. However, these
regularized transition probabilities are perhaps more believable -- simply
because certain transitions have not been previously observed doesn't mean they
should be given probability 0.

\subsubsection{Sticky HDP-HMMs}
\label{sec:sticky_hdp_hmm}

Next we describe a Bayesian nonparametric variation of sticky HMMs first
proposed in \citep{fox2008hdp}. There are two sources of motivation for such a
model,
\begin{itemize}
\item It is appealing to allow the complexity of the fitted model to increase as
  more data becomes available. This drove the original HDP-HMM proposal of
  \citep{teh2006hierarchical}.
\item State persistence -- i.e., stickiness -- can yield better fitting and more
  interpretable HMMs. This turns out to be especially relevant in the
  nonparametric setting, where there is a danger of overfitting by introducing
  many short-lived states.
\end{itemize}

The proposal of \citep{teh2006hierarchical} replaces the usual sticky-HMM
$\Dir\left(\alpha\right)$ on rows of $P$ with an $\text{HDP}\left(\alpha,
\gamma, H\right)$ prior\footnote{This refers to a Hierarchical Dirichlet Process
  prior with component concentration $\alpha$, shared concentration $\gamma$,
  and baes measure $H$, explained in more detail below.}. Note that a
$\DP\left(\alpha, H\right)$ prior on its own would not be sufficient, since two
separate draws from a $\DP\left(\alpha, H\right)$ prior with a continuous base
measure would have distinct atoms almost surely. Hence, it would be impossible
to align the rows $P_{k}$ of the transition matrix $P$ according to a set of
common states. By using a $\DP\left(\alpha, H\right)$ base measure instead, the
$\text{HDP}\left(\alpha, \gamma, H\right)$ prior allows sharing of states across
$P_{k}$.

More formally, an HDP-HMM models a sequence $x_{1}, \dots, x_{T}$, emitted from
a latent Markov chain $z_t$,
\begin{align*}
  x_t \vert z_t &\sim P_{\theta_{z_t}} \\
  z_t \vert z_{t - 1} &\sim P_{z_{t - 1}} \text{ for } t > 1 \\
  z_1 &\sim \pi_1
\end{align*}
with a prior on the rows of the transition matrix given by
\begin{align*}
  P_k \vert Q &\sim \DP\left(\alpha, Q\right) \\
  Q &\sim \DP\left(\gamma, H\right).
\end{align*}
Since each of the transition matrix rows $P_k$ are centered around the base
measure $Q$, the rows are i.i.d., and there is no stickiness.

To induce state persistence, first consider a stick breaking representation
$\DP$ base measure $Q = \sum_{j = 1}^{\infty} \beta_j \delta_{\theta_j}$ where
the sequence $\left(\beta_j\right)$ is drawn from a
$\text{GEM}\left(\gamma\right)$ distribution\footnote{The
  GEM$\left(\gamma\right)$ distribution \citep{gnedin2001characterization} can
  be described by a stick breaking constrution. Specifically,
  $\left(\beta_{j}\right)$ can be constructed by taking $\beta_{1} = V_1$,
  $\beta_2 = \left(1 - V_1\right)V_2, \dots, \beta_{j} = \left[\prod_{k = 1}^{j
      - 1}\left(1 - V_k\right)\right]V_j$ where $V_j \sim \Bet\left(1,
  \gamma\right)$ independently. }. Cosider a modified base measure,
$Q_k = \sum_{j = 1}^{\infty} \left(\beta_j + \indic{j =
    k}\kappa\right)\delta_{\theta_j}$, which places $\kappa$ additional mass on
  the atom for the $k^{th}$ latent state. Using $Q_k$ as a base measure for
  $P_k$ encourages the $k^{th}$ state to return to itself. For completeness, the
  prior for rows of the transition matrix now has the form,
\begin{align*}
  P_k \vert Q_k &\sim \DP\left(\alpha, Q_k\right) \\
  Q_k \vert \left(\beta_j\right), \left(\theta_j\right) &\sim \sum_{j = 1}^{\infty} \left(\beta_j + \indic{k = j}\kappa\right)\delta_{\theta_j} \\
  \left(\beta_j\right) &\sim \text{GEM}\left(\gamma\right) \\
  \theta_j &\sim H \text{ for }j = 1, 2, \dots.
\end{align*}

\cite{fox2009bayesian} describes two inferential approaches for this model: an
exact one that extends Chinese Restaurant Franchise analogy developed by
\citep{teh2006hierarchical} and an approximate blocked Gibbs sampler. Here, we
describe the blocked Gibbs sampler, since while it is only approximate, it mixes
more rapidly\footnote{Implementations of both approaches are available at
  \url{https://github.com/krisrs1128/tsc\_microbiome/tree/master/src/hmm}}. This
  is because the exact sampler can only swap cluster labels $z_t$ one at a time,
  which means it must pass through low probability valleys in the posterior
  where a cluster is awkwardly split in half before reaching a potentially
  higher mode.

The idea of the blocked Gibbs sampler is to sample all the assignments $z_t$
simultaneously using the same forwards-backwards variant used for inference of
the sticky HMM in Section \ref{sec:sticky_hmms}. A difficulty in directly
applying this strategy is that we no longer have a finite $K$ number of states
from which to sample. To get around this issue, \cite{fox2008hdp} employs
\cite{ishwaran2002exact}'s ``weak-limit'' approximation,

\begin{align*}
  \text{GEM}\left(\gamma\right) \approx \Dir\left(\frac{\gamma}{L}, \dots, \frac{\gamma}{L}\right),
\end{align*}
where $L$ is truncation level chosen large enough so that the number of clusters
visible in the observed data modeled by an exact $\DP$ would usually be smaller
than $L$. Note that there is a tradeoff here between statistical and
computational efficiency -- larger $L$ brings us closer to the exact model but
requires more involved computation.

With this approximation in hand, we can now describe a tractable sampler. The
full parameter set consists of $\{\left(\theta_k\right)_{k = 1}^{L},
\left(z_t\right)_{t = 1}^{T}, \left(\beta_{k}\right)_{k = 1}^{L},
\left(P_{k}\right)_{k = 1}^{L} \}$. To facilitate sampling, three sets of
auxiliary variables $\left(m_{jk}\right), \left(w_k\right)$, and
  $\left(\bar{m}_{jk}\right)$ are introduced. Each term is sampled one at a
  time, from its full conditional.

  Some mnemonics can help with tracking notation,
\begin{itemize}
\item $m_{jk}$ track the number of transitions from states $j$ to $k$ if there
  had been no stickiness.
\item $w_k$ counts the number of times stickiness is invoked in state $k$.
\item $m_{jk}$ counts the number of transitions from $j$ to $k$ after accounting
  for state persistence.
\end{itemize}

We simply state the conditionals required in each step of the Gibbs sampler. A
detailed derivation is provided in \citep{fox2009bayesian}. The conditional for
$\beta$ is
\begin{align*}
  \beta &\sim \Dir\left(\frac{\gamma}{L} + \bar{m}_{\cdot 1}, \dots, \frac{\gamma}{L} + \bar{m}_{\cdot L}\right),
\end{align*}
where $\bar{m}_{\cdot l} = \sum_{j = 1}^{L} \bar{m}_{jl}$. The updates for
$\left(\theta_k\right)$ and $\left(z_t\right)$ are the same as those for the
(finite) sticky HMM, described in Section \ref{sec:sticky_hmms}. Next consider
sampling the auxiliary variables. To sample $m_{jk}$, draw
\begin{align*}
  m_{jk} &\sim \sum_{n = 1}^{n_{jk}} \Ber\left(\frac{\alpha \beta_{k} + \kappa\indic{j = k}}{n + \alpha \beta_{k} + \kappa \indic{j = k}}\right)
\end{align*}
where $n_{jk}$ counts the number of observed transitions from states $j$ to $k$
among the $z_{t}$, which we have conditioned upon. The distribution of $m_{jk}$
is sometimes called a Chinese Restaurant Table distribution, because it counts
the total number of tables that have been occupied in a CRP after a certain
number of customers have arrived \citep{zhou2012augment}, which is a sum of
Bernoulli decisions to join an old table or occupy a new one.

For $w_k$, the conditional can be shown to be
\begin{align*}
  w_k &\sim \Bin\left(m_{jj}, \rho\left(\rho + \beta_j\left(1 - \rho\right)\right)^{-1}\right),
\end{align*}
where $\rho = \frac{\kappa}{\alpha + \kappa}$. $\bar{m}_{jk}$ is set
deterministically, according to
\begin{align*}
  \bar{m}_{jk} &= \begin{cases}
    m_{jk} &\text{for } j \neq k\\
    m_{jj} - w_{j \cdot} &\text{ for } j = k.
  \end{cases}
\end{align*}
Cycling these updates provides a blocked Gibbs sampler for the sticky HDP-HMM.

\subsubsection{Example}
\label{subsubsec:sticky_hdp_hmm_example}

\begin{figure}
  \centering
  \includegraphics[width=\textwidth]{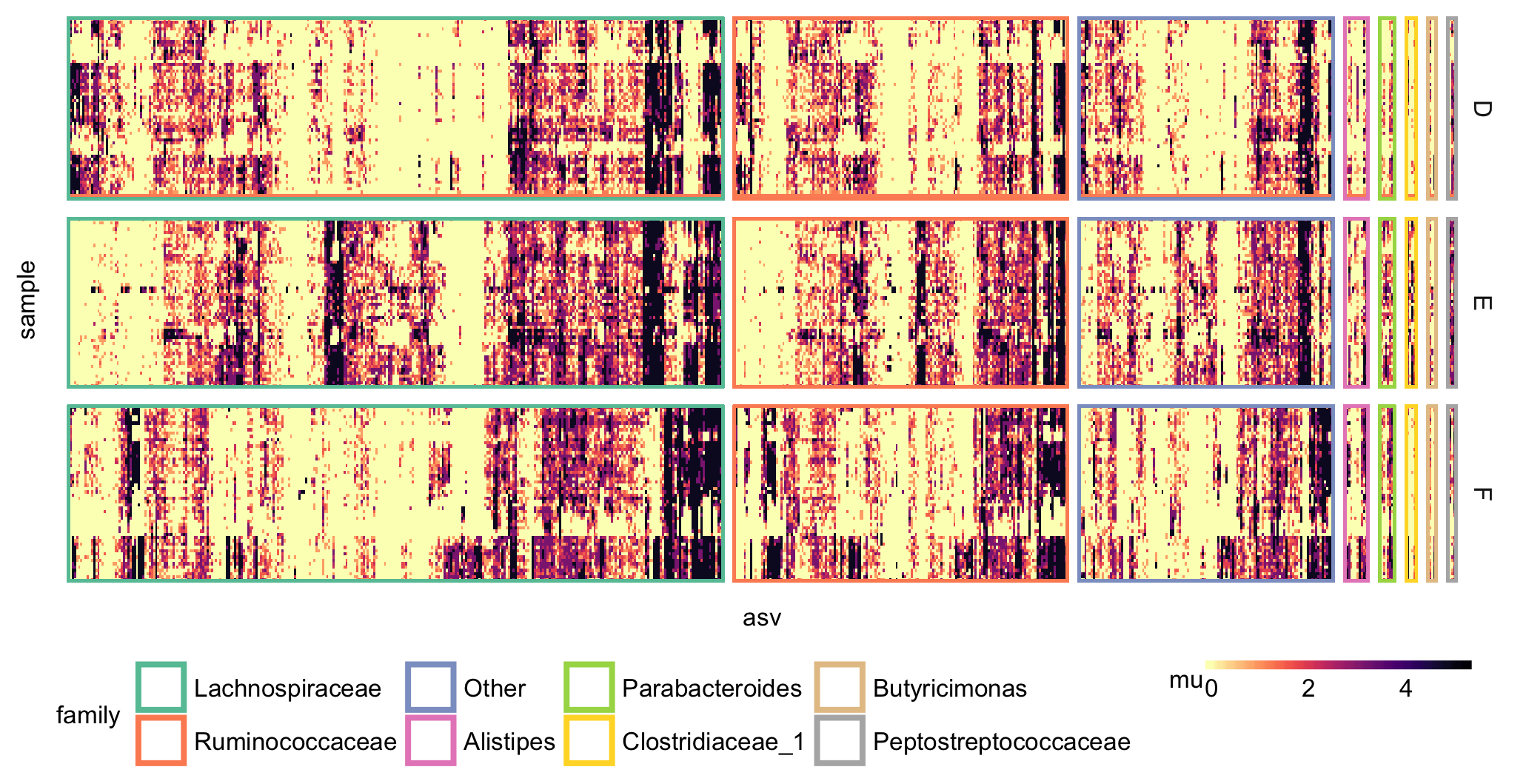}
  \caption{
    The analog of Figure \ref{fig:hmm_mode} for the sticky HDP-HMM. As before,
    each column corresponds to a single species, each row is a timepoint, and
    panels represent different individuals. Color intensity gives the emission
    mean of the associated states for that $\times$ by species combination, over
    many gibbs sampling iterations. With a choice of $\alpha = 10^{-4}$ and
    $\kappa = 0.1$, the sticky HDP-HMM introduces many more states than Figure
    \ref{fig:hmm_mode}, corresponding to a less smoothed out representation of
    the antibiotics data. \label{fig:hdp_mode} }
\end{figure}

An application of the HDP-HMM to the antibiotics data is summarized in Figure
\ref{fig:hdp_mode}. We ran a Normal-Inverse Gamma emissions version of the
sampler for 2000 iterations, initializing the model using the model using means
and covariances from clusters found with $K$-means. We set $\alpha = 10^{-4}$,
$\gamma = 10^{-6}$, $\kappa = 0.1$, and $\L = 10$.

As in the original parametric HMM displayed in Figure \ref{fig:hmm_mode}, Figure
\ref{fig:hdp_mode} clearly highlights the antibiotic regimes across subjects D
and F, as well as the differential recovery rates among some species in subject
F. However, with our choice of $\alpha$, there tend to be many transient states,
represented by the increased number of intermediate colors between light yellow
and dark purple. We can understand this figure as a kind of smoothing using a
low bandwidth. Decreasing $\alpha$ and increasing $\kappa$ leads to a higher
concentration in few states, and a more strongly smoothed representation of the
data.

Estimated transition probabilities are printed below. While in theory the
transition matrix is infinite dimensional, the weak limit approximation provides
$K$ clusters plus one additional category containing the rest of the mass from
the DP for that row. States are ordered from lowest to highest emission mean.

\begin{verbatim}
       9     6     8     5     3     4    10     2     1     7
 9 0.016 0.008 0.007 0.003 0.002 0.002 0.003 0.006 0.004 0.007
 6 0.011 0.014 0.027 0.013 0.006 0.009 0.010 0.002 0.017 0.018
 8 0.012 0.019 0.035 0.029 0.015 0.010 0.017 0.000 0.027 0.034
 5 0.011 0.036 0.023 0.030 0.025 0.011 0.026 0.000 0.023 0.034
 3 0.009 0.037 0.010 0.019 0.027 0.020 0.041 0.000 0.021 0.037
 4 0.007 0.030 0.008 0.008 0.016 0.022 0.057 0.000 0.014 0.040
10 0.001 0.008 0.006 0.016 0.020 0.037 0.284 0.000 0.059 0.176
 2 0.010 0.011 0.007 0.002 0.005 0.022 0.081 0.020 0.035 0.088
 1 0.005 0.021 0.011 0.009 0.011 0.014 0.051 0.000 0.020 0.008
 7 0.002 0.005 0.002 0.002 0.003 0.005 0.014 0.000 0.005 0.004
\end{verbatim}

As in the other HMM-based methods, the sticky HDP-HMM places most mass in the
transition matrix along the diagonal, corresponding to self-transitions. As in
the sticky HMM, self-transitions are somewhat more regularized than in the HMM
estimated through EM. Transitions from the highest to the lowest states are more
common than transitions in the opposite directions. This corresponds to the fact
that rapid drops after antibiotic time courses are somewhat more common than
rapid recoveries.

In addition to summaries of the model results, it is informative to investigate
properties of the sampling routine, to diagnose potential defects or
limitations. Supplemental Figure \ref{fig:hdp_gibbs_samples} displays the
evolution of state assignments over iterations. The variation in state colors at
the bottom of the figure represents the initial burn-in period. Most states,
especially those corresponding to low abundance states, seem relatively fixed
across Gibbs sampling iterations. In some situations, this would be a marker of
poor mixing of the block sampler, in spite of its performing full forwards and
backwards sweeps during every sampling iteration. However, a more likely
explanation in this situation is that the abundances across species can vary
quite dramatically, so that the likelihood of alternative states is very low.
This also suggests that an alternative likelihood model may provide both better
mixing and fits than the Normal Inverse-Gamma likelihood applied here.


HMMs suppose that the observed system switches between a few regimes, but that
within regimes observations are i.i.d.. In certain situations, this is not quite
plausible, and in the next few sections we describe alternatives that mix
non-i.i.d. processes. In particular, we focus on several efforts to merge the
switching idea of HMMs with LDSs and GPs, proposed in
\citep{ghahramani1998variational, rasmussen2002infinite, fox2012multiresolution,
  linderman2016recurrent}.

\subsection{Infinite Mixtures of Gaussian Process Experts}
\label{subsec:imgpe}

\begin{figure}
  \centering
  \includegraphics[width=0.9\textwidth]{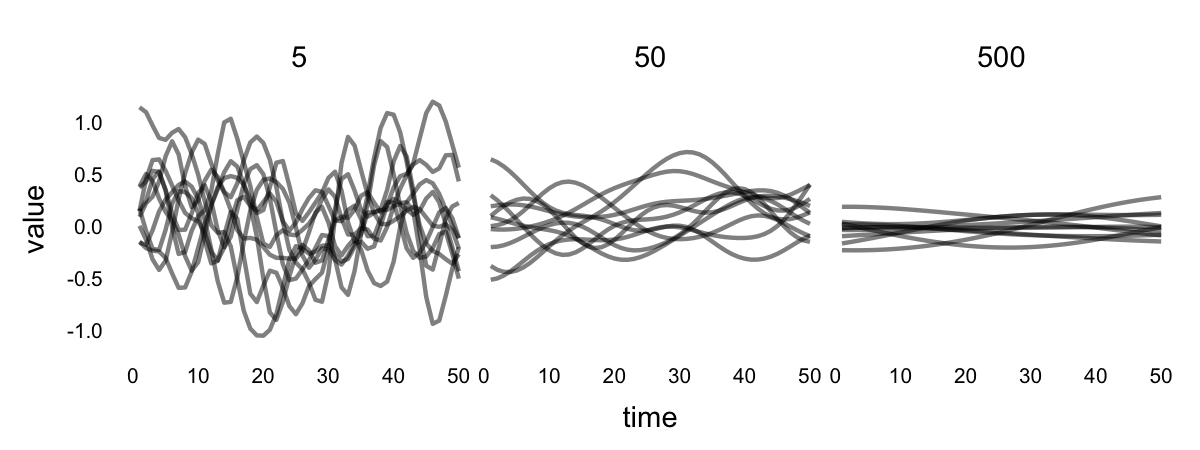}
  \caption{Each panel displays draws from the GP prior for a single kernel
    bandwidth. While different kernels can model different degrees of
    smoothness, for any single bandwidth, dynamics are relatively
    homogeneous. \label{fig:gp_bandwidths} }
\end{figure}

GPs, as described in Section \ref{subsec:gaussian_processes}, provide an
alternative to LDS and HMM models for analyzing temporal data. However, in their
original form, they are of only limited utility for the regime detection
problem, because they assume a type of homogeneity in dynamics. In particular,
once the kernel bandwidth for a GP is specified, it will tend to have
fluctuations on the order of that bandwidth throughout its entire domain, see
Figure \ref{fig:gp_bandwidths}. This makes it difficult to model differential
dynamics -- e.g., gradual evolution in some domains and rapid changes in
others\footnote{The need to specified a unified bandwidth is analogous to the
  situation in spline smoothing, which motivated the development of wavelet
  methods \citep{donoho1995adapting}. That there is such a connection is not
  surprising, considering the parallel nature of GPs and splines
  \citep{kimeldorf1970correspondence}} which is characteristic of longitudinal
microbiome data sets.

To adapt GPs to settings with more heterogeneous dynamics, a variety of mixture
\citep{tresp2001mixtures, rasmussen2002infinite}, multiscale
\citep{fox2012multiresolution, samostring}, and time-varying
\citep{paciorek2003nonstationary, heinonen2016non} approaches have been
proposed. Here, we describe the implementation and application of the Infinite
Mixture of Gaussian Process Experts (IMGPE) \citep{rasmussen2002infinite}. Code
is available at
\url{https://github.com/krisrs1128/tsc_microbiome/tree/master/src/igp}.

In this approach, timepoints are assigned to distinct GPs, each with their own
kernel parameters. Let $\left(y_{i}\right)$ be a single observed time series,
with $y_{i}$ measured at time $t_{i}$. We partition timepoints into (latent)
distinct classes, with the class indicator for $t_i$ denoted by $z_i$. The set
of $y_i$ associated with a particular $z_i = k$ are assumed to have been drawn
from a GP with a class specific kernel $\kappa_{\theta_k}$. We will apply a
one-dimensional Gaussian covariance kernel with $\theta = \{v_0, v_1, \sigma_f^2\}$,
\begin{align*}
  \kappa_\theta\left(x_i, x_i^\prime\right) &= v_0 \exp{\frac{1}{\sigma_f^2} \left(x_i - x_i^\prime\right)^2} + v_1.
\end{align*}

The associated complete data likelihood is
\begin{align*}
 y_i \vert \left(\kappa_k\right), \left(c_i\right) &\sim \prod_{k \in \text{unique}\left(c_i\right)} \Gsn\left(0, K_{\theta_k}\left(t^k\right)\right)
\end{align*}
where $K_{\theta_k}\left(t^k\right)$ denotes the covariance matrix obtained by
evaluating the $k^{th}$ kernel at the timepoints $t^k$ for which $z_i = k$,
\begin{align*}
  K_{\theta_k}\left(t^k\right) &= \begin{pmatrix}
    \kappa_{\theta_k}\left(t^k_1, t^k_1\right) & \dots  & \kappa_{\theta_k}\left(t^k_1, t^k_{n_k}\right) \\
    \vdots & & \vdots \\
    \kappa_{\theta_k}\left(t^k_{n_k}, t^k_1\right) & \dots  & \kappa_{\theta_k}\left(t^k_{n_k}, t^k_{n_k}\right) \\
  \end{pmatrix}.
\end{align*}

To complete specification of the model, priors must be placed on the
$\left(z_i\right)$ and $\left(\theta_k\right)$. \cite{rasmussen2002infinite}
propose placing a Chinese Restaurant Process (CRP) prior on the class indicators,
separate wide Gaussian priors on the logged kernel parameters, and a gamma
hyperprior on the CRP diversity parameter $\alpha$,
\begin{align*}
  \left(z_i\right)_i^n &\sim \CRP\left(\alpha\right) \\
  \log \theta_k &\sim \Gsn\left(0, \tau^2 I_3\right) \text{ for } k = 1, 2, \dots \\
  \alpha &\sim \Gam\left(a, b\right)
\end{align*}
The use of a $\CRP\left(\alpha\right)$ prior explains the name \textit{infinite}
mixture of GP experts. In our experiments, we find it sufficient to fix $\alpha$
and treat it as a tuning parameter. Further, we find that Logistic, instead of
Gaussian, priors on $\log \theta_k$ allow improved detection of different regime
dynamics, likely due to its heavier tails. Inference with this alternative prior
can be done similarly to the Gaussian case.

We next describe an algorithm for fitting this model. There are two main parts,
inference of the latent classes $z_i$ and learning of the class kernel
parameters $\theta_k$. For inference of the $z_i$, a collapsed Gibbs sampler can
be applied, in the spirit of \cite{neal2000markov}'s Algorithm 3. In
particular, the conditionals are of the form
\begin{align}
  p\left(z_i = k \vert z_{-i}, \left(y_i\right), \left(\theta_k\right)\right) &\propto p\left(\left(y_i\right) \vert \left(z_i\right),  \left(\theta_k\right)\right)
  p\left(z_i = k \vert z_{-i}, \left(\theta_k\right)\right) \nonumber \\
  &= p\left(\left(y_i\right) \vert \left(z_i\right), \left(\theta_i\right)\right)p\left(z_i = k \vert z_{-i}\right), \label{eq:igp_conditional}
\end{align}
which is tractable because the likelihood $p\left(\left(y_i\right) \vert
\left(z_i\right), \left(\theta_i\right)\right)$ decouples across classes $\{i :
z_i = k\}$ while the CRP predictive has a simple form, according to the
Polya-Urn sampling scheme,
\begin{align*}
  p\left(z_i\vert z_{-i}\right) &\propto \begin{cases}
    k &\text{with probability } \frac{n_{-i, k}}{n - 1 + \alpha} \\
    K + 1 &\text{with probability } \frac{\alpha}{n - 1 + \alpha}
    \end{cases},
\end{align*}
where $n_{-i, k}$ counts the number of $z_i$ equal to $k$, excluding $z_i$ and
$K + 1 = \max{z_{-i}} + 1$ corresponds to the case of introducing a previously
unobserved class. Note that we have conditioned on the infinite number of
$\theta_k$'s, though only at most $n$ need to be tracked at any moment during
the computation.

For learning the kernel hyperparameters $\theta_k$, a separate Hamiltonian Monte
Carlo (HMC) sampler is used for each of the observed GPs, using the same setup
as \cite{rasmussen2006gaussian} for individual GPs. In particular, for the
timepoints $t^k$ associated with a single class, the unnormalized posterior over
$\theta_k$ can be evaluated,
\begin{align*}
  &\log \Gsn\left(y^k \vert 0, K_{\theta_k}\left(t_k\right)\right) +
  \log \text{Logist}\left(\log v_0 \vert a_0, b_0\right) +
  \log \text{Logist}\left(\log v_1 \vert a_1, b_1\right) +\\
  &\log \text{Logist}\left(\log \sigma_f^2 \vert a_f, b_f\right),
\end{align*}
along with its gradients with respect to $\log v_0, \log v_1$, and $\log
\sigma_f^2$. This can be input to the generic HMC sampler in order to propogate
forwards from the current state in a way that samples from the posterior of
these parameters.

Finally, inference and learning are combined in every iteration of the mixture
of GPs algorithm. That is, in each iteration
\begin{itemize}
\item Perform a full Gibbs sampling sweep of $z_1, \dots, z_n$ using the
  conditionals in equation \ref{eq:igp_conditional}, with fixed $\theta_k$. If
  new classes $k^\ast$ are drawn, sample an associated $\log \theta_k^\ast$ from
  its prior.
\item For a given configuration of $z_1, \dots, z_n$, propogate the current
  values of $\left(\log \theta_k\right)$ forwards according to HMC dynamics for
  some number of leapfrog iterations (we use $5$ in our experiments below, each
  with stepsize $\eps = 0.005$).
\end{itemize}

While \citep{rasmussen2006gaussian} sample $\alpha$ and introduce some
refinements to the Gibbs sampling update, we find this setup sufficient for our
application, it also yields a cleaner implementation.

Stepping back from this model description, we note that this method was
proposed in the context of a single time series with several regimes. An
extension to collections of related time series, as in the microbiome context,
seems natural but as yet unstudied.

\subsubsection{Example}
\label{subsubsec:igp_mix_example}

We now apply this method to a single species abundance time series from the
antibiotics data set \citep{dethlefsen2011incomplete}. We apply it to an
Enterobacteria species, labeled in this data by Unc09aa7, which was chosen
because it includes two differently structured spikes, along with long stretches
of zeros during the antibiotics time courses. The sampler was run for 1000
iterations, which takes about 30 minutes on a relatively standard
machine\footnote{1.4GhZ, 4GB RAM.}. We manually set $\alpha = 0.15$ after
comparing the number of clusters identified with different levels of $\alpha$.
The sampler was initialized by assigning all timepoints to a single class and by
drawing a random $\theta_1$ from the prior.

\begin{figure}
  \centering
  \includegraphics[width=\textwidth]{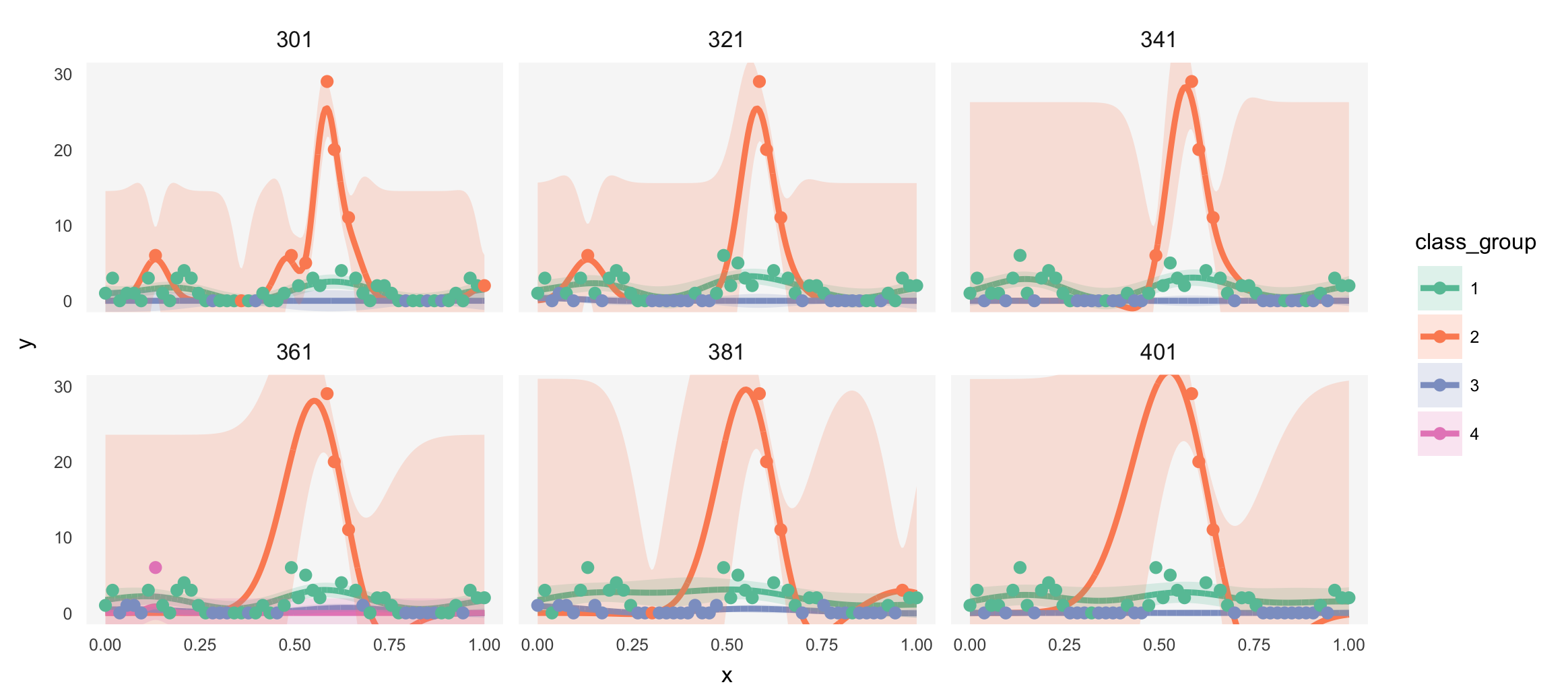}
  \caption{Each panel represents displays the posteriors at one iteration of the
    sampler. Each point is shaded according to its sampled stated $z_i$. Lines
    and shaded shaded bands represent means and variance in the osteriors for
    the processes associated with different $z_i$. The posteriors are displayed
    over the full data range, even when the associated timepoints $t^k$ occupy a
    relatively restricted range. \label{fig:igp_abt_fits}}
\end{figure}

In Figure \ref{fig:igp_abt_fits}, we display draws from the posterior over
$\left(z_i\right)$ and $\left(\theta_k\right)$ for a subset of iterations. There
tends to be some redundancy in the estimated mixture components, but there is
almost always at least one processes that fits the zero and nonzero intervals,
respectively. There is some differentiation of the very high points in the
second peak from other points, including those in the earlier peak.

Interestingly, in light of our original motivation for fitting this model, the
differentiation between regimes can occur due to variations in noise levels in
addition to bandwidth. For example, the process that is nearly always zero (the
blue process) seems to have been identified because it has low variance, while
spike after the first regime seems (the red process) to be distinguished by
having a small bandwidth. The green process includes those nonzero points that
are lower than the spike.

Somewhat unsatisfyingly, the zeros that occur early in the zeros are grouped
along with the zeros during the antibiotic regime, though informally we think of
these points as ``pre-antibiotic.'' In fact, the timepoints seem to be clustered
based mostly on their $y$-axis values, rather than requiring any sense of
temporal continuity. However, this is not surprising in light of the model's
search criterion, since there is no sense in which the labels $z_i$ are required
to be close to one another when their associated $t_i$ are close -- the choice
is made entirely on the basis of cluster sizes and process likelihood. This is
well-suited to the situation where there are several processes in different
$y$-axis scales and overlapping $t_i$, but when regime assignments should be
more contiguous, the fit appears artificial.

\begin{figure}
  \centering
  \includegraphics[width=\textwidth]{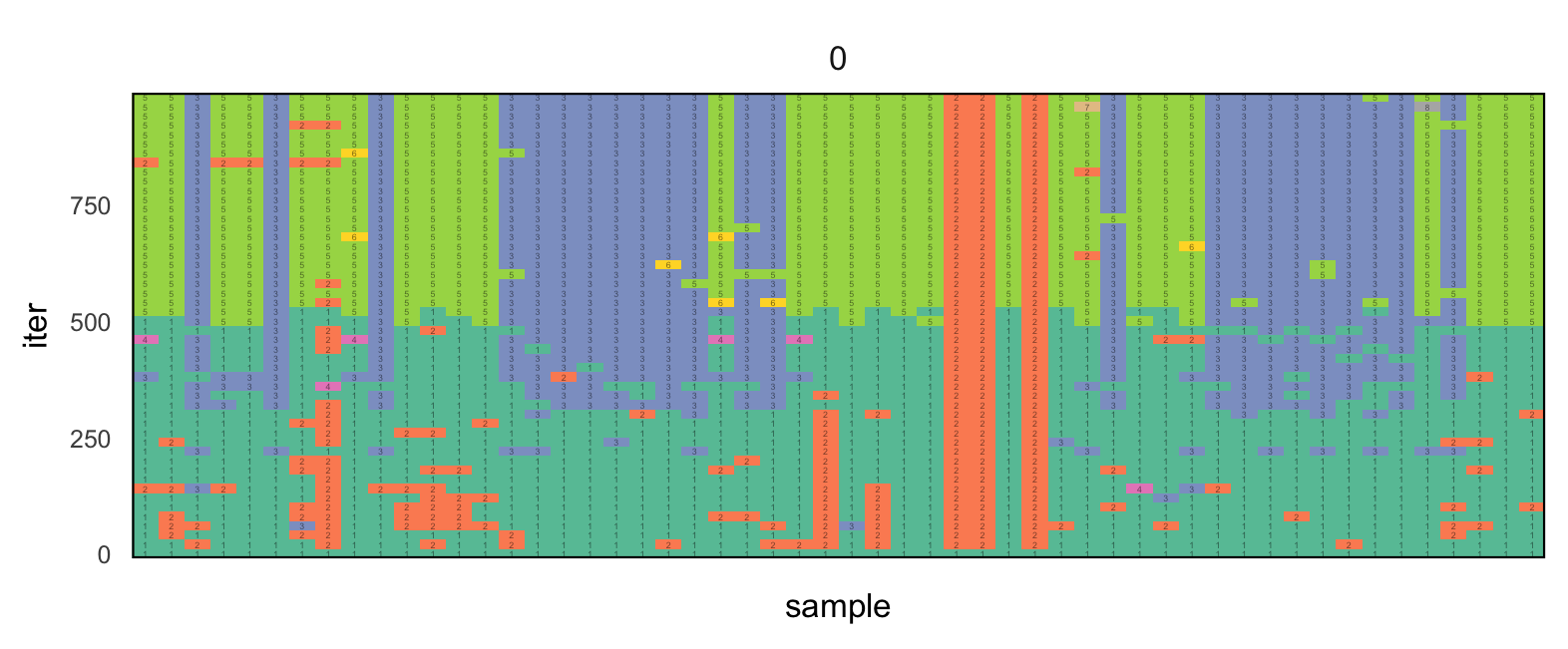}
  \caption{Each row is an iteration of the sampler, each column is a timepoint.
    Tiles are shaded by their class assignments. Note the label switching that
    occurs a few times. \label{fig:igp_abt_states} }
\end{figure}

In Figure \ref{fig:igp_abt_states}, we display the class assignments over
iterations of the sampler. This provides a more concise view of the estimated
regime assignments. The antibiotic regimes and the peaks seem to have been
distinguished from one another. This figure also illustrates some properties of
the CRP prior -- some states are introduced briefly and then deleted forever,
others may last some time before being swept out by classes with similar
parameters $\theta_k$. When computing summary statistics over multiple
iterations, this figure can be used to ensure averages are only computed over
iterations where no label-switching has occured. Alternatively, it could be used
to align similar clusters.

\begin{figure}
  \centering
  \includegraphics[width=0.7\textwidth]{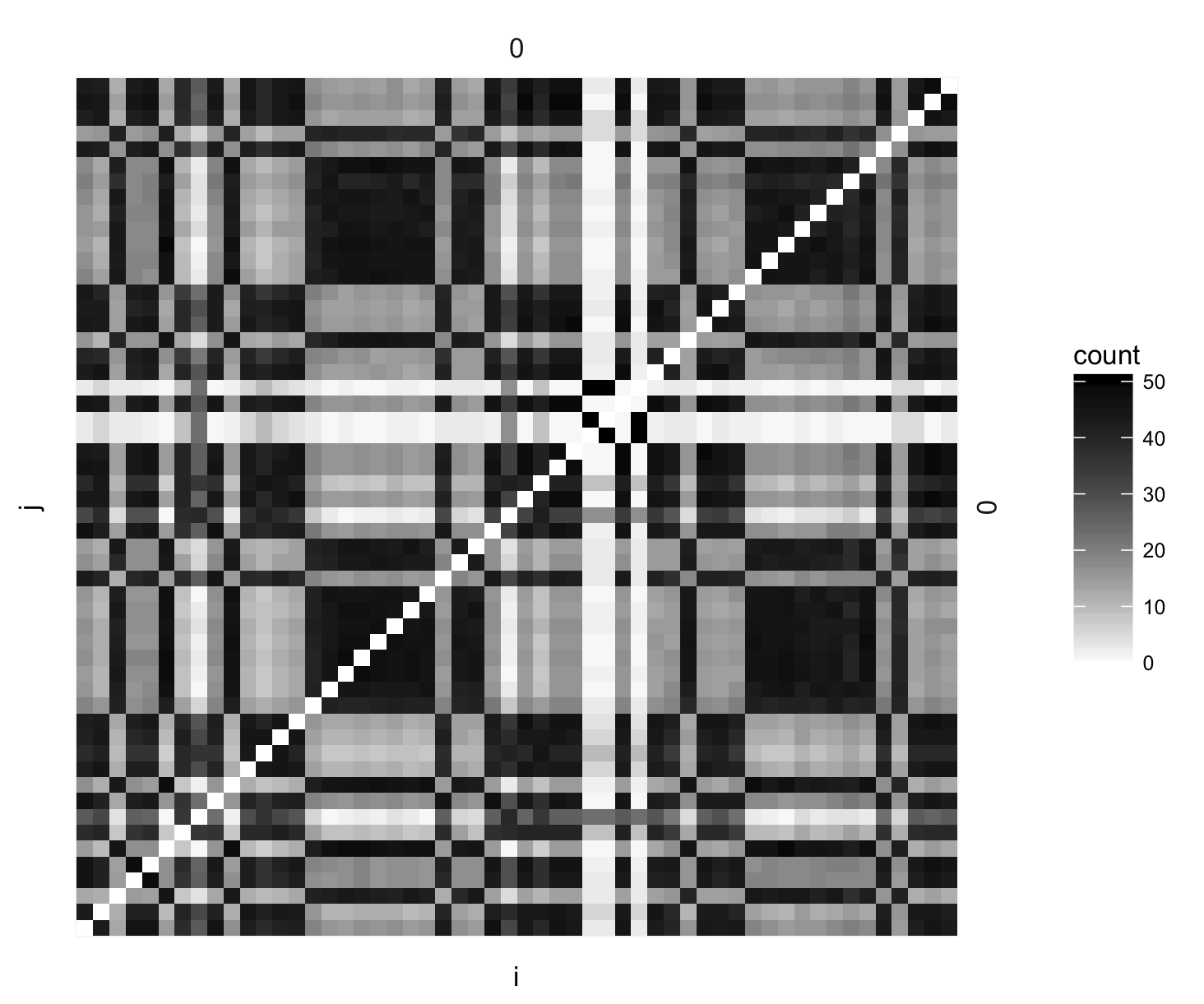}
  \caption{We can count how often two timepoints were placed in the same cluster
    (note that this statistic is invariant to label-switching). We see clear
    antibiotic regime groups. \label{fig:igp_abt_cooccurrence}}
\end{figure}

Figure \ref{fig:igp_abt_cooccurrence} provides an even more succinct
representation of the different regimes. The number of times two points are
assigned to the same state gives a measure of regime similarity, and block
structure is evidence of distinct temporal regimes.

\subsection{Switching Linear Dynamical Systems}
\label{subsubsec:switching_dynamical}

Switching Linear Dynamical Systems (SDLSs) are a blend of HMMs (Section
\ref{subsec:hmms} and LDSs (Section \ref{subsec:linear_dynamical_systems})
\citep{ghahramani1998variational, fox2009sharing, linderman2016recurrent}. The
motivation for such a model is that there may be good reason to believe that the
system exhibits regime switching behavior, but that the HMM's assumption that
samples are drawn i.i.d. conditional on latent regimes may be naive. A more
plausible assumption may be that, within a regime, observations follow
homogeneous dynamics, but that neighboring timepoints may be dependent on one
another. Consider an imaginative example from \citep{linderman2016recurrent}:
when a mouse is foraging for food, it searches slowly through a garden, but when
it notices a predator, it suddenly begins to move rapidly to evade any threat.
The dynamics of the mouse's movements are stable within each of the two regimes,
even though its average position is not\footnote{More traditional examples
  include the trajectories of fighter aircraft executing different types of
  maneuvers, the prices of stocks in response to world events, or patterns of
  EEG waveforms during different activities. We will of course be interested in
  trajectories of bacterial abundances across environmental shifts.}.

To encode this intuition in a mathematical model, suppose a time series
$\left(y_{t}\right)$ evolves according to different LDSs during different time
intervals. Following the formulation of \citep{linderman2016recurrent},
\begin{align*}
  y_{t} \vert  x_t, z_t &\sim \Gsn\left(C_{z_t}, x_t, R_{z_t}\right) \\
  x_{t} \vert x_{t - 1}, z_{t} &\sim \Gsn\left(A_{z_t}x_{t - 1}, Q_{z_t}\right) \\
  z_t \vert z_{t - 1} &\sim P_{z_{t - 1}}.
\end{align*}
$z_t \in \{1, \dots, K\}$ describes what regime the series is in at time $t$
while $x_t$ describes underlying state dynamics. The collections $\left(C_k,
R_k\right)_{k = 1}^{K}$ and $\left(A_k, Q_k\right)$ correspond to $K$ sets of
emission and state evolution parameters, respectively. We will think of
$\left(x_t\right)$ and $\left(z_t\right)$ as latent data, while $\Theta = \{P,
A_k, Q_k, C_k, R_k\}$ are parameters that must be learned. Conjugate priors are
placed on $\Theta$ -- each row $P_k$ of $P$ is given a $\Dir\left(\alpha\right)$
prior, while each $\left(A_k, Q_k\right)$ and $\left(C_k, R_k\right)$ pair is
given a MNIW prior.

Inference can be done by blocked Gibbs sampling. The conditional for $\Theta$
can be decomposed across each of the $K$ sets of parameters, and is tractable
due to conjugacy. Conditionals for $\left(z_t\right)$ and $\left(x_t\right)$ can
be derived using elementary calculations analogous to those for blocked sampling
of the sticky HMM, as detailed in Section \ref{sec:sticky_hmms}. However, a more
enlightened approach to the same updates makes use of message passing, so will
briefly review the basics of message passing.

\paragraph{Message passing basics}
\label{paragraph:message_passing}

Message passing is a device for efficiently organizing high-dimensional
integrals. It is particularly useful in probabilistic modeling settings, where
we our goal may be to evaluate the marginal of a large joint density, which can
be be decomposed into many conditionally independent components. As a concrete
example, following \cite{fox2009bayesian}, suppose we can write
\begin{align*}
p\left(x\right) &\propto \psi_{12}\left(x_1, x_2\right)\psi_{23}\left(x_2, x_3\right)\psi_{24}\left(x_2, x_4\right) \prod_{i = 1}^{4}\psi_{i}\left(x_i\right)
\end{align*}
and that we would like a closed-form expression for the marginal
$p\left(x_1\right)$. Note that we can order the necessary integrals according to
\begin{align*}
  p\left(x_1\right) &\propto \psi_1\left(x_1\right)
  \int_{X_2}\left[\psi_{12}\left(x_1, x_2\right)\psi_2\left(x_2\right)
    \int_{X_3} \psi_{23}\left(x_2, x_3\right) \psi_3\left(x_3\right) dx_3
    \int_{X_4} \psi_{24}\left(x_2, x_4\right)\psi_4\left(x_4\right) dx_4
    \right] dx_2
\end{align*}
which avoids redundant computation.

To suggest a more general lesson, define a set of messages
$m_{ji}\left(x_i\right)$, going from $j$ to $i$. These integrate over the
$j^{th}$ variable and are functions of the $i^{th}$,
\begin{align*}
  m_{32}\left(x_2\right) &= \int_{X_3} \psi_{23}\left(x_2, x_3\right)\psi_3\left(x_3\right) dx_3 \\
  m_{42}\left(x_2\right) &= \int_{X_4} \psi_{24}\left(x_2, x_4\right)\psi_4\left(x_4\right) dx_4 \\
  m_{21}\left(x_1\right) &= \int_{X_2} \psi_{12}\left(x_1, x_2\right) \psi_2\left(x_2\right) m_{32}\left(x_2\right) m_{42}\left(x_2\right) dx_2.
\end{align*}
Observe then that the marginal of interest can be concisely written as
\begin{align}
  \label{eq:message_passing_example}
  p\left(x_1\right) &\propto \psi_1\left(x_1\right) m_{21}\left(x_1\right).
\end{align}
Next consider arbitrary distributions that can be decomposed as
\begin{align*}
  p\left(x_i\right) &= \prod_{C_j} \psi_{C_j}\left(x_{C_j}\right)\psi_{j}\left(x_{C_j}, y_{j}\right),
\end{align*}
where the $y_i$ are fixed in advanced (typically this is data upon which we
condition). Each $C_j$ is a \textit{clique} in the undirected conditional
independence graph linking the $x_i$. Define messages according to
\begin{align*}
  m_{ji}\left(x_{C_i}\right) &= \int_{X_{C_{j}}} \psi_{C_i}\left(x_{C_{i}}\right) \psi_{i}\left(x_{C_{i}}, y_{i}\right) \prod_{k \in N\left(j\right)} m_{kj}\left(x_{C_{j}}\right) dx_{C_{j}},
\end{align*}
where $N\left(j\right)$ denotes the neighbors of node $j$.

It turns out that we can always write
\begin{align}
  \label{eq:message_update}
  p\left(x_{i} \vert y\right) &\propto \psi_{i}\left(x_{C_i}, y_i\right) \prod_{j \in N\left(i\right)} m_{ji}\left(x_{C_i}\right),
\end{align}
which parallels equation \ref{eq:message_passing_example}.

One interesting application of this technique is a quick derivation of the
forwards-backwards algorithm for HMMs. Denote observations by $\left(x_t\right)$
and latent states by $\left(z_t\right)$. Note that we can write the joint as a
product of cliques made from latent states and their associated emission,
\begin{align*}
  p\left(x, z\right) &= \left[p\left(z_1\right) \prod_{t = 1}^{T - 1} p\left(z_{t + 1} \vert z_t\right)\right]
  \left[\prod_{t = 1}^{T} p\left(x_t \vert z_t\right)\right] \\
    &= \left[\psi_1\left(z_1\right)\prod_{t = 1}^{T - 1} \psi_{t, t + 1}\left(z_t, z_t +
    1\right)\right] \left[\prod_{t = 1}^{T} \psi_t\left(z_t, x_t\right)\right],
\end{align*}
where we defined $\psi_{t, t + 1}$ and $\psi_t$ so that the last two equations
match. By definition, the messages have the form
\begin{align*}
  m_{t - 1, t}\left(z_{t}\right) &= \sum_{z_{t - 1} = 1}^{K} \psi_{t - 1, t}\left(z_{t - 1}, z_t\right) \psi_{t}\left(z_t, x_t\right) m_{t - 2, t - 1}\left(z_{t - 1}\right) \\
  &= \sum_{z_{t - 1} = 1}^{K} p\left(z_t \vert z_{t - 1}\right)p\left(x_t \vert z_t\right)m_{t - 2, t - 1}\left(z_{t - 1}\right)
\end{align*}
for those going left-to-right and
\begin{align*}
  m_{t + 1, t}\left(z_t\right) &= \sum_{z_{t + 1} = 1}^{K} p\left(z_{t + 1}\vert z_t\right)p\left(x_t \vert z_t\right)m_{t + 2, t + 1}\left(z_{t + 1}\right),
\end{align*}
for those going right-to-left.

According to the message passing update in equation \ref{eq:message_update}, the
marginals of interest have the form
\begin{align*}
  p\left(z_{t} \vert y_{1:T}\right) &\propto p\left(x_t \vert z_t\right) m_{t - 1, t}\left(z_t\right)m_{t + 1, t}\left(z_t\right)
\end{align*}
which has the exactly the same form as found during the derivation of the
forwards-backwards algorithm in Section \ref{subsec:hmms}.


\paragraph{Message passing for SLDS}

As demonstrated in the HMM forwards-backwards calculation, the message passing
abstraction can be used to compute complex marginals in a way that avoids what
can otherwise become tedious calculation. Consider again inference for the SLDS.
We claimed that we block sample both $\left(z_t\right)$ and $\left(x_t\right)$.
The idea for both steps is to perform a forwards message passing sweep followed
by backwards sampling\footnote{This method is sometimes called
  Forwards-Filtering Backwards-Sampling \citep{carter1994gibbs}.}.

We provide details for block sampling $\left(x_t\right)$ conditional on
$\left(z_t\right), \left(y_t\right)$, and $\Theta$. The key identity for
backwards sampling is
\begin{align*}
  p\left(x_{1:T} \vert z_{1:T}, y_{1:T}, \Theta\right) &= p\left(x_{T} \vert z_{1:T}, y_{1:T}, \Theta\right) \prod_{t = 1}^{T - 1} p\left(x_{t} \vert x_{t + 1:T}, z_{1:T}, y_{1:T}, \Theta\right),
\end{align*}
and the idea is to sample $x_{T}$ given all the $z_{t}$ and $y_{t}$'s, and then
proceed forwards, sampling $x_{t}$ given in addition all the later $x_{t +
  1:T}$. In the following, we suppress notation for dependence on $\Theta$. A
single message passing forwards pass is sufficient for sampling each of these
densities. To see this, note that jointly
\begin{align*}
  p\left(x_{1:T} \vert y_{1:T}, z_{1:T}\right) &\propto
  \prod_{t = 1}^{T} \Gsn\left(y_t \vert C_{z_t} x_t, R_{z_t}\right) \prod_{t = 2}^{T} \Gsn\left(x_{t} \vert A_{z_t} x_{t - 1}, Q_{z_t}\right) \\
    &\propto\prod_{t = 1}^{T} \psi_{t}\left(x_{t}, y_{t}, z_{t}\right) \psi_{t - 1, t}\left(x_{t - 1}, x_{t}, z_{t}\right)
\end{align*}
and that the associated messages have the form
\begin{align*}
  m_{t - 1, t}\left(x_t\right) &= \int \psi\left(x_{t} , y_t, z_t\right)\psi\left(x_{t - 1}, x_t, z_t\right) m_{t - 2, t - 1}\left(x_{t - 1}\right) dx_{t - 1}
\end{align*}
which can be computed in closed-form, by Gaussian marginalization.

Once these messages have been computed in a forwards filtering pass, the
densities required for backwards sampling can be found as
\begin{align*}
  p\left(x_t \vert x_{t + 1:T}, y_{1:T}, z_{1:T}\right) &=
  \int_{X_1 \times \dots \times X_{t - 1}} p\left(x_{1:t} \vert x_{t + 1: T}, y_{1:T}, z_{1:T}\right) dx_{1:(t - 1)} \\
  &\propto \int_{X_1 \times \dots X_{t - 1}} \prod_{i = 1}^{t} \psi\left(x_i, y_i, z_i\right) \prod \psi\left(x_{i - 1}, x_i, z_i\right) dx_{1:\left(t - 1\right)} \\
  &\propto \int_{X_{1}} \psi\left(x_{t - 1}, y_{t - 1}, z_{t - 1}\right)\psi\left(x_{t - 2}, x_{t - 1}, z_{t - 1}\right) m_{t - 2, t - 1}\left(x_{t - 1}\right) d x_{t - 1},
\end{align*}
using the messages computed in the forwards pass.

A similar calculation provides the block-sampling update for
$\left(z_{t}\right)$ given all other variables (in a way it is simpler, since it
is a discrete sum rather than an integral). The updates for $\Theta$ are
available by conjugacy. Together, these three updates allow efficient block
Gibbs sampling of the SLDS.

\subsubsection{Example}
\label{subsubsec:slds_example}

We next provide an example application of SLDS to the antibiotics data. We use
an implementation provided by the pyslds package, available at
\url{https://github.com/mattjj/pyslds}. The purpose of this analysis is to
determine both a regime segmentation of timepoints for each series, along with a
characterization of the dynamics within individual regimes. Our code is
available at
\url{https://github.com/krisrs1128/tsc_microbiome/tree/master/src/slds}.

It is necessary to decide whether to share the parameters $\Theta$ across all
species or not. There is a tradeoff between the two alternatives,
\begin{itemize}
\item If we share regime parameters across all species, then information can be
  shared across species with similar dynamics.
\item If we estimate parameters for each species separately, we do not need to
  be concerned that some species have very different overall abundances or
  dynamics.
\end{itemize}
Ideally, it would be possible to limit the sets of parameters that are available
to each species, so that only similar species shared information. This type of
partial sharing was proposed in \citep{fox2009sharing} for a similar multiple
time series model, for example.

In this example, we fit species specific models, however. In addition to the
tradeoff above, our decision is based on (1) the fact that we have already seen
the second approach in Section \ref{subsubsec:hmm_example} and (2) the practical
limitation that the pyslds package only provides state estimation for individual
sequences.

Note that the state identities $z_{it}$ will no longer be aligned across
sequences $i$, since estimation is performed across each species $i$ separately.
Instead of attempting to align the estimated state sequences, which is
complicated even in the (rare) case that states are actually shared across all
species, we cluster the estimated parameter sequences.

In more detail, we study the parameters $\Theta_{ik} := \{C_{ik}, R_{ik},
A_{ik}, Q_{ik}\}$ associated with the dynamic state of species $i$ at time $t$,
where $z_{it} = k$. Not only does this allow us to avoid the label-switching
problem, it provides a way of inspecting gradients of variation in dynamic
regimes, according to smooth variation in the associated parameters.

Before applying this model, we $\asinh$ transform the raw count data, as in our
other applications. We further center each species' series, since the marginal
mean for each species must always be zero.

After fitting an SLDS model, we have clustered species according the posterior
mean of the parameters $\left(\Theta_{ik}\right)_{k =1}^{K}$ after laying them
out across timepoints. In more detail, we represent species $i$ by the vector
$\left(\Theta_{i z_{i1}}, \dots, \Theta_{i z_{iT}}\right)$, hierarchically
cluster them using a Euclidean distance, and then arrange species according to
the order of leaves on the resulting tree.

\begin{figure}
  \centering
  \includegraphics[width=\textwidth]{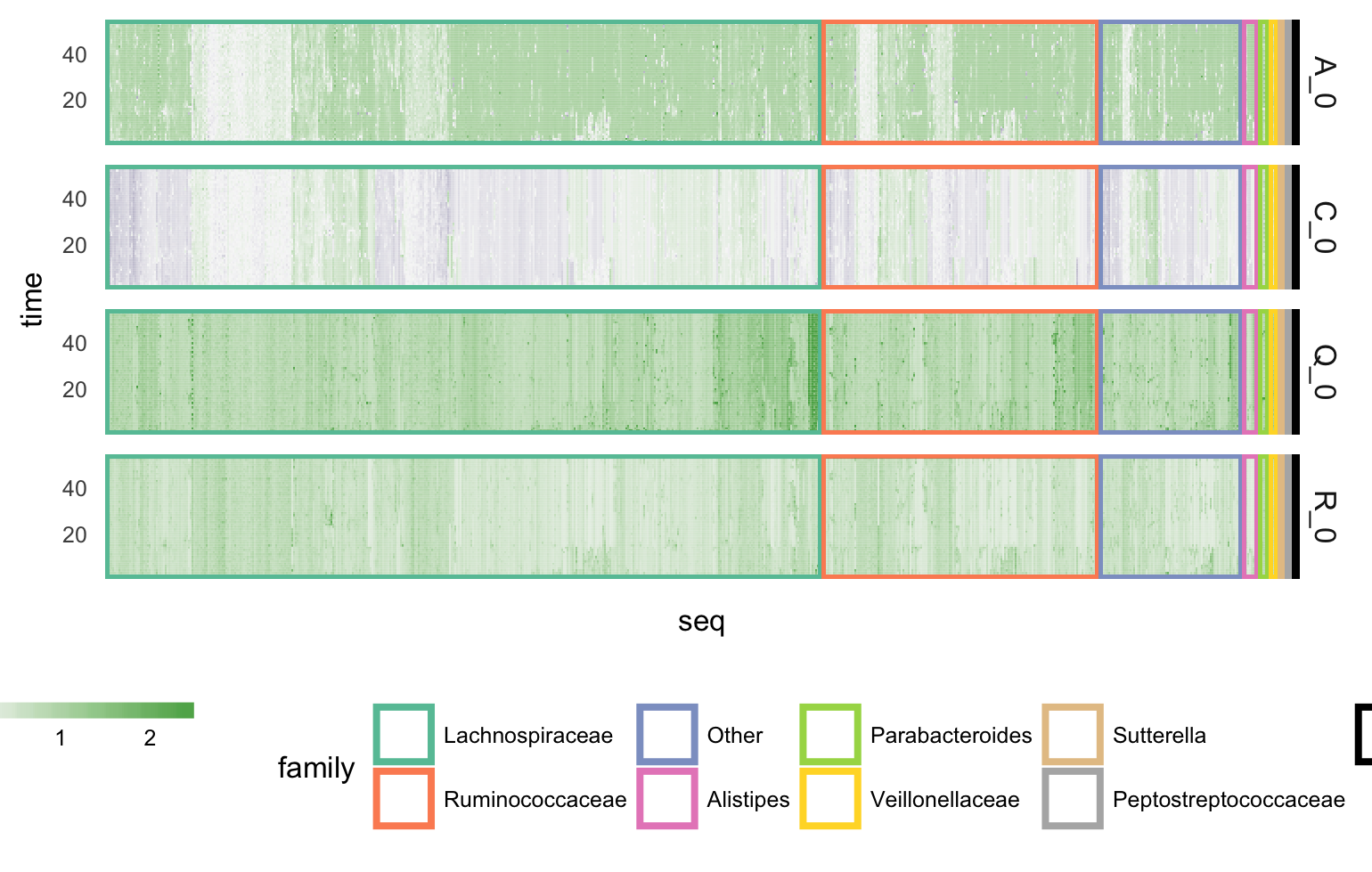}
  \caption{Heatmap of SLDS posterior means associated with parameters, across
    species and timepoints. Each row of panels is associated with one type of
    SLDS parameter. Rows and columns of cells correspond to timepoints and
    species, respectively. The border around panels provides the taxonomic
    identity for individual species. Cells are shaded in according to their
    parameter estimates, and species are sorted by a hierarchical clustering.
    Blue, white, and green correspond to negative, zero, and positive parameter
    estimates, respectively. Estimates that are very large or small (above 2.1
    or less than -1.1) have been thresholded, to keep the plot from being
    dominated by outliers. \label{fig:slds_parameter_heatmap}
  }
\end{figure}

The resulting estimates are displayed in Figure
\ref{fig:slds_parameter_heatmap}. $A_k$ and $Q_k$ describe the autocorrelation
and noise of underlying state dynamics. The noise generally seems about the same
size as the autocorrelation. The autocorrelations $A_{k}$ seem to increase after
the antibiotics time courses, which corresponds to the decrease in overall
abundance during those time intervals, but the noise levels seem generally
stable across timepoints.

The emissions parameters are somewhat more complex. Within each taxa, about a
third of species have negative, approximately zero, and positive emission
matrices $C_{k}$ each. Species that have positive (negative) emissions tend to
have positive (negative) emissions across all timepoints. During the antibiotics
time courses, the emissions matrices shrink towards zero, which fits zero counts
during this regime. Similarly, when the abundance decreases, the noise levels
$R_k$ shrink.

Now that these parameters have been estimated, it would be possible to simulate
series according to different regimes. This could be an alternative device for
interpreting fitted parameters.

\subsection{BASIC changepoint detection}

The methods considered so far have always made two modeling choices,
\begin{itemize}
\item There can be no partial sharing in regime switching behavior over series.
  That is, either series are modeled independently, in which case there is no
  sharing of regime changes, or, at the opposite extreme, all series must change
  simultaneously, with no exceptions for any less typical series.
\item Parameters associated with switches are modeled explicitly, and it is
  possible to generatively simulate according to different regimes, once the
  overall model is fit.
\end{itemize}

However, neither choice is necessary, and \cite{fan2015empirical} describe an
algorithm, Bayesian Analysis of Simultaneous Changepoints (BASIC), that discards
them, with the goal of improving interpretability and model estimation
properties. Indeed, allowing shared changepoints for some, but not all, time
series can improve model fit and interpretability. Further, collapsing
regime-specific parameters can facilitate estimation, according to certain
theoretical studies \citep{liu1994collapsed}. In addition to detailing the model
and inference, \cite{fan2015empirical} also describe an Empirical Bayesian
mechanism for prior elicitation.

We summarize the BASIC model and the key observations used for inference, before
discussing an application to the antibiotics data in Section
\ref{subsec:basic_example}.

\subsubsection{Model}
\label{subsec:basic_model}

Let $\left(x_{it}\right)_{i = 1}^{n}$ denote the abundance of species $i$ at
time $t$. We imagine species $i$'s abundances are drawn i.i.d. from a likelihood
model with fixed parameters until the next changepoint. Certain timepoints are
thought to have higher propensity for containing changepoints, across all
species. More formally, the assumed model has the form,
\begin{align*}
  x_{it} \vert \theta_{it} &\sim p\left(x_{it} \vert \theta_{it}\right) \\
  \theta_{it} \vert z_{it}, \theta_{i,t - 1} &\sim z_{it}\pi_{\Theta}\left(\theta\right) + \left(1 - z_{it}\right) \delta_{\theta_{i, t - 1}}\left(\theta\right) \\
  z_{it} \vert q_{t} &\sim \Ber\left(q_{t}\right) \\
  q_t &\sim \pi_{Q}\left(q\right).
\end{align*}

Here, $z_{it} = 1$ means a changepoint occurs in species $i$ at time $t$. The
inference algorithm proposed for this model is general enough to accomodate
arbitrary conjugate likelihood-prior pairs, only requiring closed-form
expressions for the marginal likelihoods of time segments for individual
species,
\begin{align*}
  P_i\left(t, s\right) &= \int \prod_{r = t}^{s} p\left(x_{ir} \vert \theta\right) \pi_{\Theta}\left(\theta\right) d\theta.
\end{align*}

For example, in our applications, we will consider Gaussian-Gaussian and
Beta-Bernoulli pairs, to model abundance and presence-absence data,
respectively.

\subsubsection{Inference}
\label{subsubsec:basic_inference}

\cite{fan2015empirical} propose sampler with block Gibbs and Metrpolis-Hastings
elements to sample from $p\left(\left(z_{it}\right) \vert
\left(x_{it}\right)\right)$, the posterior after having marginalized out all the
$\theta_{it}$ and $q_t$. Then, an empirical bayes approach for selecting
hyperparameters in the priors $\pi_{Q}$ and $\pi_{\Theta}$ is described.

The sampler iterates row and column block Gibbs sampling followed by a
Metropolis-Hastings corrected jittering step to refine changepoint positions.
For rowwise block Gibbs, the goal is to sample each row $i$ according to
\begin{align*}
  p\left(z_{i\cdot} \vert \left(x_{it}\right), \left(z_{-i \cdot}\right)\right).
\end{align*}
For notational convenience, we will write $z_{i} := z_{i\cdot}$ for the $i^{th}$
row of $z$s and $z_{-i} := z_{-i\cdot}$ for the matrix made up of all but the
$i^{th}$ row. In a similar spirit, we write $z$ and $x$ for the full matrix of
$z$s and $x$s, instead of the more cumbersome $\left(z_{it}\right)$ and
$\left(x_{it}\right)$. Further, define variables representing the probability of
a changepoint occurring at time $t$ in sequence $i$, after having observed all
other sequences,
\begin{align}
  \label{eq:basic_cit}
  c_{i}\left(t\right) &:= p\left(z_{it} \vert z_{-i}\right),
\end{align}
as well as the likelihood of sequence $i$ from time $t$ to the end, after having
observed a changepoint at time $t$ and changepoints for all other sequences,
\begin{align*}
  Q_{i}\left(t\right) &:= p\left(x_{i, t:T} \vert z_{it} = 1, z_{-i}\right).
\end{align*}

Then the posterior probability of no changepoints in sequence $i$ exactly until
time $t$ can be expressed as
\begin{align}
  p\left(z_{i, 1:\left(t - 1\right)} = 0, z_{it} = 1 \vert \left(x_{it}\right), z_{-i}\right) &= P_{i}\left(1, t\right)\frac{Q_{i}\left(t\right)}{Q_{i}\left(s\right)}\left[\prod_{r = s + 1}^{t - 1} \left(1 - c_{i}\left(r\right)\right)\right] c_{i}\left(t\right), \label{eq:first_changepoint}
\end{align}
using Bayes' rule.

For subsequent segments between changepoints, similar reasoning yields
\begin{align}
  p\left(z_{i, \left(s + 1\right):\left(t - 1\right)} = 0, z_{it} = 1 \vert z_{is} = 1, x, z_{i, 1:\left(t - 1\right)}, z_{-i}\right) &= P_{i}\left(s, t\right)\frac{Q_{i}\left(t\right)}{Q_{i}\left(s\right)} \left[\prod_{r = s + 1}^{t - 1} \left(1 - c_{i}\left(r\right)\right)\right]c_{i}\left(t\right). \label{eq:subsequent_changepoints}
\end{align}
Hence, if $P_{i}\left(s, t\right)$, $c_{i}\left(t\right)$ and
$Q_{i}\left(t\right)$ are available, then for all $s, t$, then the row $z_{i}$
can be sampled by identifying the first changepoint according to equation
\ref{eq:first_changepoint} and the remaining ones using equation
\ref{eq:subsequent_changepoints}.

To find $c_{i}\left(t\right)$, observe that
\begin{align*}
  p\left(z_{it} = 1 \vert z_{-i}\right) &\propto p\left(z_{-i} \vert z_{it} = 1\right) p\left(z_{it} = 1\right) \\
  &= q_{t}^{N_{-i}\left(t\right)} \left(1 - q_{t}\right)^{n - N_{-i}\left(t\right)} \pi_{Q}\left(q_{t}\right),
\end{align*}
where we set $N_{-i}\left(t\right) := \{\# \text{changepoints at time $t$,
excluding sequence } i\} = \sum_{i^\prime \neq i} z_{i^\prime t}$.

To compute $Q_{i}\left(t\right)$ efficiently, we can use a dynamic programming
routine. Starting from $Q_{i}\left(T, T + 1\right) = P_{i}\left(T, T +
1\right)$, we can make a forwards pass through the sequence, according to the
recursion,
\begin{align*}
  Q_{i}\left(t\right) &= \sum_{s = t + 1}^{T} \left[\prod_{r = t + 1}^{s - 1} \left(1 - c_{i}\left(r\right)\right)\right]c_{i}\left(s\right)P_{i}\left(t, s\right)Q_{i}\left(s\right) +
  \left[\prod_{r = s + 1}^{T}\left(1 - c_{i}\left(r\right)\right)\right]P_{i}\left(t, T + 1\right).
\end{align*}
The first term corresponds to the case that there is an intermediate changepoint
at time $s$, for some $s > t$, and the second is the case that $t$ is the last
changepoint in sequence $i$.

This completes specification of the rowwise-sampling step. For columnwise
sampling, similar dynamic programming ideas apply. We denote the
conditional probability that a certain entry $z_{it} = 1$, after having observed
$z$ at all other timepoints,
\begin{align*}
  c_{t}\left(i\right) :&= p\left(z_{it} = 1 \vert z_{-t}, q_{t}\right),
\end{align*}
which is the column analog of equation \ref{eq:basic_cit}. For the likelihood of
the $i^{th}$ sequence over the interval containing $t$, take $t$ to be either a
changepoint or not,
\begin{align*}
  A_{t}\left(i\right) := P_{i}\left(r_{t}\left(i\right), t\right)P_{i}\left(t, s_{t}\left(i\right)\right) \\
  B_{t}\left(i\right) := P_{i}\left(r_{t}\left(i\right), s_{t}\left(i\right)\right),
\end{align*}
where $r_{t}\left(i\right)$ and $s_{t}\left(i\right)$ denote the times for the
changepoints immediately preceding and following $t$, in the $i^{th}$ sequence.

With this notation, it then follows that
\begin{align*}
  p\left(z_{it} = 1 \vert z_{1:(i - 1)}, z_{-t}, x\right) &= \frac{A_{t}\left(i\right)c_{t}\left(i\right)}{A_{t}\left(i\right)c_t\left(i\right) + B_t\left(i\right)\left(1 - c_t\left(i\right)\right)},
\end{align*}
which can be used to sample the $t^{th}$ column of $z$, from the top down. Since
$A_{t}\left(i\right)$ and $B_{t}\left(i\right)$ can both be computed from the
marginal likelihoods $P_{i}\left(r, s\right)$, which are assumed available
analytically, for sampling columns $z_t$, it is sufficient to have an efficient
way of computing the $c_{t}\left(i\right)$. However, using Bayes' rule and the
fact that a changepoint either does or doesn't occur in sequence $i$ at time
$t$, it can be shown that $c_{t}\left(i\right) = \frac{\int q
  \tilde{\pi}_{it}^{c}\left(q\right) dq}{\int \tilde{\pi}_{it}^{c}\left(q\right)
  dq}$, where
\begin{align*}
  \tilde{\pi}_{it}^{c}\left(q\right) &:= \left[\prod_{i^\prime = i + 1}^{n} \left(A_{t}\left(i^\prime\right)q + B_{t}\left(i^\prime\right)\left(1 - q\right)\right)q^{N_{-t}\left(i\right)}\left(1 - q\right)^{i - 1 - N_{-t}\left(i\right)}\right] \pi_{Q}\left(q\right),
\end{align*}
is the unnormalized likelihood times prior. Note the coefficients in the product
$\prod_{i^\prime = i + 1}^{n} \left[A_{t}\left(i^\prime\right)q +
B_{t}\left(i^\prime\right)\left(1 - q\right)\right]$ can be computed efficiently by
starting with $i = n$ and then updating the coefficients in front of each term
$q^k \left(1 - q\right)^{n - i - k}$ as $i$ is decremented to 1.

Next, consider the choice of priors $\pi_{Q}$ and $\pi_{\Theta}$, for the
changepoint probabilities and likelihood parameters, respectively. In practice,
it can be hard to elicit these priors, and a common strategy is to define
parameterized families of prior $\pi_{Q}^{\alpha}$ and $\pi_{\Theta}^{\beta}$.
Inference can then proceed by placing an additional layer at the top of the
hierarchical model (the fully Bayesian approach), or alternatively choosing the
members of these families that maximize the marginal likelihood (the Empirical
Bayesian approach). \cite{fan2015empirical} describe an interesting Empirical
Bayesian (EB) approach that uses rich families of priors $\pi_{Q}^{w, \nu}$ and
$\pi_{\Theta}^{\eta}$,

In more detail, that the marginal likelihood has the form
\begin{align}
 \label{eq:pi_q_marginal_lik}
 \log p\left(x, z \vert w, nu\right) &= \log p\left(x, \vert z, \eta\right) + \log p\left(z \vert \left(w_k\right)\right) \\
 &= \sum_{i = 1}^{n} \sum_{r, s \in \S\left(z_{i}\right)} \log P_{i}\left(r, s \vert \eta \right) + \sum_{i = 1}^{n} \log p\left(z_{i} \vert \left(w_{k}\right)\right)
\end{align}

In a typical EB analysis, we might consider a Beta family for $\pi_{Q}$, since
$z_{it} \sim \Ber\left(q\right)$. That is, we would suppose $\pi_{Q}^{a_0, b_0}
= \Bet\left(q \vert a_{0}, b_{0}\right)$ and choose $a_{0}$ and $b_0$ to
maximize the marginal likelihood in equation \ref{eq:pi_q_marginal_lik}.
However, since the marginal likelihood is not directly available, this
maximization would be done iteratively, replacing intractable expectations with
Monte Carlo samples of the $z_i$ drawn from a current EB estimate of the prior,
then redrawing $z_i$, and so on until convergence.

A nonparametric alternative, proposed by \citep{fan2015empirical}, considers
$\pi_{Q}^{w, \nu}\left(q\right) = \sum_{k = 1}^{K} w_{k} \nu_{k}\left(q\right)$,
for some weights $w_{k}$ and predefined basis functions $\nu_{k}$. For example,
$\nu_{k}\left(q\right) = \delta_{1 / k}\left(q\right)$ defines a grid of point
masses over the interval $\left[0, 1\right]$. Then, the EB approach optimizes
the marginal likelihood over $\left(w_k\right)$. This problem does not have an
analytical solution, but it is convex, so can be input to generic convex
optimization software. This provides another example for how increased
computational resources have made certain modeling restrictions, like requiring
a Beta hyperprior for $q$, unecessary.


\subsubsection{Example}
\label{subsec:basic_example}

We now apply the BASIC algorithm to the antibiotics data set, since we imagine
many species may respond to the introduction and removal antibiotics in similar
ways and in sync with one another. We use an implementation available from
\url{https://web.stanford.edu/~zhoufan/software.html}, and all code for this
example is available at
\url{https://github.com/krisrs1128/tsc_microbiome/tree/master/src/changepoint}.

We first filter down to samples that are present in at least 20\% of samples,
and then we perform an $\asinh$ transformation. We decompose our analysis into
two parts -- an abundance and a presence-absence component. In the continuous
abundance analysis, we use a Normal model with changing means and variances.
This allows us to model changes in mean and variance of abundance series over
time, and is useful in situations where antibiotics might decrease abundances
without completely wiping out a species. In the second component, we consider
whether the probability of a species being present in a sample changes across
antibiotics time courses. This is accomplished by using a Beta-Bernoulli
prior-likelihood pair in the BASIC model.

\begin{figure}
  \centering
  \includegraphics[width=\textwidth]{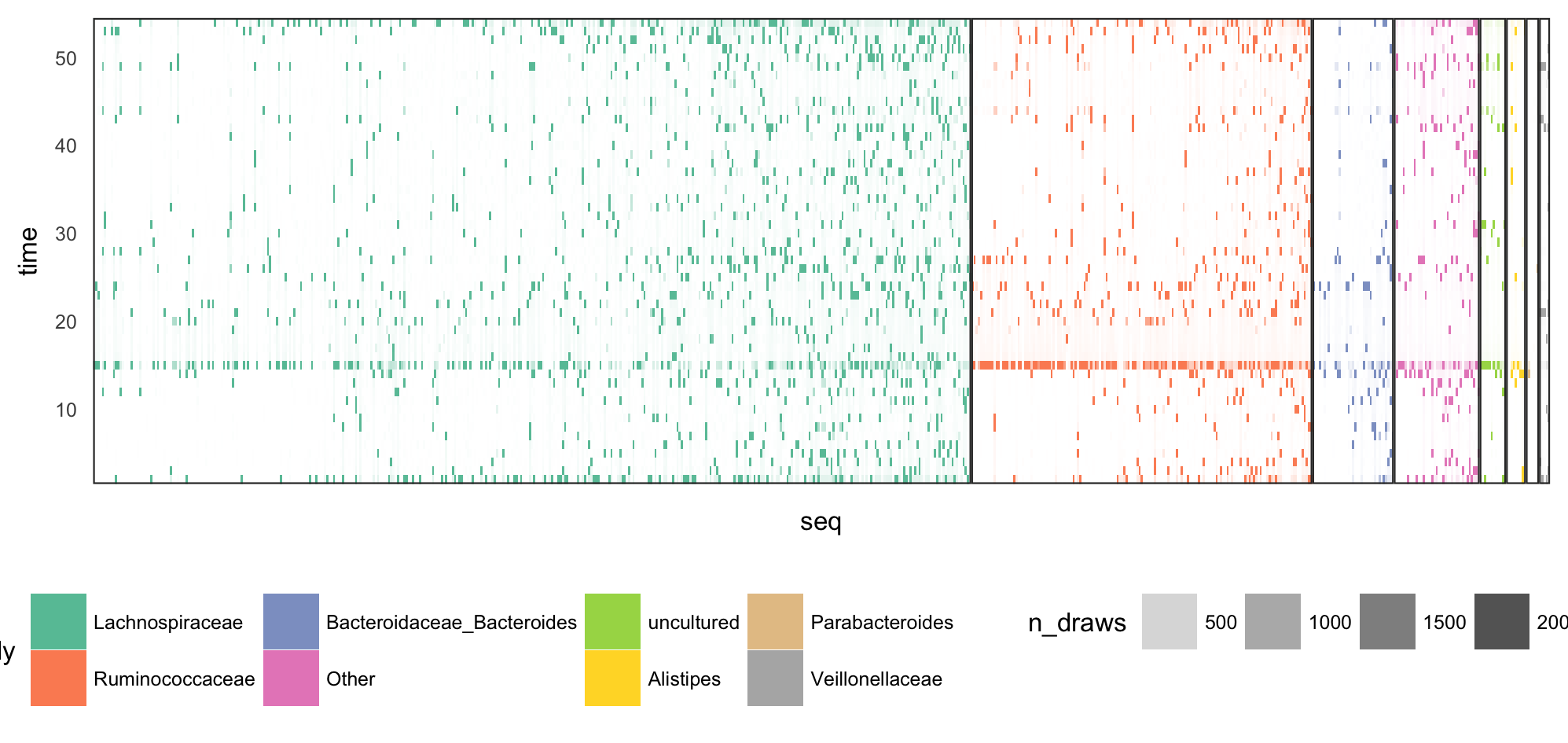}
  \caption{The estimated changepoints according to the abundance analysis, using
    a Gaussian-Gaussian likelihood in the BASIC model. Each column gives the
    sequence for a species, with earliest times at the bottom and latest times
    at the top. Species are sorted according from left to right in order of
    decreasing abundance, and colors in each tile indicate taxonomic membership.
    The darkness of the tile reflects the number of posterior samples in which
    that tile had $z_{it} = 1$.
    \label{fig:basic_heatmap} }
\end{figure}

Figure \ref{fig:basic_heatmap} displays the samples from the posterior for
$p\left(z \vert x\right)$ within the abundance model. A very clear horizontal
band marks the start of the first antibiotic time course, while other antibiotic
events seem to have less association with estimated changepoints. There is
little differentiation between different taxonomic families. One exception is
that some of the more abundant Ruminococcus appear to be more strongly affected
by the second antibiotic time course than the more abundant Lachnospiraceae.

Note that the number of tiles estimated to be changepoints increases near the
right of each panel, meaning that rarer species are estimated to have a large
number of changepoints. It is not entirely clear why this would arise, though we
suspect it is an artifact of model misspecification, and the fact that even
after $\asinh$-transforming the raw species abundances are quite right-skewed
(not to mention nonnegative). Developing an alternative likelihood model that
may be more suitable for jointly studying abundances of both rare and common
species is beyond the scope of this work, however.

\begin{figure}
  \centering
  \includegraphics[width=\textwidth]{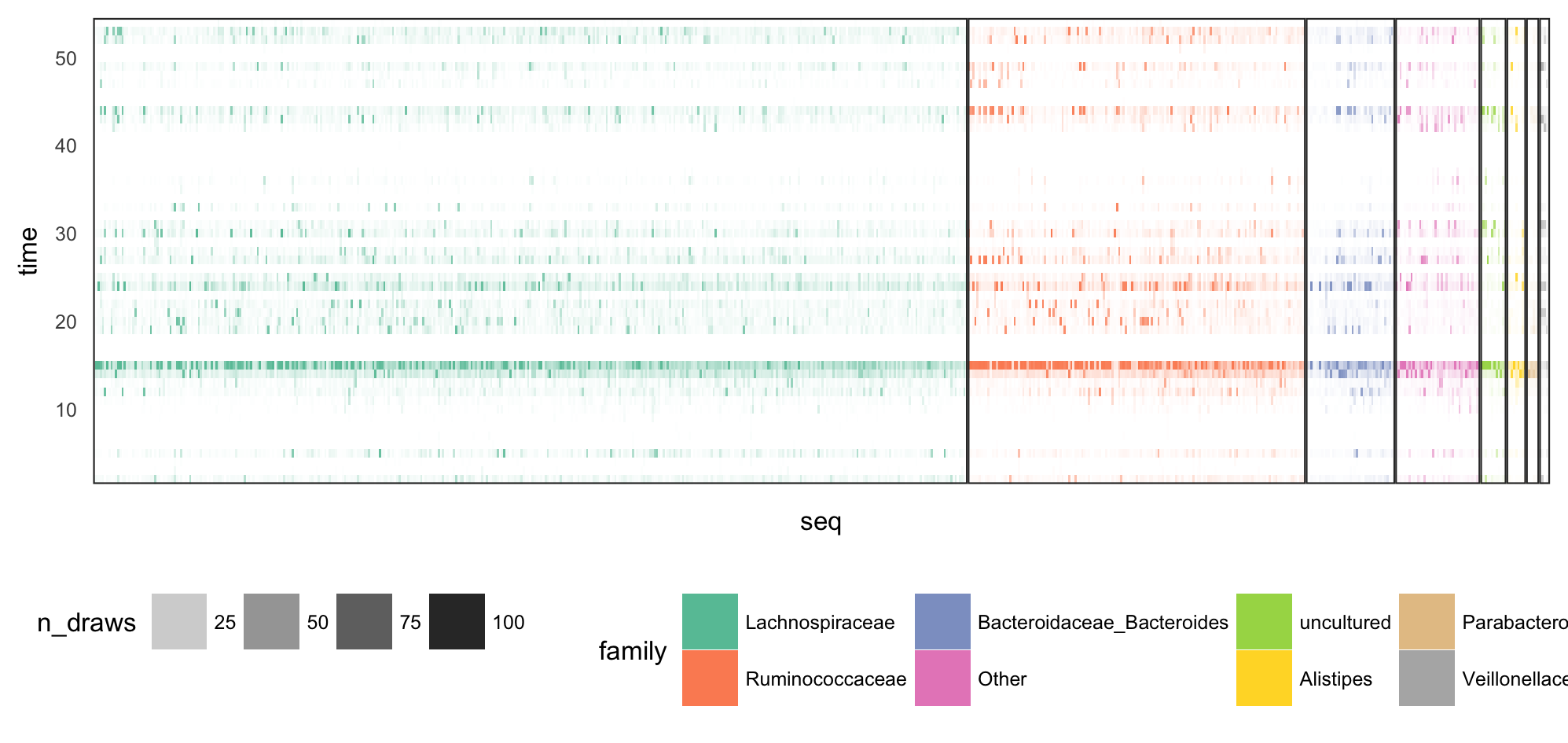}
  \caption{
    The analog of Figure \ref{fig:basic_heatmap} for the Bernoulli likelihood
    model applied to the presence-absence data. In contrast to Figure
    \ref{fig:basic_heatmap}, most species have comparable estimated
    changepoints.
    \label{fig:basic_bern_heatmap}
  }
\end{figure}

In Figure \ref{fig:basic_bern_heatmap} we display the estimated changepoints the
Beta-Bernoulli BASIC model applied to the binarized presence-absence version of
the abundance data. Evidently, the changepoint behavior between high and low
abundance species is similar when viewed through this transformation. This is
somewhat surprising, because it suggests that even the high overall abundance
species can drop down to zero during the antibiotics time courses.

The many horizontal bands above the main first time course band suggests that
there is some ambiguity in recovery times. Further, there does not appear to be
evidence for any differential recovery across species, which had been visible in
our complementary analysis. Finally, as in Figure \ref{fig:basic_heatmap}, the
effect from the second antibiotic time course is much less pronounced than that
from the first.

\begin{figure}
  \centering
  \includegraphics[width=\textwidth]{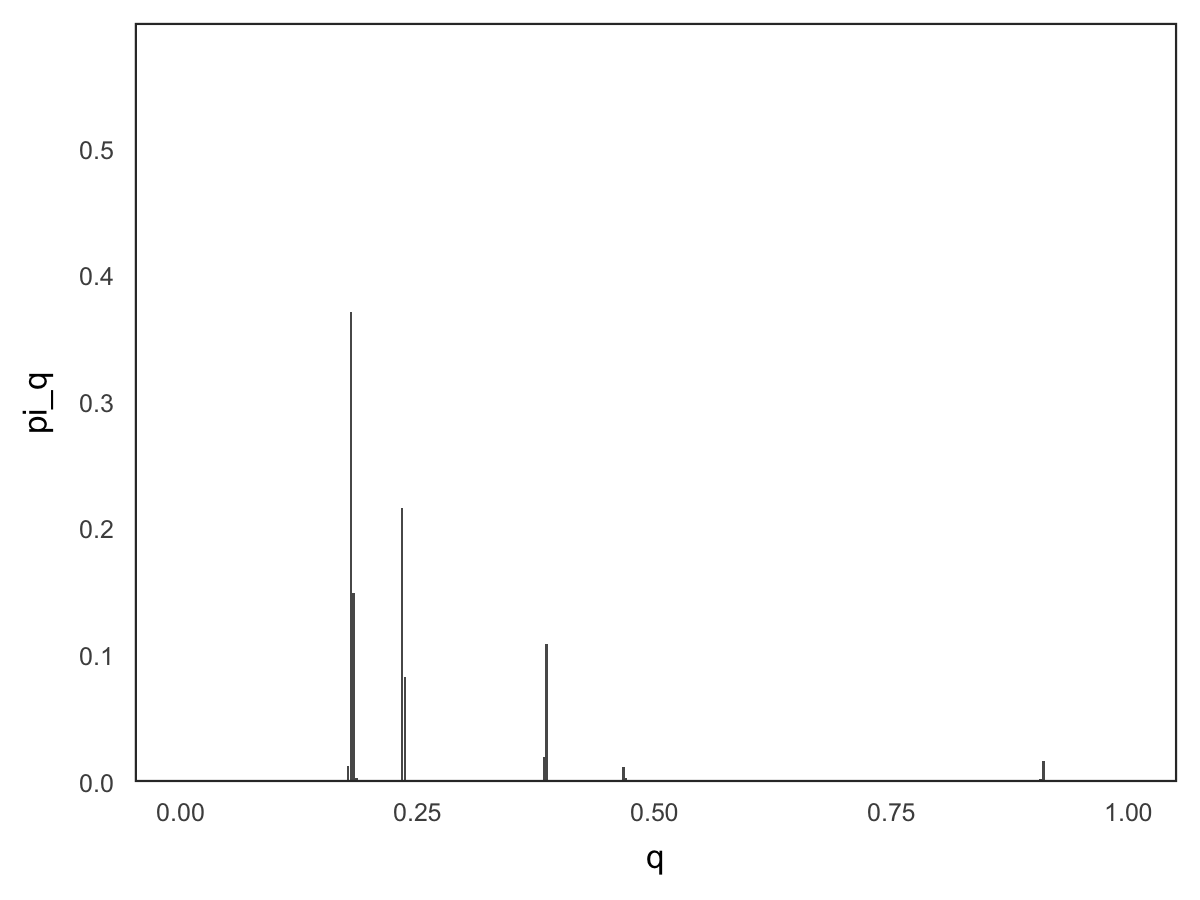}
  \caption{The prior $\pi_{Q}$ optimized using EB. Evidently, most weight falls
    on probabilities between 0.2 and 0.5, modulating the overall frequency of
    changepoints.
    \label{fig:changepoint_eb_prior} }
\end{figure}

Figure \ref{fig:changepoint_eb_prior} displays the EB optimized weights for the
prior $\pi_{Q}^{\hat{w}, \nu}$, when using $\nu_{k}\left(q\right) = \delta_{k /
  K}\left(\delta\right)$. On the one hand, it is nice that this approach
learns a flexible prior. However, it seems unusual that this prior exhibits such
strong discontinuity, placing all its mass in a few clumps. It would be
interesting to see what happens when using smooth bump functions for the $\nu_k$
instead of these discrete $\delta$s.

\section{Conclusion}

This ends our tour of techniques for modeling dynamic regimes. We have seen a
variety of methods, from classical to modern, standard to exotic, and have
provided both methodological summaries and practical applications to a real
microbiome study. We have found the regime detection problem to be one with rich
and interesting structure, which is reflected in the diversity of approaches
that have been proposed historically. Bridging our theoretical and applied
discussions, we have made implementations for all examples -- including data
preparation, model fitting, and figure generation -- easily accessible through
our repository, \url{https://github.com/krisrs1128/tsc\_microbiome}. We hope
that this review of the literature on regime detection will increase awareness
about connections between communities working on the various instances of this
problem, facilitate flexible and informed microbiome data analysis, and inspire
the development of improved methods that can identify regime structure in real
scientific problems.

\bibliographystyle{plainnat}
\bibliography{bibliography.bib}

\section{Appendix}
\label{sec:appendix}

\subsection{Supplemental Figures}
\label{subsec:supplemental_figures}

\begin{figure}
  \centering
  \includegraphics[width=0.9\textwidth]{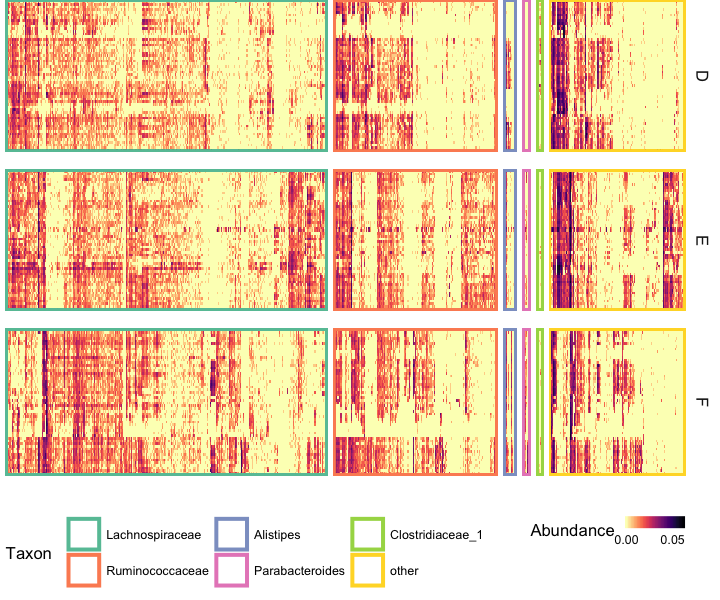}
  \caption{
    The analog of Figures \ref{fig:heatmap-euclidean} and
    \ref{fig:heatmap-jaccard}, obtained by using a distance that mixes between
    the two: $d\left(x_i, x_{i^{\prime}}\right) = 0.5 d_{\text{Euc}}\left(x_{i},
    x_{i^{\prime}}\right) + 0.5d_{\text{Jac}}\left(x_{i}, x_{i^\prime}\right)$.
    Species now must have both nearby abundances and similar zero patterns, in
    order to appear as neighboring columns in this heatmap.
    \label{fig:heatmap-mix} }
\end{figure}

\begin{figure}
  \centering
  \includegraphics[width=0.9\textwidth]{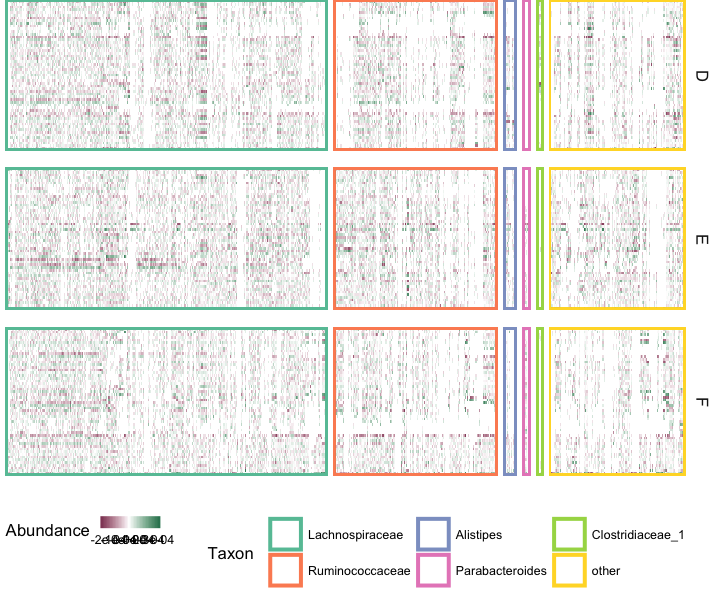}
  \caption{
    A version of Figure \ref{fig:heatmap-euclidean} obtained after clustering on
    the first-differenced series. The shading of cells now corresponds to
    increases (green) or decreases (red) in abundance between neighboring
    timepoints. Species are now considered similar as long as they have similar
    changes from time to time, even when their absolute abundances may not be
    close.
    \label{fig:heatmap-innovations} }
\end{figure}

\begin{figure}
  \centering
  \includegraphics[width=0.9\textwidth]{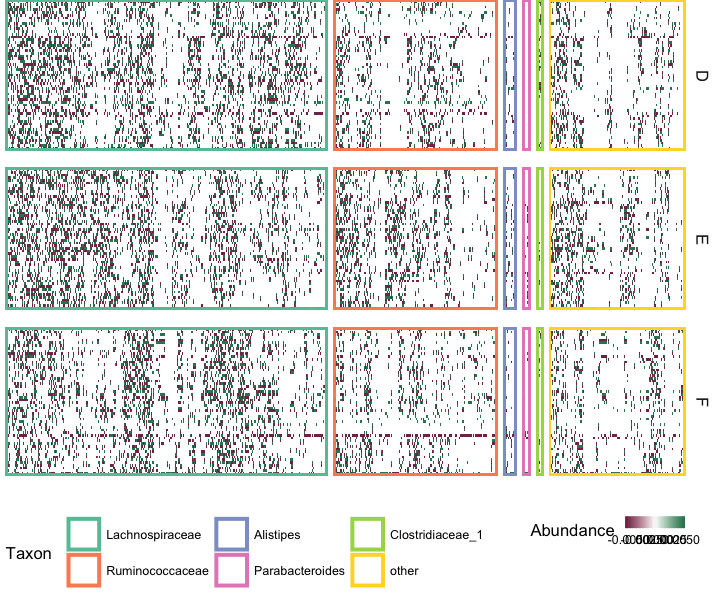}
  \caption{
    The idea in Figure \ref{fig:heatmap-innovations} of comparing first
    differenced series can be coarsened to simply studying whether series
    increased or decreased between neighboring timepoints, ignoring the actual
    value. By computing Manhattan distances between these coarsened differences,
    we can achieve an effect similar to viewing Jaccard distances on the
    original abundances.
    \label{fig:heatmap-innovations-bin} }
\end{figure}

\begin{figure}
  \centering
  \includegraphics[width=0.9\textwidth]{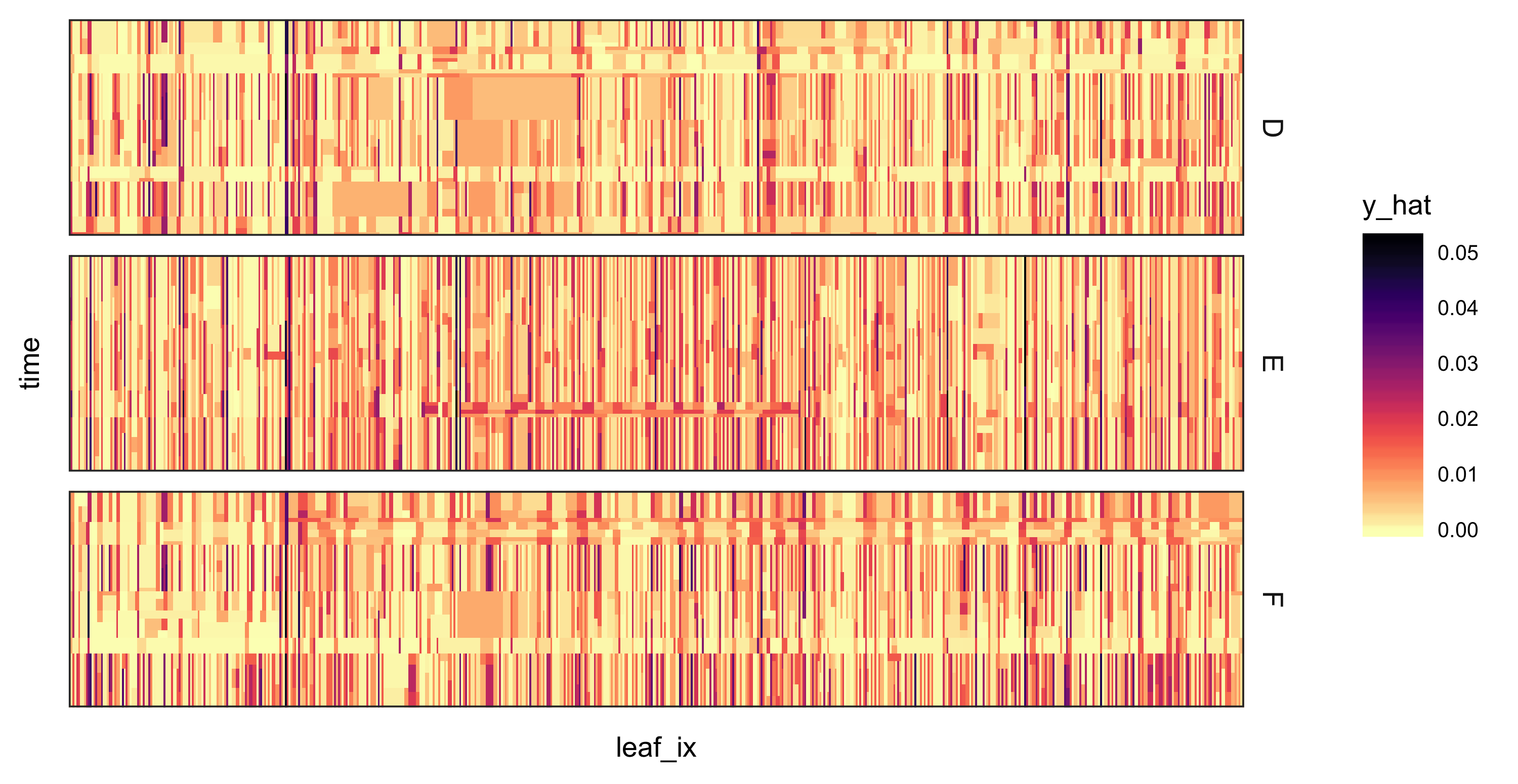}
  \caption{A version of Figure \ref{fig:rpart_complex} made when the penalty $k$
    for having complex trees is very low. The resulting partition is generally
    too rough to be useful, though certain structures -- like the increase in
    abundance among some species during the first antibiotic time course for
    subject D -- to become more visible. In a purely supervised setting, this
    model would be quite overfit, and though we are now applying models for
    exploratory analysis, a similar principle applies.
    \label{fig:rpart_simple}
  }
\end{figure}

\begin{figure}
  \centering
  \includegraphics[width=0.9\textwidth]{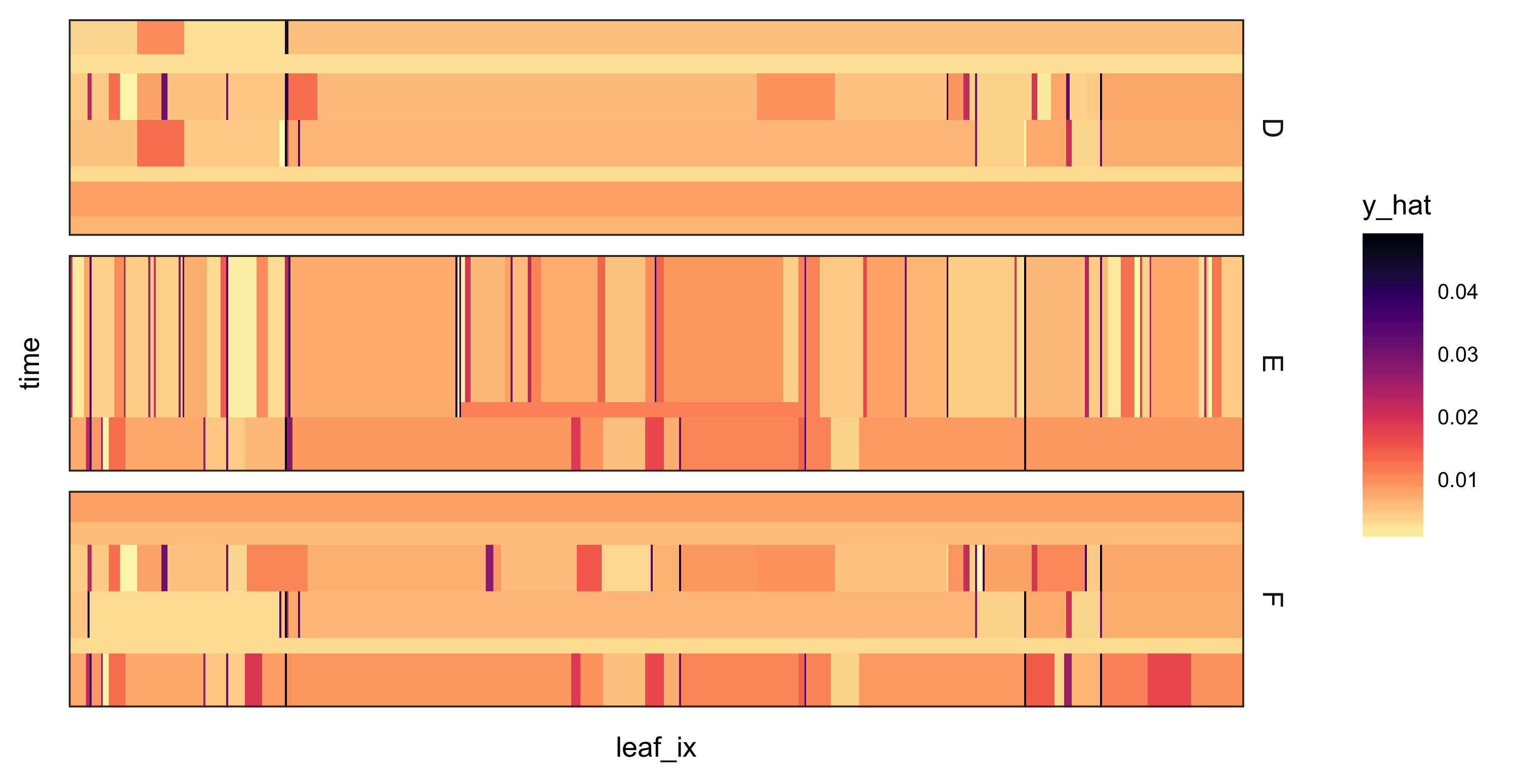}
  \caption{
    A version of Figure \ref{fig:rpart_complex} with much more penalty on the
    complexity of the learned partitions. While Supplementary Figure
    \ref{fig:rpart_simple} displayed a very overfit tree, the tree here seems
    quite underfit, as only a few leaves seem to have survived the pruning
    process. Between this figure and the two mentioned before, we can get a
    sense of the degree to which model tuning can affect visual interpretation.
    \label{fig:rpart_complex_2} }
\end{figure}

\begin{figure}
  \centering
  \includegraphics[width=0.9\textwidth]{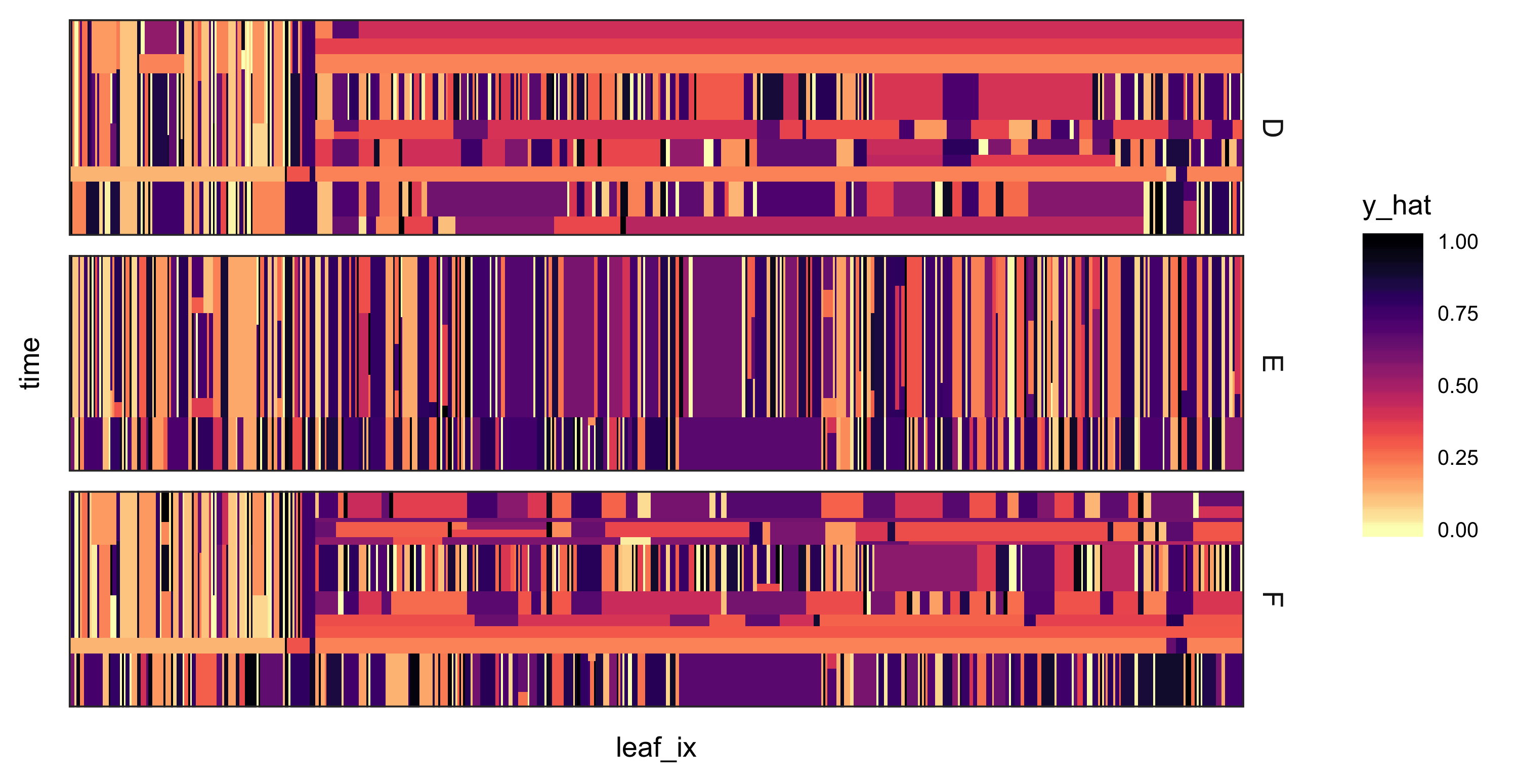}
  \caption{CART partitions across species and timepoints. Each column here
    corresponds to an Amplicon Sequence Variant (ASV), and rows are timepoints.
    The three subjects (D, E, and F) are laid out side by side. Each rectangle
    in the figure represents the leaf node for a CART model fitted on
    presence-absence data, shaded by the fitted probability of being present in
    a given species by timepoint combination.
    \label{fig:rpart_binary_simple}}
\end{figure}

\begin{figure}
  \centering
  \includegraphics[width=0.9\textwidth]{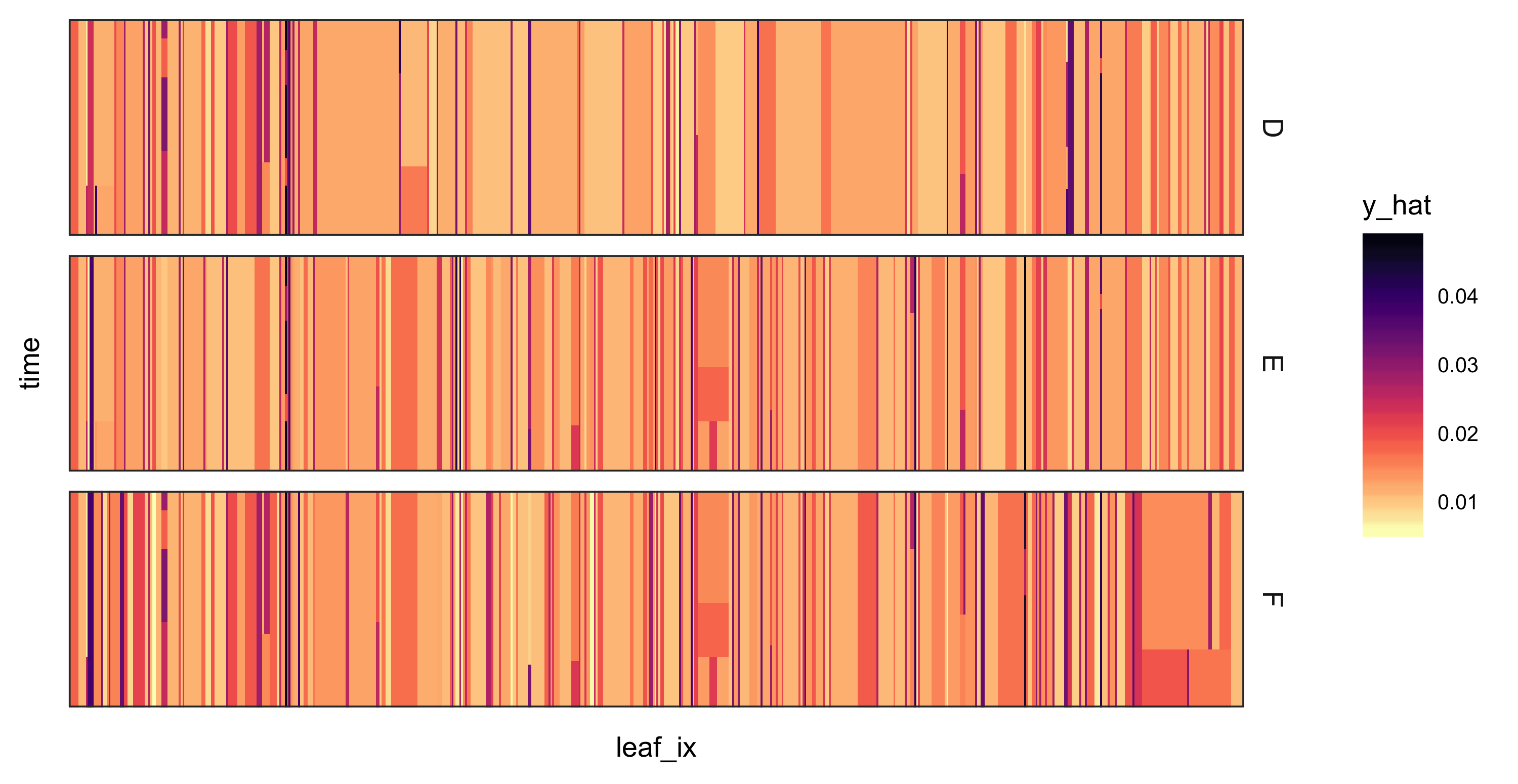}
  \caption{CART partitions when predicting abundances for species $\times$
    timepoint combinations, using a model trained on only nonzero cells of the
    species abundance matrix. The resulting fit can be interpreted as the
    predicted abundance conditional on being present, as in a standard hurdle
    model. \label{fig:rpart_conditional} }
\end{figure}

\begin{figure}
  \centering
  \includegraphics[width=\textwidth]{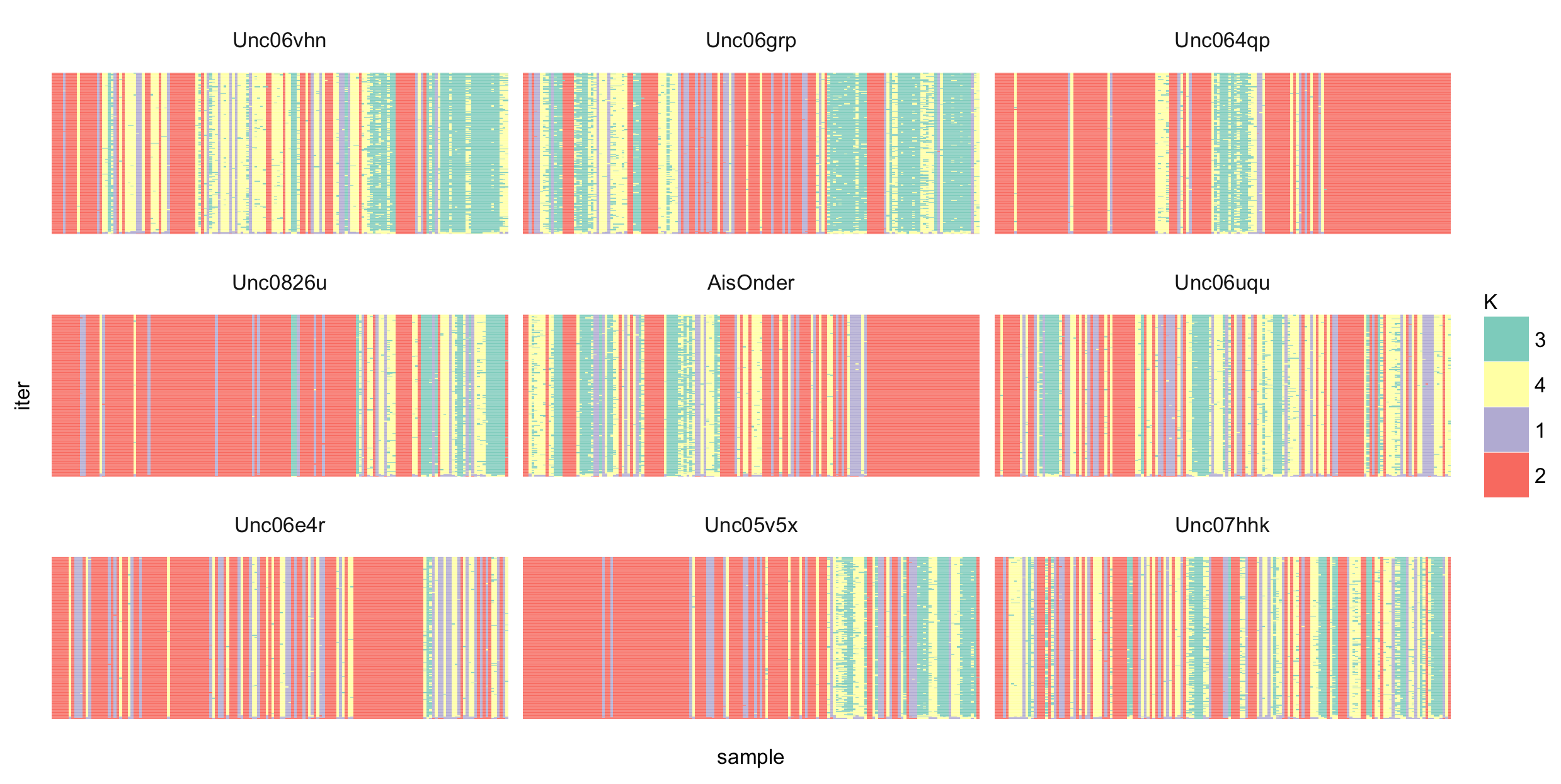}
  \caption{State mixing for the sticky HMM. Each panel represents a separate
    species. Within each panel, columns indicate timepoints, while rows give
    states $z_{it}$ at every $20^{th}$ Gibbs sampling iteration, sorted from
    early (bottom) to late (top). Each color represents a different state. After
    an initial burn-in, most timepoints seem to have a relatively fixed
    state. \label{fig:bayes_gibbs_samples} }
\end{figure}

\begin{figure}
  \centering
  \includegraphics[width=\textwidth]{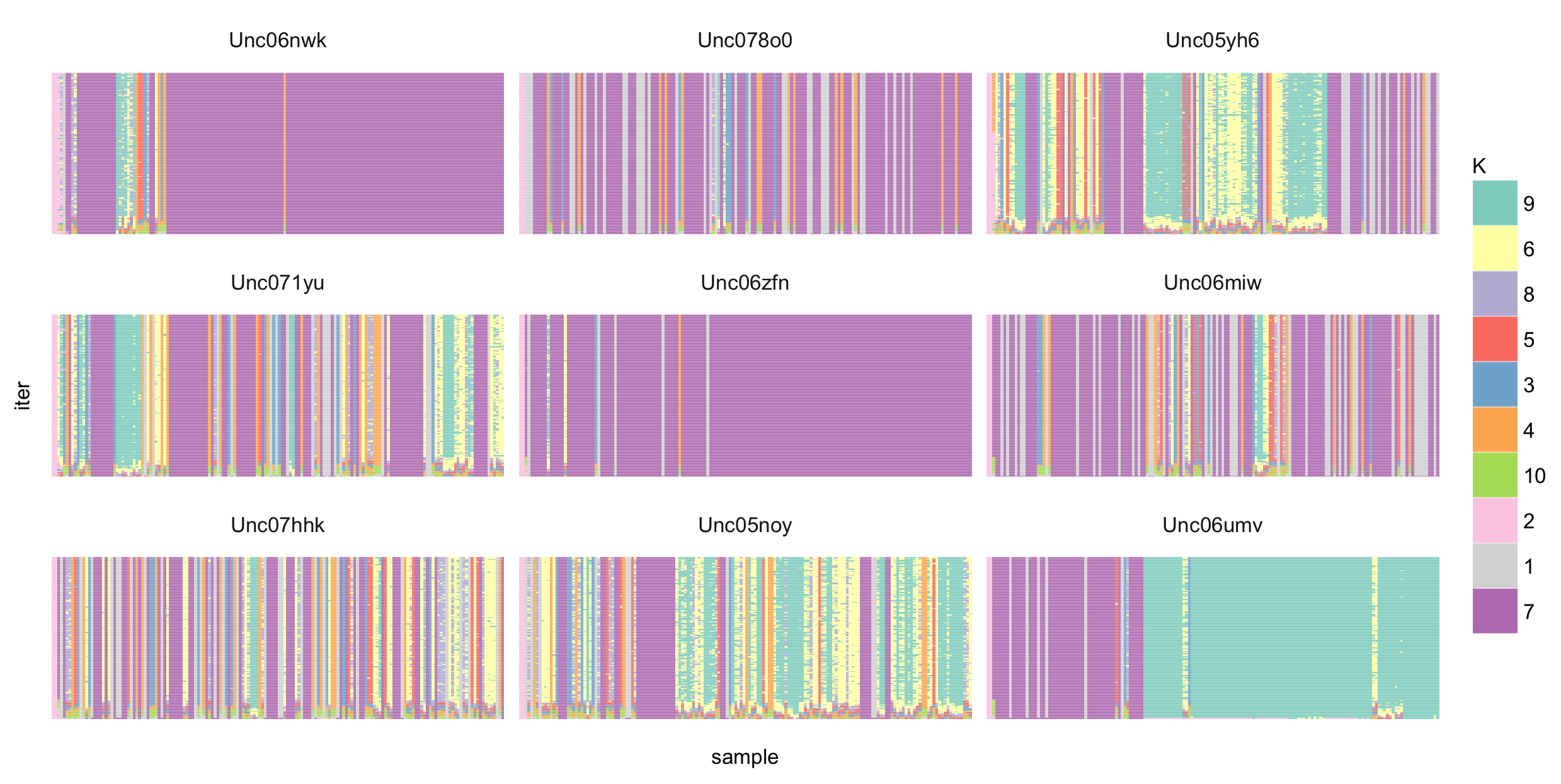}
  \caption{
    A summary of state mixing for the sticky HDP-HMM, read in the same as Figure
    \ref{fig:bayes_gibbs_samples}. There are more states here than for the
    sticky HMM, but nonetheless some timepoints are relatively fixed across
    Gibbs sampling iterations.
    \label{fig:hdp_gibbs_samples} }
\end{figure}

\end{document}

%% file: preamble.tex
\usepackage{amsmath}

\newcommand{\BIT}{\begin{itemize}}
\newcommand{\EIT}{\end{itemize}}
\newcommand{\BNUM}{\begin{enumerate}}
\newcommand{\ENUM}{\end{enumerate}}

\newcommand\mbb[1]{\mathbb{#1}}

\def\mrm#1{\mathrm{#1}}
\def\reals{\mathbb{R}} 
\def\simplex{\mathcal{S}} 
\renewcommand{\exp}[1]{\operatorname{exp}\left(#1\right)} 
\def\indic#1{\mbb{I}\left({#1}\right)} 

\providecommand{\asinh}{\mathop\mathrm{asinh}}
\providecommand{\diag}{\mathop\mathrm{diag}}

\def\E{\mathbb{E}} 
\def\Earg#1{\E\left[{#1}\right]}

\def\Cov{\mrm{Cov}} 
\def\Covarg#1{\Cov\left[{#1}\right]}

\newcommand{\eps}{\epsilon}
\def\absarg#1{\left|#1\right|}

\def\logarg#1{\log\left(#1\right)}
\newcommand\independent{\protect\mathpalette{\protect\independenT}{\perp}}
\def\independenT#1#2{\mathrel{\rlap{$#1#2$}\mkern2mu{#1#2}}}
\def\Gsn{\mathcal{N}}
\def\Ber{\textnormal{Ber}}
\def\Bin{\textnormal{Bin}}

\def\Mult{\textnormal{Mult}}

\def\Gam{\textnormal{Gam}}

\def\Dir{\textnormal{Dir}}

\def\Bet{\textnormal{Beta}}

\def\DP{\textnormal{DP}}
\def\CRP{\textnormal{CRP}}




%% file: chapter_arxiv.bbl
\begin{thebibliography}{43}
\providecommand{\natexlab}[1]{#1}
\providecommand{\url}[1]{\texttt{#1}}
\expandafter\ifx\csname urlstyle\endcsname\relax
  \providecommand{\doi}[1]{doi: #1}\else
  \providecommand{\doi}{doi: \begingroup \urlstyle{rm}\Url}\fi

\bibitem[Breiman et~al.(1984)Breiman, Friedman, Stone, and
  Olshen]{breiman1984classification}
Leo Breiman, Jerome Friedman, Charles~J Stone, and Richard~A Olshen.
\newblock \emph{Classification and regression trees}.
\newblock CRC press, 1984.

\bibitem[Camilleri et~al.(2014)Camilleri, Camilleri, and
  Fabri]{camilleri2014automatic}
Tracey~A Camilleri, Kenneth~P Camilleri, and Simon~G Fabri.
\newblock Automatic detection of spindles and k-complexes in sleep eeg using
  switching multiple models.
\newblock \emph{Biomedical Signal Processing and Control}, 10:\penalty0
  117--127, 2014.

\bibitem[Carter and Kohn(1994)]{carter1994gibbs}
Chris~K Carter and Robert Kohn.
\newblock On gibbs sampling for state space models.
\newblock \emph{Biometrika}, 81\penalty0 (3):\penalty0 541--553, 1994.

\bibitem[Costello et~al.(2012)Costello, Stagaman, Dethlefsen, Bohannan, and
  Relman]{costello2012application}
Elizabeth~K Costello, Keaton Stagaman, Les Dethlefsen, Brendan~JM Bohannan, and
  David~A Relman.
\newblock The application of ecological theory toward an understanding of the
  human microbiome.
\newblock \emph{Science}, 336\penalty0 (6086):\penalty0 1255--1262, 2012.

\bibitem[De~Jong(1997)]{de1997scan}
Piet De~Jong.
\newblock The scan sampler for time series models.
\newblock \emph{Biometrika}, 84\penalty0 (4):\penalty0 929--937, 1997.

\bibitem[Dethlefsen and Relman(2011)]{dethlefsen2011incomplete}
Les Dethlefsen and David~A Relman.
\newblock Incomplete recovery and individualized responses of the human distal
  gut microbiota to repeated antibiotic perturbation.
\newblock \emph{Proceedings of the National Academy of Sciences}, 108\penalty0
  (Supplement 1):\penalty0 4554--4561, 2011.

\bibitem[Donoho and Johnstone(1995)]{donoho1995adapting}
David~L Donoho and Iain~M Johnstone.
\newblock Adapting to unknown smoothness via wavelet shrinkage.
\newblock \emph{Journal of the american statistical association}, 90\penalty0
  (432):\penalty0 1200--1224, 1995.

\bibitem[Doucet et~al.(2000)Doucet, De~Freitas, Murphy, and
  Russell]{doucet2000rao}
Arnaud Doucet, Nando De~Freitas, Kevin Murphy, and Stuart Russell.
\newblock Rao-blackwellised particle filtering for dynamic bayesian networks.
\newblock In \emph{Proceedings of the Sixteenth conference on Uncertainty in
  artificial intelligence}, pages 176--183. Morgan Kaufmann Publishers Inc.,
  2000.

\bibitem[Fan and Mackey(2015)]{fan2015empirical}
Zhou Fan and Lester Mackey.
\newblock Empirical bayesian analysis of simultaneous changepoints in multiple
  data sequences.
\newblock \emph{arXiv preprint arXiv:1508.01280}, 2015.

\bibitem[Faust et~al.(2015)Faust, Lahti, Gonze, de~Vos, and
  Raes]{faust2015metagenomics}
Karoline Faust, Leo Lahti, Didier Gonze, Willem~M de~Vos, and Jeroen Raes.
\newblock Metagenomics meets time series analysis: unraveling microbial
  community dynamics.
\newblock \emph{Current opinion in microbiology}, 25:\penalty0 56--66, 2015.

\bibitem[Fox and Dunson(2012)]{fox2012multiresolution}
Emily Fox and David~B Dunson.
\newblock Multiresolution gaussian processes.
\newblock pages 737--745, 2012.

\bibitem[Fox et~al.(2009)Fox, Jordan, Sudderth, and Willsky]{fox2009sharing}
Emily Fox, Michael~I Jordan, Erik~B Sudderth, and Alan~S Willsky.
\newblock Sharing features among dynamical systems with beta processes.
\newblock In \emph{Advances in Neural Information Processing Systems}, pages
  549--557, 2009.

\bibitem[Fox et~al.(2008)Fox, Sudderth, Jordan, and Willsky]{fox2008hdp}
Emily~B Fox, Erik~B Sudderth, Michael~I Jordan, and Alan~S Willsky.
\newblock An hdp-hmm for systems with state persistence.
\newblock pages 312--319, 2008.

\bibitem[Fox et~al.(2011)Fox, Sudderth, Jordan, and Willsky]{fox2011sticky}
Emily~B Fox, Erik~B Sudderth, Michael~I Jordan, and Alan~S Willsky.
\newblock A sticky hdp-hmm with application to speaker diarization.
\newblock \emph{The Annals of Applied Statistics}, pages 1020--1056, 2011.

\bibitem[Fox(2009)]{fox2009bayesian}
Emily~Beth Fox.
\newblock \emph{Bayesian nonparametric learning of complex dynamical
  phenomena}.
\newblock PhD thesis, Massachusetts Institute of Technology, 2009.

\bibitem[Friedman(2017)]{stat315bnotes}
Jerome Friedman.
\newblock \emph{Statistics 315B Lecture Notes}.
\newblock 2017.

\bibitem[Fr{\"u}hwirth-Schnatter(2006)]{fruhwirth2006finite}
Sylvia Fr{\"u}hwirth-Schnatter.
\newblock \emph{Finite mixture and Markov switching models}.
\newblock Springer Science \& Business Media, 2006.

\bibitem[Ghahramani and Hinton(1998)]{ghahramani1998variational}
Zoubin Ghahramani and Geoffrey~E Hinton.
\newblock Variational learning for switching state-space models.
\newblock 1998.

\bibitem[Gnedin and Kerov(2001)]{gnedin2001characterization}
Alexander Gnedin and Sergei Kerov.
\newblock A characterization of gem distributions.
\newblock \emph{Combinatorics, Probability and Computing}, 10\penalty0
  (3):\penalty0 213--217, 2001.

\bibitem[Heinonen et~al.(2016)Heinonen, Mannerstr{\"o}m, Rousu, Kaski, and
  L{\"a}hdesm{\"a}ki]{heinonen2016non}
Markus Heinonen, Henrik Mannerstr{\"o}m, Juho Rousu, Samuel Kaski, and Harri
  L{\"a}hdesm{\"a}ki.
\newblock Non-stationary gaussian process regression with hamiltonian monte
  carlo.
\newblock In \emph{Artificial Intelligence and Statistics}, pages 732--740,
  2016.

\bibitem[Hostetler and Andreas(1983)]{hostetler1983nonlinear}
Larry Hostetler and Ronald Andreas.
\newblock Nonlinear kalman filtering techniques for terrain-aided navigation.
\newblock \emph{IEEE Transactions on Automatic Control}, 28\penalty0
  (3):\penalty0 315--323, 1983.

\bibitem[Ishwaran and Zarepour(2002)]{ishwaran2002exact}
Hemant Ishwaran and Mahmoud Zarepour.
\newblock Exact and approximate sum representations for the dirichlet process.
\newblock \emph{Canadian Journal of Statistics}, 30\penalty0 (2):\penalty0
  269--283, 2002.

\bibitem[Kimeldorf and Wahba(1970)]{kimeldorf1970correspondence}
George~S Kimeldorf and Grace Wahba.
\newblock A correspondence between bayesian estimation on stochastic processes
  and smoothing by splines.
\newblock \emph{The Annals of Mathematical Statistics}, 41\penalty0
  (2):\penalty0 495--502, 1970.

\bibitem[Lee(2009)]{lee2009optimal}
Hsiang-Tai Lee.
\newblock Optimal futures hedging under jump switching dynamics.
\newblock \emph{Journal of Empirical Finance}, 16\penalty0 (3):\penalty0
  446--456, 2009.

\bibitem[Linderman et~al.(2016)Linderman, Miller, Adams, Blei, Paninski, and
  Johnson]{linderman2016recurrent}
Scott~W Linderman, Andrew~C Miller, Ryan~P Adams, David~M Blei, Liam Paninski,
  and Matthew~J Johnson.
\newblock Recurrent switching linear dynamical systems.
\newblock \emph{arXiv preprint arXiv:1610.08466}, 2016.

\bibitem[Liu(1994)]{liu1994collapsed}
Jun~S Liu.
\newblock The collapsed gibbs sampler in bayesian computations with
  applications to a gene regulation problem.
\newblock \emph{Journal of the American Statistical Association}, 89\penalty0
  (427):\penalty0 958--966, 1994.

\bibitem[Manrique and Shephard(1998)]{manrique1998simulation}
Aurora Manrique and Neil Shephard.
\newblock Simulation-based likelihood inference for limited dependent
  processes.
\newblock \emph{The Econometrics Journal}, 1\penalty0 (1):\penalty0 174--202,
  1998.

\bibitem[Min and Agresti(2002)]{min2002modeling}
Yongyi Min and Alan Agresti.
\newblock Modeling nonnegative data with clumping at zero: a survey.
\newblock \emph{Journal of the Iranian Statistical Society}, 1\penalty0
  (1):\penalty0 7--33, 2002.

\bibitem[Neal(2000)]{neal2000markov}
Radford~M Neal.
\newblock Markov chain sampling methods for dirichlet process mixture models.
\newblock \emph{Journal of computational and graphical statistics}, 9\penalty0
  (2):\penalty0 249--265, 2000.

\bibitem[Paciorek(2003)]{paciorek2003nonstationary}
Christopher~Joseph Paciorek.
\newblock \emph{Nonstationary Gaussian processes for regression and spatial
  modelling}.
\newblock PhD thesis, PhD thesis, Carnegie Mellon University, Pittsburgh,
  Pennsylvania, 2003.

\bibitem[Quinonero-Candela et~al.(2007)Quinonero-Candela, Rasmussen, and
  Williams]{quinonero2007approximation}
Joaquin Quinonero-Candela, Carl~Edward Rasmussen, and Christopher~KI Williams.
\newblock Approximation methods for gaussian process regression.
\newblock \emph{Large-scale kernel machines}, pages 203--224, 2007.

\bibitem[Rabiner and Juang(1986)]{rabiner1986introduction}
Lawrence Rabiner and B~Juang.
\newblock An introduction to hidden markov models.
\newblock \emph{ieee assp magazine}, 3\penalty0 (1):\penalty0 4--16, 1986.

\bibitem[Rasmussen and Ghahramani(2002)]{rasmussen2002infinite}
Carl~E Rasmussen and Zoubin Ghahramani.
\newblock Infinite mixtures of gaussian process experts.
\newblock pages 881--888, 2002.

\bibitem[Rasmussen and Williams(2006)]{rasmussen2006gaussian}
Carl~Edward Rasmussen and Christopher~KI Williams.
\newblock \emph{Gaussian processes for machine learning}, volume~1.
\newblock MIT press Cambridge, 2006.

\bibitem[Samo and Roberts()]{samostring}
Yves-Laurent~Kom Samo and Stephen~J Roberts.
\newblock String gaussian processes.

\bibitem[Stein et~al.(2013)Stein, Bucci, Toussaint, Buffie, R{\"a}tsch, Pamer,
  Sander, and Xavier]{stein2013ecological}
Richard~R Stein, Vanni Bucci, Nora~C Toussaint, Charlie~G Buffie, Gunnar
  R{\"a}tsch, Eric~G Pamer, Chris Sander, and Jo{\~a}o~B Xavier.
\newblock Ecological modeling from time-series inference: insight into dynamics
  and stability of intestinal microbiota.
\newblock \emph{PLoS computational biology}, 9\penalty0 (12):\penalty0
  e1003388, 2013.

\bibitem[Teh(2006)]{teh2006hierarchical}
Yee~Whye Teh.
\newblock A hierarchical bayesian language model based on pitman-yor processes.
\newblock In \emph{Proceedings of the 21st International Conference on
  Computational Linguistics and the 44th annual meeting of the Association for
  Computational Linguistics}, pages 985--992. Association for Computational
  Linguistics, 2006.

\bibitem[Teh and Jordan(2010)]{teh2010hierarchical}
Yee~Whye Teh and Michael~I Jordan.
\newblock Hierarchical bayesian nonparametric models with applications.
\newblock \emph{Bayesian nonparametrics}, 1, 2010.

\bibitem[Tresp(2001)]{tresp2001mixtures}
Volker Tresp.
\newblock Mixtures of gaussian processes.
\newblock In \emph{Advances in neural information processing systems}, pages
  654--660, 2001.

\bibitem[Wan and Van Der~Merwe(2000)]{wan2000unscented}
Eric~A Wan and Rudolph Van Der~Merwe.
\newblock The unscented kalman filter for nonlinear estimation.
\newblock In \emph{Adaptive Systems for Signal Processing, Communications, and
  Control Symposium 2000. AS-SPCC. The IEEE 2000}, pages 153--158. Ieee, 2000.

\bibitem[Weatherley and Mora(2002)]{weatherley2002relationship}
D~Weatherley and P~Mora.
\newblock Relationship between stress heterogeneity and mode-switching dynamics
  of earthquake analogue automata.
\newblock In \emph{AGU Fall Meeting Abstracts}, 2002.

\bibitem[Wei(1999)]{wei1999bayesian}
Steven~X Wei.
\newblock A bayesian approach to dynamic tobit models.
\newblock \emph{Econometric Reviews}, 18\penalty0 (4):\penalty0 417--439, 1999.

\bibitem[Zhou and Carin(2012)]{zhou2012augment}
Mingyuan Zhou and Lawrence Carin.
\newblock Augment-and-conquer negative binomial processes.
\newblock In \emph{Advances in Neural Information Processing Systems}, pages
  2546--2554, 2012.

\end{thebibliography}
